\documentclass[nofootinbib,preprintnumbers]{revtex4}
\usepackage{graphicx}
\usepackage{dcolumn}
\usepackage{amsmath,amssymb,epsfig}
\usepackage{paralist}
\usepackage{comment}
\usepackage{graphicx}
\usepackage{multirow}
\usepackage{wrapfig}

\renewcommand{\vec}[1]{\boldsymbol{\mathrm{#1}}}

\newcommand{\editnote}[2]{}

\allowdisplaybreaks

\begin{document}

\title{General relativistic laser interferometric observables of \\the GRACE-Follow-On mission}

\author{Slava G. Turyshev$^{1}$, Mikhail V. Sazhin$^2$, and Viktor T. Toth$^3$}

\affiliation{\vskip 3pt
$^1$Jet Propulsion Laboratory, California Institute of Technology, 4800 Oak Grove Drive, Pasadena, CA 91109-0899, USA
}%

\affiliation{\vskip 3pt
$^2$Sternberg Astronomical Institute, Lomonosov Moscow State University,  Moscow, Russia
}%

\affiliation{\vskip 3pt
$^3$Ottawa, ON  K1N 9H5, Canada
}%

\date{\today}

\begin{abstract}

We develop a high-precision model for laser ranging interferometric (LRI) observables of the GRACE Follow-On (GRACE-FO) mission. For this, we study the propagation of an electromagnetic wave in the gravitational field in the vicinity of an extended body, in the post-Newtonian approximation of the general theory of relativity. We present a general relativistic model for the phase of a plane wave that accounts for contributions of all the multipoles of the gravitating body, its angular momentum, as well as the contribution of tidal fields produced by external sources. We develop a new approach to model a coherent signal transmission in the gravitational field of the solar system that relies on a relativistic treatment of the phase. We use this approach to describe high-precision interferometric measurements on GRACE-FO and formulate the key LRI observables, namely the phase and phase rate of a coherent laser link between the two spacecraft. We develop a relativistic model for the LRI-enabled range between the two GRACE-FO spacecraft, accurate to less than 1~nm, and a high-precision model for the corresponding range rate, accurate to better than 0.1~nm/s. We also formulate high-precision relativistic models for the double one-way range (DOWR) and DOWR-enabled  range rate observables originally used on GRACE and now studied for interferometric measurements on GRACE-FO.  Our formulation justifies the basic assumptions behind the design of the GRACE-FO mission and highlights the importance of achieving nearly circular and nearly identical orbits for the GRACE-FO spacecraft.
\end{abstract}

\pacs{03.30.+p, 04.25.Nx, 04.80.-y, 06.30.Gv, 95.10.Eg, 95.10.Jk, 95.55.Pe}

\maketitle


\section{Introduction}
\label{sec:intro}

The Gravity Recovery and Climate Experiment (GRACE) was a very successful 2002 US-German space mission. A pair of satellites spent nine years mapping the gravitational field of the Earth using a highly accurate microwave ranging system between the two spacecraft, which were flying in nearly identical orbits. GRACE demonstrated the feasibility of monitoring temporal variations in the Earth's gravitational field and thus detect, for instance, both seasonal variations and long term changes in the hydrosphere.

GRACE's planned successor, tentatively called the GRACE Follow-On (GRACE-FO) mission, is scheduled for launch in 2017. In addition to a microwave ranging system similar to that of GRACE and operating in the Ka-band, GRACE-FO will be equipped with a laser ranging interferometer (LRI) instrument \cite{Pierce-etal:2008}, a system using heterodyne optical interferometry. The LRI is expected to provide range with an accuracy of 1~nm and corresponding estimates for range rate.

The LRI experiment on board GRACE-FO will rely on two-way measurements, designating one spacecraft as master and the other as transponder (we shall call them GRACE-FO-{\rm A} and GRACE-FO-{\rm B}, correspondingly) and utilizing a phase-locked active laser transponder with a frequency offset \cite{Sheard_etal:2012}. Phase locking in the transponder is expected to eliminate transponder laser frequency noise, leaving the transmitter laser frequency noise and phase noise due to pointing errors, experimental features that are captured in the instrument design \cite{Pierce-etal:2008,Sheard_etal:2012}, as the two major noise sources.

Although the original GRACE model formulation was based on Newtonian arguments (see details in \cite{Kim:2000}), from the early stages of the GRACE-FO mission development it became clear that at the level of accuracy expected from GRACE-FO, general relativistic effects may become significant. The two-way nature of the experiment reduces the contribution of some of these effects to the GRACE-FO observables, yet their influence on the modeling of these quantities must be carefully analyzed. The mathematical model of the ultimate LRI observables---time series data obtained at the master spacecraft, recording continuous changes in the phase difference between the local laser oscillator and the laser beam returned by the transponder spacecraft---must take into account general relativistic contributions to the length of the signal path and differences between coordinate time and proper time.

In this paper, we focus on the formulation of a relativistic model for computing and processing the observables of the GRACE-FO mission. We rely on a previously developed theory of relativistic proper reference frames within a system of $N$ extended bodies and the motion of light and test particles in the vicinity of an extended body \cite{Turyshev-Toth:2013}. The organization of the paper is as follows: In Sec.~\ref{sec:em-phase} we discuss light propagation in the vicinity of the extended Earth and derive a general relativistic solution for the phase of an electromagnetic wave. In Sec.~\ref{sec:LRI-obs} we discuss the process of forming the inter-satellite LRI observables of GRACE-FO and derive a model for the phase and the relevant LRI-enabled range observable. We also develop a relativistic model for the frequency observable and a related model for interferometric range rate. We develop equations to model observable fluctuations in the phase rate and range acceleration. In Sec.~\ref{sec:dowr} we discuss the possibility of extracting dual-one-way (DOWR) style observables from the interferometric data, similar to that used by the GRACE-FO microwave ranging system. We develop the relevant equations and models for range, range-rate and range acceleration observations. We conclude with a set of recommendations and an outlook in Sec.~\ref{sec:sonc}.

In an attempt to streamline the presentation of results and to keep the main body of the paper focused, we present some relevant details in the form of appendices. In Appendix~\ref{sec:GR-background} we discuss the post-Newtonian approximation of general relativity, introduce the metric tensor in this formulation, and discuss its properties. In Appendix \ref{sec:ranges} we present details of the derivation of instantaneous distances between the spacecraft. In Appendix \ref{sec:need-for-J2} we emphasize the need to include properly the quadrupole moment of the Earth into relativistic coordinate transformations, equations of light propagation and equations of the motion of geocentric satellites. Finally, in Appendix \ref{sec:useful-rels} we introduce some useful relations that help in the evaluation of the magnitudes of various expressions that involve combinations of orbital parameters of the GRACE-FO spacecraft.

The notational conventions employed in this paper are those used
in
\cite{Landau-Lifshitz:1988}. Letters from the second half of the Latin alphabet, $m, n,...=0...3$ denote spacetime indices. Greek letters $\alpha, \beta,...=1...3$ denote spatial indices. The metric $\gamma_{mn}$ is that of Minkowski spacetime with $\gamma_{mn}={\rm diag}(+1,-1,-1,-1)$ in the Cartesian representation. We employ the Einstein summation convention with indices being lowered or raised using $\gamma_{mn}$. We use powers of $G$ and negative powers of $c$ as bookkeeping devices for order terms. Other notations are explained as they occur.

\section{Phase of an electromagnetic wave in the vicinity of the Earth}
\label{sec:em-phase}

\subsection{The Geocentric Coordinate Reference System}
\label{sec:GCRS}

In the vicinity of the Earth, we utilize a standard coordinate system: the Geocentric Coordinate Reference System (GCRS), centered at the Earth's center of mass is used to track orbits in the vicinity of the Earth. Recently, we developed a new perturbative solution of the gravitational $N$-body problem in general relativity \cite{Turyshev-Toth:2013} and presented a formulation of the proper reference frame associated with an extended and rotating gravitating body. Based on this formulation, we determined the metric tensor $g^{\rm E}_{mn}$ of the non-rotating GCRS \cite{Turyshev:2012nw,Turyshev-Toth:2013}. We denote the coordinates of this reference frame as $\{x^m_{\rm E}\}\equiv(x^0=ct, {\vec x})$ and present the metric tensor $g^{\rm E}_{mn}$ in the following form:
{}
\begin{eqnarray}
g^{\rm E}_{00}&=& 1-\frac{2}{c^2}w_{\rm E}+\frac{2}{c^4}w^2_{\rm E}+O(c^{-6}), \quad
g^{\rm E}_{0\alpha}= -\gamma_{\alpha\lambda}\frac{4}{c^3}w^\lambda_{\rm E}+O(c^{-5}),\quad
g^{\rm E}_{\alpha\beta}= \gamma_{\alpha\beta}+\gamma_{\alpha\beta}\frac{2}{c^2} w_{\rm E}+O(c^{-4}),~~~
\label{eq:gab-E}
\end{eqnarray}
where $w_{\rm E}$ is the scalar harmonic potential that is given by
\begin{eqnarray}
w_{\rm E}&=&U_{\rm E}+ u^{\tt tidal}_{\rm E}+{\cal O}(c^{-3}).~~~
\label{eq:pot_loc-w_0}
\end{eqnarray}
The scalar potential $w_{\rm E}$ is formed as a linear superposition of the gravitational potential $U_{\rm E}$ of the isolated Earth and the tidal potential $u^{\tt tidal}_{\rm E}$ produced by all the solar system bodies (excluding the Earth itself or $b\not={\rm E}$) evaluated at the origin of the GCRS. The Earth's gravitational potential $U_{\rm E}$ at a location defined by spherical coordinates $(r\equiv|{\vec x}|,\phi,\theta)$ is given by
{}
\begin{eqnarray}
U_{\rm E}&=&G\int \frac{\sigma(t,{\vec x'})d^3x'}{|{\vec x}-{\vec x'}|}+{\cal O}(c^{-4})=\frac{GM_{\rm E}}{r}\Big(1+\sum_{\ell=2}^\infty\sum_{k=0}^{+\ell}\Big(\frac{R_{\rm E}}{r}\Big)^\ell P_{\ell k}(\cos\theta)(C^{{\rm E}}_{\ell k}\cos k\phi+S^{{\rm E}}_{\ell k}\sin k\phi)\Big)+
{\cal O}(c^{-4}),
\label{eq:pot_w_0}
\end{eqnarray}
where $\sigma(t,{\vec x'})$ is the relativistic mass density inside the Earth (see discussion in Appendix~\ref{sec:GR-background}),  $M_{\rm E}$ is the Earth's mass, $R_{\rm E}$ is its radius, $P_{\ell k}$ are the Legendre polynomials, while $C^{{\rm E}}_{\ell k}$ and $S^{{\rm E}}_{\ell k}$ are normalized spherical harmonic coefficients that characterize the Earth. While the GRACE-FO experiment is designed to model gravitational harmonics of the Earth at a fine spatial resolution, at the level of sensitivity of the project, only the lowest order harmonics are affected by relativistic contributions, and time-dependent contributions to relativistic effects due to the elasticity of the Earth can be ignored. Insofar as the tidal potential $u^{\tt tidal}_{\rm E}$ is concerned, for GRACE-FO it is sufficient to keep only its Newtonian contribution (primarily due to the Sun and the Moon) which can be given as usual:
{}
\begin{eqnarray}
u^{\tt tidal}_{\rm E}&=&\sum_{b\not={\rm E}}\Big(U_b({\vec r}^{}_{b{\rm E}}+\vec{x})-U_b({\vec r}^{}_{b{\rm E}}) - \vec{x}\cdot {\vec \nabla} U_b ({\vec r}^{}_{b{\rm E}})\Big)\simeq\sum_{b\not={\rm E}}\frac{GM_b}{2r^3_{b{\rm E}}}\Big(3(\vec{n}^{}_{b{\rm E}}\cdot\vec{x})^2-\vec{x}^2\Big)+
{\cal O}(r^{-4}_{b{\rm E}},c^{-2}),
\label{eq:u-tidal-E}
\end{eqnarray}
where $U_b$ is the Newtonian gravitational potential of body $b$, $\vec{r}_{b{\rm E}}$ is the vector connecting the center of mass of body $b$ with that of the Earth, and $\vec \nabla U_b$ denotes the gradient of the potential. Note that in Eq.~(\ref{eq:u-tidal-E}) we omitted relativistic tidal contributions of ${\cal O}(c^{-2})$ that are produced by the external gravitational potentials. These are of the order of $10^{-16}$ compared to $U_{\rm E}$ and, thus, negligible even at the level of accuracy of the GRACE-FO LRI. We present only the largest term in the tidal potential, which is of the order of $\sim r^{-3}_{b{\rm E}}$; however, using the explicit form of this potential on the left side of  Eq.~(\ref{eq:u-tidal-E}), one can easily evaluate this expression to any order needed to solve a particular problem.

Finally, the contribution of the body's rotation is captured by the vector harmonic  potential, $w^\alpha_{\rm E}$, defined as:
{}
\begin{eqnarray}
w^\alpha_{\rm E}
&=&G\int \frac{\sigma^\alpha(t,\vec{x}')d^3x'}{|\vec{x}-\vec{x}'|}+{\cal O}(c^{-2})=-\frac{GM_{\rm E}}{2r^3}[{\vec x}\times{\vec S}_{\rm E}]^\alpha+{\cal O}(r^{-4}, c^{-2}),
\label{eq:pot_loc-w_a+}
\end{eqnarray}
where $\sigma^\alpha(t,\vec{x}')$ is the relativistic current density of the matter distribution inside the rotating Earth. Also, in (\ref{eq:pot_loc-w_a+}) we explicitly account only for the largest rotational moment,  ${\vec S}_{\rm E}$, which is the Earth's spin moment (angular momentum per unit of mass). The contribution of other vector harmonics due to rotation of the Earth is negligible.

The metric tensor (\ref{eq:gab-E}) with the gravitational potentials (\ref{eq:pot_loc-w_0})--(\ref{eq:pot_loc-w_a+}) represents spacetime in the GCRS, which we choose to formulate the relativistic model for GRACE-FO observables. Details on the formulation of the GCRS are in \cite{Soffel:2003cr,Turyshev:2012nw,Turyshev-Toth:2013}.

\subsection{Geometric optics approximation for the wave propagation in the vicinity of the Earth}
\label{sec:geom-optics}

The phase of an electromagnetic wave is a scalar function that is invariant under a set of general coordinate transformations. In the geometric optics approximation, the phase $\varphi$ is found as a solution to the eikonal equation \cite{Fock-book:1959,Landau-Lifshitz:1988,Sazhin:1998,Kopeikin:2009,Kopeikin-book-2011}:
\begin{equation}
g^{mn}\partial_m\varphi\partial_n\varphi=0,
\label{eq:eq_eik}
\end{equation}
which is a direct consequence of Maxwell's equations. Its solution describes the wavefront of an electromagnetic wave propagating in curved spacetime. The solution's geometric properties are defined by the metric tensor $g_{mn}$ which is derived as the solution of Einstein's field equations. In the vicinity of the Earth this tensor is given by Eqs.~(\ref{eq:gab-E})--(\ref{eq:u-tidal-E}).

To solve Eq.~(\ref{eq:eq_eik}), we introduce a covector of the electromagnetic wavefront in curved spacetime, $K_m = \partial_m\varphi$. We use $\lambda$ to denote an affine parameter along the trajectory of a light ray being orthogonal to the wavefront $\varphi$ (note that the dimension of $\lambda$ is (length)$^2$). The vector $K^m = dx^m/d\lambda = g^{mn}\partial_n\varphi$ is tangent to the light ray. Equation~(\ref{eq:eq_eik}) states that $K^m$ simply is null or $g_{mn}K^mK^n = 0$. Therefore, the light rays are null geodesics \cite{Landau-Lifshitz:1988} described by
\begin{equation}
\frac{dK_m}{d\lambda} = \frac{1}{2}\partial_m g_{kl}K^kK^l.
\label{eq:eq_eik-K}
\end{equation}
Since the eikonal and light ray equations, given by Eqs.~(\ref{eq:eq_eik}) and (\ref{eq:eq_eik-K}) respectively, have equivalent physical content in the general theory of relativity, one can use either of them to study the properties of an electromagnetic wave. However, the eikonal equation offers a more straightforward way to study the propagation of a wave.

To find a solution of Eq.~(\ref{eq:eq_eik}), we expand the eikonal $\varphi$ with respect to the gravitational constant $G$ assuming that the unperturbed solution is a plane wave. The expansion may be given as
\begin{equation}
\varphi(t,{\vec x}) = \varphi_0+\int k_m dx^m+\varphi_G (t,{\vec x})+{\cal O}(G^2),
\label{eq:eq_eik-phi}
\end{equation}
where $\varphi_0$ is an integration constant and $k_m = k^0(1, {\vec k})$ is a constant (with respect to the Minkowski metric) null vector (i.e., $\gamma_{mn}k^mk^n=0$) along the direction of propagation of the unperturbed electromagnetic plane wave. The wave direction is given by the vector ${\vec k}\equiv k^\epsilon$, which is the unit vector along the ray's path, $|{\vec k}|=1$. Furthermore, $k^0=\omega/c$, where $\omega$ is the constant angular frequency of the unperturbed wave, and $\varphi_G$ is the perturbation of the eikonal of first order in $G$, which is yet to be determined. Also, as a consequence of Eq.~(\ref{eq:eq_eik-phi}), the wave vector of an electromagnetic wave in curved spacetime, $K^m(t,{\vec x})$, admits a series expansion with respect to $G$ in the form
\begin{equation}
K^m(t,{\vec x})=\frac{dx^m}{d\lambda}\equiv g^{mn}\partial_n\varphi=k^m+k_G^m(t,{\vec x})+{\cal O}(G^2),
\label{eq:K}
\end{equation}
where $k^m_G(t,{\vec x})=\gamma^{mn}\partial_n\varphi_G(t,{\vec x})$ is the first order perturbation of the wave vector with respect to $G$.

To solve Eqs.~(\ref{eq:eq_eik}) and (\ref{eq:eq_eik-phi}) for $\varphi_G$ in the GCRS, we first substitute (\ref{eq:eq_eik-phi}) into (\ref{eq:eq_eik}). Then, defining $h^{mn}=g^{mn}-\gamma^{mn}$ (as discussed in (\ref{eq:gmn})) with $g_{mn}$ given by Eqs.~(\ref{eq:gab-E})--(\ref{eq:u-tidal-E}) and keeping only first order terms in $G$ , we obtain an ordinary differential equation to determine $\varphi_G$:
\begin{equation}
\frac{d\varphi_G}{d\lambda}= -\frac{1}{2}h^{mn}k_mk_n = -\frac{2k_0^2}{c^2}w_{\rm E}- \frac{4k_0^2}{c^3}(k_\epsilon w^\epsilon_{\rm E})+{\cal O}(G^2),
\label{eq:eq_eik-phi-lamb}
\end{equation}
where ${d\varphi_G}/{d\lambda}= k_m\partial^m\varphi $.  Eq.~(\ref{eq:eq_eik-phi-lamb}) alternatively can also be obtained by integrating the null geodesic equation (\ref{eq:eq_eik-K}). Substituting the scalar and vector potentials $w_{\rm E}$ and  $w^\lambda_{\rm E}$ from Eqs.~(\ref{eq:pot_loc-w_0})--(\ref{eq:u-tidal-E}), we obtain:
{}
\begin{eqnarray}
\frac{d\varphi_G}{d\lambda}&=& \frac{d\varphi^{\rm E}_G}{d\lambda}+\frac{d\varphi^{\rm S}_G}{d\lambda}+\frac{d\varphi^{\rm tidal}_G}{d\lambda},
\label{eq:eq_eik-phi-lamb-a}
\end{eqnarray}
where the three terms of the relativistic phase due to the mass multipole moments of Earth's gravity, $\varphi^{\rm E}_G$ (determined by the potential (\ref{eq:pot_w_0})), the contribution due to the Earth's rotation, $\varphi^{\rm S}_G$ (due to the potential (\ref{eq:pot_loc-w_a+})), and the tidal gravitational field of external bodies in the GCRS, $\varphi^{\rm tidal}_G$ (due to (\ref{eq:u-tidal-E})), are determined from the following equations:
{}
\begin{eqnarray}
\frac{d\varphi^{\rm E}_G}{d\lambda}&=& -\frac{2k_0^2G}{c^2}\int \frac{\sigma_{\rm E}(t,{\vec x'})d^3x'}{|{\vec x}-{\vec x'}|}+{\cal O}(G^2),
\label{eq:eq_eik-phi-lamb-E}\\
\frac{d\varphi^{\rm S}_G}{d\lambda}&=& -\frac{4k_0^2G}{c^3}k_\epsilon\int \frac{\sigma^\epsilon_{\rm E}(t,{\vec x'})d^3x'}{|{\vec x}-{\vec x'}|}+
{\cal O}(G^2),
\label{eq:eq_eik-phi-lamb-S}\\
\frac{d\varphi^{\rm tidal}_G}{d\lambda}&=&
-\sum_{b\not={\rm E}}\frac{GM_b}{c^2}\frac{k_0^2}{r^3_{b{\rm E}}}\Big(3(\vec{n}^{}_{b{\rm E}}\cdot\vec{x})^2-\vec{x}^2\Big)+{\cal O}(r^{-4}_{b{\rm E}}, G^2).
\label{eq:eq_eik-phi-lamb-T}
\end{eqnarray}

We represent the light ray's trajectory, correct to the Newtonian order, as
\begin{equation}
\{x^m\}\equiv\Big(x^0=ct, ~~{\vec x}(t)={\vec x}_{\rm 0}+{\vec k} c(t-t_0)\Big)+{\cal O}(G).
\label{eq:light-traj}
\end{equation}
This representation allows us to express the Newtonian part of the wave vector $K^m$ presented by Eq.~(\ref{eq:K}) as follows:
$k^m= {dx^m}/{d\lambda} =k^0\big(1, {\vec k}\big)+{\cal O}(G)$, where $k^0$ is immediately derived as $k^0={cdt}/{d\lambda}+{\cal O}(G)$ and $|{\vec k}|=1$. Keeping in mind that $k^m$ is constant, we establish an important relationship:
\begin{equation}
d\lambda= \frac{cdt}{k^0}+{\cal O}(G),
\label{eq:eq_eik-relat}
\end{equation}
which we will use to integrate (\ref{eq:eq_eik-phi-lamb-E}).
This expression allows including contributions from all multipoles of the Earth's mass distribution, as given in Eq.~(\ref{eq:pot_w_0}). Using (\ref{eq:light-traj}) we present the right-hand side of Eq.~({\ref{eq:eq_eik-phi-lamb-E}) as:
\begin{eqnarray}
\frac{d\varphi^{\rm E}_G}{d\lambda}&=& -\frac{2k_0^2G}{c^2}\int \frac{\sigma_{\rm E}(t,{\vec x'})d^3x'}{|{\vec x}_{\rm 0}+{\vec k} c(t-t_{\rm 0})-{\vec x'}|}.
\label{eq:eq_eik-phi-lamb-E+}
\end{eqnarray}
In this and preceding equations, we presented the density $\sigma(t,\vec{x}')$ as a time-dependent quantity, reflecting on the fact that the GRACE-FO is intended, among other things, to study the temporal evolution of the Earth's gravitational field due to shifting masses. During the $\sim 1$~ms propagation time of a light signal between the two spacecraft, however, changes in mass distribution inside the Earth are completely negligible, and on these timescales, we can safely assume that the matter distribution is static. In other words, the characteristic time for changes occurring inside the Earth are much longer than the light transit time. This allows us to integrate Eq.~(\ref{eq:eq_eik-phi-lamb-E+}) as if the density is static, allowing us to treat $\sigma(t,{\vec x'})$ as a time-independent quantity. Under these assumptions and relying on Eqs.~(\ref{eq:light-traj})--(\ref{eq:eq_eik-relat}), we integrate Eq.~(\ref{eq:eq_eik-phi-lamb-E+}) with respect to time from $t_{\rm 0}$ to $t$ and write a plane wave solution that includes the Earth's gravity contribution to the waveform in the following form:
\begin{eqnarray}
\varphi^{\rm E}_G(t,{\vec x}) &=& -\frac{2k_0G}{c^2}\int \sigma_{\rm E}(t,{\vec x'})\ln\Big[\frac{|{\vec x}-{\vec x'}|+{\vec k} \cdot({\vec x}-{\vec x'})}{|{\vec x_{\rm 0}}-{\vec x'}|+{\vec k} \cdot({\vec x_{\rm 0}}-{\vec x'})}\Big]d^3x'+{\cal O}(G^2).~~~
\label{eq:eik-sol-all}
\end{eqnarray}
The resulting solution extends all previously obtained solutions for the gravitational delay of light by accounting for the contributions from all the multipoles of the extended body. As the density $\sigma(t,{\vec x'})$ may be thought of as a collection of a large number of elementary mass monopoles, the integral in (\ref{eq:eik-sol-all}) is a sum of the corresponding Shapiro delays produced by each of these elementary masses integrated over the mass distribution.

Eq.~(\ref{eq:eik-sol-all}) allows one to  calculate the contributions to the wavefront due to any mass multipole moment. However, for the purposes of the analysis of the GRACE-FO mission, we will only keep contributions from the largest multipoles, namely, the monopole $M_{\rm E}$ and quadrupole moments $J^{\epsilon\lambda}_{\rm E}$. (In fact, we have derived the expression and estimated the magnitude of the octopole term, $J_{\rm E}^{\alpha\beta\epsilon}$, which turned to be negligible for GRACE-FO). To do this, we expand this expression under the integral sign. Using the fact that outside the body of the Earth $|{\vec x}'|<|{\vec x}|$, we obtain:
{}
\begin{equation}
\ln\Big(|{\vec x}-{\vec x'}|+{\vec k} \cdot({\vec x}-{\vec x'})\Big)=\ln\big(r+{\vec k} \cdot{\vec x}\big)+\frac{(n_\epsilon+k_\epsilon)x'^{\epsilon}}{r+{\vec k} \cdot{\vec x}}-
{\textstyle\frac{1}{2}}\Big[\frac{(n_\epsilon+k_\epsilon)(n_\lambda+k_\lambda)}{(r+{\vec k}\cdot{\vec x})^2}+\frac{1}{r}\frac{\gamma_{\epsilon\lambda}+n_\epsilon n_\lambda}{(r+{\vec k}\cdot{\vec x})}\Big]x'^{\epsilon}x'^{\lambda}+{\cal O}(x'^3).
\label{eq:log-expand}
\end{equation}
Similarly, one can develop an expression for $\ln\big(|{\vec x}_0-{\vec x'}|+{\vec k} \cdot({\vec x}_0-{\vec x'})\big)$, when ${\vec x}_0$ is outside the body and $|{\vec x}'|<|{\vec x}_0|$.

We can now integrate (\ref{eq:eik-sol-all}) over the body's volume using a spherical harmonics expansion of the Earth's gravity potential (\ref{eq:pot_w_0}) produced by the density of matter distribution inside the Earth, $\sigma_{\rm E}(t,{\vec x})$, where the mass, $M_{\rm E}$, dipole moment, $d^{\epsilon}_{\rm E}$, quadrupole moment, $J^{\epsilon\lambda}_{\rm E}$, and spin moment, $S^{\alpha\beta}_{\rm E}$, of the Earth's gravitational field are defined as:
\begin{eqnarray}
M_{\rm E}=\int d^3x'\sigma_{\rm E}(t,{\vec x}'), ~~~
d^{\epsilon}_{\rm E}&=&M_{\rm E}^{-1}\int d^3x'\sigma_{\rm E}(t,{\vec x}')x'^\epsilon \equiv0,~~~
J^{\epsilon\lambda}_{\rm E}=M_{\rm E}^{-1}\int d^3x'\sigma_{\rm E}(t,{\vec x}')\big(3x'^\epsilon x'^\lambda
+\gamma^{\epsilon\lambda}r'^2\big),
\nonumber\\
S^{\alpha\beta}_{\rm E}&=&M_{\rm E}^{-1}\int d^3x'\sigma_{\rm E}(t,{\vec x}')\big(v'^\alpha x'^\beta-v'^\beta x'^\alpha\big)=
\gamma_{\mu\nu}\epsilon^{\alpha\beta}_{~~\mu}S^{\mu}_{\rm E},
\label{eq:mass-spin-moments}
\end{eqnarray}
where $\epsilon_{\alpha\mu\nu}$ is the fully antisymmetric Levi-Civita symbol, $\epsilon_{123}=1$, and $S^{\mu}_{\rm E}$ is the spin moment of the Earth.

As a result, the expression describing the contribution of mass multipoles of Earth's gravity to the waveform has the following form:
{}
\begin{eqnarray}
\varphi^{\rm E}_G(t,{\vec x}) &=& -\frac{2GM_{\rm E}}{c^2}k_0\Big\{\ln\Big[\frac{r+{\vec k}\cdot{\vec x}}{r_0+{\vec k}\cdot{\vec x}_0}\Big]-\nonumber\\
&-&{\textstyle\frac{1}{6}}\Big[\frac{(n_\epsilon+k_\epsilon)(n_\lambda+k_\lambda)}{(r+{\vec k}\cdot{\vec x})^2}+\frac{1}{r}\frac{\gamma_{\epsilon\lambda}+n_\epsilon n_\lambda}{(r+{\vec k}\cdot{\vec x})}-
\frac{(n_{0\epsilon}+k_\epsilon)(n_{0\lambda}+k_\lambda)}{(r_0+{\vec k}\cdot{\vec x}_0)^2}
-\frac{1}{r_0}\frac{\gamma_{\epsilon\lambda}+n_{0\epsilon} n_{0\lambda}}{(r_0+{\vec k}\cdot{\vec x}_0)} \Big]J_{\rm E}^{\epsilon\lambda}\Big\}+{\cal O}(G^2).~~~~
\label{eq:eik-sol-E-J}
\end{eqnarray}
The solution for the contribution of the quadrupole moment to the relativistic delay generalizes similar expressions obtained by other means and under much simplifying assumptions on $J_{\rm E}^{\epsilon\lambda}$, notably \cite{LePoncinLafitte:2007tx,Teyssandier:2007vg}. Although Eq.~(\ref{eq:log-expand}) naturally yields an introduction of a moment of inertia, $I^{\epsilon\lambda}_{\rm E}=M_{\rm E}^{-1}\int d^3x'\sigma_{\rm E}(t,{\vec x}')x'^\epsilon x'^\lambda$, we, nevertheless, for consistency reasons \cite{Fock-book:1959,Landau-Lifshitz:1988},  have introduced the quadrupole mass moment $J^{\epsilon\lambda}_{\rm E}$ in (\ref{eq:mass-spin-moments}). As the expression in front of $x'^{\epsilon}x'^{\lambda}$ in Eq.~(\ref{eq:log-expand}) is trace-free, the quadrupole term in (\ref{eq:eik-sol-E-J}) has an additional factor of ${\textstyle\frac{1}{3}}$.

Similarly to the approach that led to the solution (\ref{eq:eik-sol-E-J}), by assuming a constant rotation,  we represent the mass current as $\sigma^\alpha_{\rm E}(t,{\vec x})=\sigma_{\rm E}(t,{\vec x})v^\alpha_{\rm E}(t,{\vec x})$, and now can integrate Eq.~(\ref{eq:eq_eik-phi-lamb-S}) as:
{}
\begin{eqnarray}
\varphi^{\rm S}_G(t,{\vec x}) &=& -\frac{4k_0G}{c^3}k_\epsilon\int \sigma_{\rm E}(t,{\vec x'})v^\epsilon_{\rm E}(t,{\vec x'})\ln\Big[\frac{|{\vec x}-{\vec x'}|+{\vec k} \cdot({\vec x}-{\vec x'})}{|{\vec x_{\rm 0}}-{\vec x'}|+{\vec k} \cdot({\vec x_{\rm 0}}-{\vec x'})}\Big]d^3x'+{\cal O}(G^2).~~~
\label{eq:eik-sol-all-S}
\end{eqnarray}
Assuming further a constant rotation with frequency $\omega^\alpha_{\rm E}$ allows us to express the velocity field inside the Earth as $v^\alpha_{\rm E}({\vec x})=\epsilon^\alpha_{\lambda\mu}\omega_{\rm E}^\lambda x^\mu$. Also, using the fact that $({\vec k}\cdot[{\vec k}\times{\vec S}_{\rm E}])=0$, we can determine the contribution due to the spin moment of the Earth, $\varphi^{\rm S}_G$, in the form
{}
\begin{eqnarray}
\varphi^{\rm S}_G(t,{\vec x})&=& -\frac{2GM_{\rm E}}{c^3}k_0\Big({\vec k}\cdot\Big[{\vec S}_{\rm E}\times\big(\frac{{\vec n}}{r+{\vec k}\cdot{\vec x}}-\frac{{\vec n}_0}{r_0+{\vec k}\cdot{\vec x}_0}\big)\Big]\Big)+
{\cal O}(r^{-2}, G^2).
\label{eq:eq_eik-phi-lamb-S2}
\end{eqnarray}
Although, analogously to Eq.~(\ref{eq:eik-sol-all}), we can use (\ref{eq:eik-sol-all-S}) to extend (\ref{eq:eq_eik-phi-lamb-S2}) to include contributions from current moments of an arbitrary order, we will limit ourselves to the contribution of the lowest, first order moment only (i.e., spin moment), as the effect of higher order current moments on the relativistic delay of light in the solar system is negligibly small.

The contribution of external gravitational fields can be obtained by integrating (\ref{eq:eq_eik-phi-lamb-T}) as follows:
{}
\begin{eqnarray}
\varphi^{\rm tidal}_G(t,{\vec x})&=& -\sum_{b\not={\rm E}}\frac{GM_b}{2c^2}\frac{\gamma_{\epsilon\lambda}+3n^{}_{b{\rm E}\epsilon}n^{}_{b{\rm E}\lambda}}{r^3_{b{\rm E}}}
k_0\big({\vec k}\cdot({\vec x}-{\vec x}_0)\big)\Big(x^\epsilon x^\lambda_0+x^\lambda x^\epsilon_0+{\textstyle\frac{2}{3}}(x^\epsilon-x^\epsilon_0)(x^\lambda-x^\lambda_0)\Big)+{\cal O}(r^{-4}_{b{\rm E}}, G^2).~~~~
\label{eq:eq_eik-phi-lamb-T1}
\end{eqnarray}

We can now write the post-Minkowskian expansion for the phase of an electromagnetic wave that propagates in the vicinity of the extended and rotating gravitating body. In the body's proper reference frame (a formulation that accounts for the presence of the external gravity field produced by the external bodies of the $N$-body system \cite{Turyshev:2012nw,Turyshev-Toth:2013}), collecting all the appropriate contributions  coming from the Earth's mass distribution $\varphi^{\rm E}_G$, Earth's rotation $\varphi^{\rm S}_G$, and external gravity $\varphi^{\rm tidal}_G$, the total phase Eq.~(\ref{eq:eq_eik-phi}) has the form:
{}
\begin{equation}
\varphi(t,{\vec x}) = \varphi_0+\int k_m dx^m+\varphi^{\rm E}_G(t,{\vec x})+\varphi^{\rm S}_G(t,{\vec x})+\varphi^{\rm tidal}_G(t,{\vec x})+{\cal O}(G^2),
\label{eq:eq_eik-phi+}
\end{equation}
which, with the help of solutions represented by Eqs.~(\ref{eq:eik-sol-E-J}), (\ref{eq:eq_eik-phi-lamb-S2}), and
(\ref{eq:eq_eik-phi-lamb-T1}), can be given as
{}
\begin{eqnarray}
\varphi(t,{\vec x}) &=& \varphi_0+k_0\Big(c(t-t_0)-{\vec k}\cdot ({\vec x}-{\vec x}_0)-\nonumber\\
&-& \frac{2GM_{\rm E}}{c^2}\Big\{\ln\Big[\frac{r+{\vec k}\cdot {\vec x}}{r_0+{\vec k}\cdot {\vec x}_0}\Big]+
\frac{1}{c}\Big({\vec k}\cdot\big[{\vec S}_{\rm E}\times\big(\frac{{\vec n}}{r+{\vec k}\cdot{\vec x}}-\frac{{\vec n}_0}{r_0+{\vec k}\cdot{\vec x}_0}\big)\big]\Big)+\nonumber\\
&&\hskip 35pt-
{\textstyle\frac{1}{6}}\Big[\frac{(n_\epsilon+k_\epsilon)(n_\lambda+k_\lambda)}{(r+{\vec k}\cdot{\vec x})^2}+\frac{1}{r}\frac{\gamma_{\epsilon\lambda}+n_\epsilon n_\lambda}{(r+{\vec k}\cdot{\vec x})}-
\frac{(n_{0\epsilon}+k_\epsilon)(n_{0\lambda}+k_\lambda)}{(r_0+{\vec k}\cdot{\vec x}_0)^2}
-\frac{1}{r_0}\frac{\gamma_{\epsilon\lambda}+n_{0\epsilon} n_{0\lambda}}{(r_0+{\vec k}\cdot{\vec x}_0)} \Big]J_{\rm E}^{\epsilon\lambda}\Big\}-\nonumber\\
&-& \sum_{b\not={\rm E}}\frac{GM_b}{2c^2}\frac{\gamma_{\epsilon\lambda}+3n^{}_{b{\rm E}\epsilon}n^{}_{b{\rm E}\lambda}}{r^3_{b{\rm E}}}
\big({\vec k}\cdot({\vec x}-{\vec x}_0)\big)\left(x^\epsilon x^\lambda_0+x^\lambda x^\epsilon_0+{\textstyle\frac{2}{3}}(x^\epsilon-x^\epsilon_0)(x^\lambda-x^\lambda_0)\right)\Big)+{\cal O}(G^2).~~~~~
\label{eq:phase_t}
\end{eqnarray}

Eq.~(\ref{eq:phase_t}) extends the well-known expression for relativistic delay. In addition to the classic Shapiro gravitational time delay due to a mass monopole (represented by the logarithmic term), it also includes contributions due to quadrupole  (the term multiplied by $J^{\epsilon\lambda}_{\rm E}$) and spin (the term multiplied by ${\vec S}_{\rm E}$) moments of the extended and rotating Earth, as well as terms due to tidal gravity of external bodies of the solar system that are present in the GCRS.

\subsection{Estimating the magnitudes of various terms}
\label{sec:phasemag}

We can now evaluate the magnitudes of the terms involved in
Eq.~(\ref{eq:phase_t}) in the context of the GRACE-FO mission, which will help us to simplify this general expression for the relativistic phase in GCRS. To do this, we consider signal propagating between the two spacecraft GRACE-FO-A and GRACE-FO-B that follow two worldlines, ${\vec x}_{\rm A}(t)$ and ${\vec x}_{\rm B}(t)$, correspondingly. The signal transmission begins at spacecraft $A$ at geocentric coordinates $(ct_{\rm A},{\vec x}_{\rm A})$. The signal is received by spacecraft $B$ at $(ct_{\rm B},{\vec x}_{\rm B})$. To describe the relevant geometry we introduce the  geocentric Euclidean vector between the two events, ${\vec R}_{\rm AB} = {\vec x}_{\rm B} - {\vec x}_{\rm A}$, the distance between them, ${R}_{\rm AB}=|{\vec R}_{\rm AB}|$, and the unit vector in the direction between them, ${\vec N}_{\rm AB} = {\vec R}_{\rm AB}/R_{\rm AB}$. Geocentric positions of the spacecraft described by familiar quantities: ${\vec n}_{\rm A} = {\vec x}_{\rm A}/r_{\rm A}$, ${\vec n}_{\rm B} = {\vec x}_{\rm B}/r_{\rm B}$, where $r_{\rm A}=|{\vec x}_{\rm A}|, r_{\rm B}=|{\vec x}_{\rm B}|$. The unperturbed direction of the wave propagation along the unit vector connecting the two points ${\vec x}_{\rm A}(t)$ and ${\vec x}_{\rm B}(t)$ and defined by Eq.~(\ref{eq:light-traj}), is given by ${\vec k}={\vec N}_{\rm AB}+{\cal O}(G)$. Using these definitions we establish the following exact relation \cite{Turyshev:2012nw}:
\begin{equation}
\frac{r_{\rm B}+{\vec k}\cdot{\vec x}_{\rm B}}{r_{\rm A}+{\vec k}\cdot{\vec x}_{\rm A}}=\frac{r_{\rm A}+r_{\rm B}+R_{\rm AB}}{r_{\rm A}+r_{\rm B}-R_{\rm AB}}.
\label{eq:eq_eik-phi+v++}
\end{equation}

As a result, the expression for the phase (\ref{eq:phase_t}) at point $(ct_{\rm B}, {\vec x}_{\rm B})$ with $(ct_0, {\vec x}_0)=(ct_{\rm A}, {\vec x}_{\rm A})$ has the form:
\begin{eqnarray}
\varphi(t_{\rm B},{\vec x}_{\rm B}) &=& \varphi_0+k_0\Big(c(t_{\rm B}-t_{\rm A})-{\vec k}\cdot({\vec x}_{\rm B}-{\vec x}_{\rm A})-\nonumber\\
&-& \frac{2GM_{\rm E}}{c^2}\Big\{\ln\Big[\frac{r_{\rm A}+r_{\rm B}+R_{\rm AB}}{r_{\rm A}+r_{\rm B}-R_{\rm AB}}\Big]+
\frac{1}{c}\Big({\vec k}\cdot\Big[{\vec S}_{\rm E}\times\big(\frac{{\vec n}_{\rm B}}{r_{\rm B}+{\vec k}\cdot{\vec x}_{\rm B}}-\frac{{\vec n}_{\rm A}}{r_{\rm A}+{\vec k}\cdot{\vec x}_{\rm A}}\big)\Big]\Big)+\nonumber\\
&-&
{\textstyle\frac{1}{6}}\Big[\frac{(n_{\rm B\epsilon}+k_\epsilon)(n_{\rm B\lambda}+k_\lambda)}{(r_{\rm B}+{\vec k}\cdot{\vec x}_{\rm B})^2}
+\frac{1}{r_{\rm B}}\frac{\gamma_{\epsilon\lambda}+n_{\rm B\epsilon} n_{\rm B\lambda}}{(r_{\rm B}+{\vec k}\cdot{\vec x}_{\rm B})}-
\frac{(n_{\rm A\epsilon}+k_\epsilon)(n_{\rm A\lambda}+k_\lambda)}{(r_{\rm A}+{\vec k}\cdot{\vec x}_{\rm A})^2}
-\frac{1}{r_{\rm A}}\frac{\gamma_{\epsilon\lambda}+n_{\rm A\epsilon} n_{\rm A\lambda}}{(r_{\rm A}+{\vec k}\cdot{\vec x}_{\rm A})} \Big]J_{\rm E}^{\epsilon\lambda}\Big\}-\nonumber\\
&-& \sum_{b\not={\rm E}}\frac{GM_b}{2c^2}\frac{\gamma_{\epsilon\lambda}+3n^{}_{b{\rm E}\epsilon}n^{}_{b{\rm E}\lambda}}{r^3_{b{\rm E}}}R_{\rm AB}\big(x^\epsilon_{\rm B} x^\lambda_{\rm A}+x^\lambda_{\rm B} x^\epsilon_{\rm A}+{\textstyle\frac{2}{3}}R^\epsilon_{\rm AB} R^\lambda_{\rm AB}\big)\Big)+{\cal O}(G^2).~~~~~
\label{eq:phase_AB}
\end{eqnarray}

We can now estimate the sizes of the terms involved in (\ref{eq:phase_AB}). We assume that both GRACE-FO spacecraft follow identical nearly circular orbits with of $e=0.001$ with other mission parameters summarized in Table~\ref{tb:params}. Although the actual spacecraft orbits are not going to be identical (primarily due to launch vehicle orbit insertion errors, actual behavior of the spacecraft, etc), we will use these values to evaluate the order of the terms in the model (\ref{eq:phase_AB}).

We start with the Shapiro term. Assuming the instantaneous range between the two spacecraft $d_{\rm AB}= 270$ km, a spacecraft attitude $h_G=450$ km, and defining spacecraft's semi-major axis $a=R_\oplus+h_G$, with $R_\oplus=6371$~km, being the Earth's radius, this term  evaluates to:
{}
\begin{equation}
\frac{2GM_{\rm E}}{c^2}\ln\Big[\frac{r_{\rm A}+r_{\rm B}+R_{\rm AB}}{r_{\rm A}+r_{\rm B}-R_{\rm AB}}\Big]\approx{\textstyle\frac{1}{2}}(\gamma+1)\frac{2GM_{\rm E}}{c^2}\frac{d_{\rm AB}}{a}(1+e\cos\omega_{\rm G}t)={\textstyle\frac{1}{2}}(\gamma+1)\cdot351.2~\mu{\rm m} +0.351~\mu{\rm m}\cdot\cos\omega_{\rm G}t,
\label{eq:Shapiro}
\end{equation}
where $\omega_{\rm G}$ is the mean orbital frequency of the GRACE-FO configuration. Note that in (\ref{eq:Shapiro}) we have reinstated the Eddington parameter $\gamma$ (see details in \cite{Will_book93}). If we were to compute this term without making use of the small parameter $d_{\rm AB}/a$ in the approximation, we get a result that is slightly higher at $376.5~\mu$m.  This estimate suggests that, should GRACE-FO be able to achieve an absolute range accuracy at the order of 1 nm, this mission could yield a new estimate of $\gamma$ with an accuracy of $\sigma_\gamma=5.3\times 10^{-6}$, which is an improvement by a factor of 5 over the current best value of $\sigma_\gamma=(2.1\pm2.3)\times 10^{-5}$ reported by the Cassini mission \cite{Bertotti-Iess-Tortora:1999} (also see discussion in \cite{Turyshev-2008ufn}). Given the anticipated range accuracy of 1~nm, clearly the Shapiro relativistic delay term is quite significant and must be kept in the model for GRACE-FO observables. At the same time, for the chosen GRACE-FO orbits, the largest contributions of the Shapiro effect is constant that will be absorbed into other constant terms without affecting the science data analysis.

\begin{table*}[t!]
\vskip-15pt
\caption{Select parameters of the GRACE-FO mission, along with corresponding symbols and approximate formulae used in the text. (A more detailed list of formulae and useful relations is derived and presented in Appendix \ref{sec:useful-rels}).
\label{tb:params}}
\begin{tabular}{|l|c|c|c|}\hline
Parameter  &Symbol  & Equation&Value\\\hline\hline
Orbital altitude &$h_{\rm G}$ &&\phantom{0}450~km\phantom{/s}\\
Orbital eccentricity&$e_{\rm }$ &&\phantom{00}0.001\phantom{/s}\\
Inter-spacecraft range &$d_{\rm AB}$ &&\phantom{0}270~km\phantom{/s}\\
Inter-spacecraft range rate &$\dot d_{\rm AB}=(\vec{n}_{\rm AB}\cdot\vec{v}_{\rm AB})\approx v_{\rm AB}\, e$&(\ref{eq:nAB.vAB})&\phantom{0.}0.3~m/s\phantom{k}\\
Geocentric velocity      & $v_{\rm A0} = ({G M_{\rm E}/(R_\oplus+h_{\rm G})})^{1/2}$ &(\ref{eq:kepler-va})&\phantom{.}7.65~km/s\phantom{}\\
Mean orbital frequency & $\omega_{\rm G}=({G M_{\rm E}/(R_\oplus+h_{\rm G})^3})^{1/2}$ & (\ref{eq:E}) &\phantom{0}1.12~mHz\phantom{k}\\
Relative spacecraft velocity &$v_{\rm AB}\simeq v_{\rm A0} d_{\rm AB}/(R_\oplus+h_{\rm G})$\, &(\ref{eq:vAB})&\phantom{0}303~m/s\phantom{k}\\
Geocentric acceleration & $a_{\rm A0} =G M_{\rm E}/(R_\oplus+h_{\rm G})^2$~\, & (\ref{eq:kepler-va}) &\phantom{0.}8.57~m/s$^2$\phantom{k}\\
Relative spacecraft acceleration &$a_{\rm AB}\simeq a_{\rm A0} d_{\rm AB}/(R_\oplus+h_{\rm G})$&(\ref{eq:aAB})&\phantom{00}0.34~m/s$^2$
\phantom{k}\\
Operating wavelength &$\lambda_{\rm A0}$ && 1064 nm~~~~\\
Frequency offset &$f^{\rm off}_{\rm B}$ && \phantom{0..}6~MHz\\\hline
\end{tabular}
\end{table*}
\vskip-0pt

Next, we look at the second contribution to the delay, which is due to the Earth's rotation. Assuming the Earth's spin moment to be that of a rigidly rotating sphere of uniform density, we arrive to the value of $M_{\rm E}S_{\rm E0}=\textstyle{\frac{2}{5}}M_{\rm E}
\omega_\oplus R_\oplus^2=7.05\times 10^{33}$ kg m$^2$/s, which allows us to evaluate this term to
{}
\begin{equation}
\frac{2GM_{\rm E}}{c^3}\Big({\vec k}\cdot\Big[{\vec S}_{\rm E}\times\big(\frac{{\vec n}_{\rm B}}{r_{\rm B}+{\vec k}\cdot{\vec x}_{\rm B}}-\frac{{\vec n}_{\rm A}}{r_{\rm A}+{\vec k}\cdot{\vec x}_{\rm A}}\big)\Big]\Big)\approx\frac{2GM_{\rm E}}{c^3}\big({\vec k}\cdot\big[{\vec S}_{\rm E}\times{\vec n}_{\rm A}\big]\big)\frac{d_{\rm AB}}{a^2}=2\times10^{-10}~{\rm m},
\label{eq:spin}
\end{equation}
which is negligible for GRACE-FO and can be omitted from the model.

The third term in (\ref{eq:phase_AB}) is the contribution to the delay due to the quadrupole moment, $\Delta d_{J_2}$, given as:
{}
\begin{eqnarray}
\Delta d_{J_2}&=&
-\frac{GM_{\rm E}}{3c^2}\Big[\frac{(n_{\rm B\epsilon}+k_\epsilon)(n_{\rm B\lambda}+k_\lambda)}{(r_{\rm B}+{\vec k}\cdot{\vec x}_{\rm B})^2}
+\frac{1}{r_{\rm B}}\frac{\gamma_{\epsilon\lambda}+n_{\rm B\epsilon} n_{\rm B\lambda}}{(r_{\rm B}+{\vec k}\cdot{\vec x}_{\rm B})}-
\frac{(n_{\rm A\epsilon}+k_\epsilon)(n_{\rm A\lambda}+k_\lambda)}{(r_{\rm A}+{\vec k}\cdot{\vec x}_{\rm A})^2}
-\frac{1}{r_{\rm A}}\frac{\gamma_{\epsilon\lambda}+n_{\rm A\epsilon} n_{\rm A\lambda}}{(r_{\rm A}+{\vec k}\cdot{\vec x}_{\rm A})} \Big]J_{\rm E}^{\epsilon\lambda}.~~~
\label{eq:del-J2}
\end{eqnarray}
From (\ref{eq:light-traj}) we have ${\vec x}_{\rm B}={\vec x}_{\rm A}+{\vec k}c(t_{\rm B}-t_{\rm A})+{\cal O}(G)={\vec x}_{\rm A}+{\vec k}R_{\rm AB}+{\cal O}(G)$. Thus, Eq.~(\ref{eq:del-J2}) can be approximated in terms of the small parameter ${R_{\rm AB}}/{r_{\rm A}}$ as
\begin{eqnarray}
\Delta d_{J_2}&=&\frac{GM_{\rm E}}{3c^2}\frac{J_{\rm E}^{\epsilon\lambda}}{r^2_{\rm A}}\Big\{\big(\gamma_{\epsilon\lambda}+3n_{\rm A\epsilon} n_{\rm A\lambda}\big)\frac{R_{\rm AB}}{r_{\rm A}}-{\textstyle\frac{3}{2}}\Big[
(\gamma_{\epsilon\lambda}+5n_{\rm A\epsilon} n_{\rm A\lambda}\big)({\vec k}\cdot{\vec n}_A)-n_{\rm A\epsilon} k_{\lambda}-n_{\rm A\lambda} k_{\epsilon}\Big]\frac{R^2_{\rm AB}}{r^2_{\rm A}}\Big\}+{\cal O}(\frac{R^3_{\rm AB}}{r^3_{\rm A}}).~~~
\label{eq:del-J2-expand}
\end{eqnarray}

To estimate the magnitudes of the terms in Eq.~(\ref{eq:del-J2-expand}), we introduce a convenient quantity $j^{\epsilon\lambda}_{\rm E}=J^{\epsilon\lambda}_{\rm E}/(3R^2_\oplus J_{2\oplus})$, which essentially represents the components of the Earth's quadrupole tensor (\ref{eq:mass-spin-moments}), normalized to the Earth's oblateness, $J_{2\oplus}=1.08263\times 10^{-3}$. Such a definition implies $||j^{\epsilon\lambda}_{\rm E}|| \simeq1$. Next, accounting for the expected orbital parameters of the GRACE-FO mission and taking $R_{\rm AB}=|{\vec x}_{\rm B}(t_{\rm B}) - {\vec x}_{\rm A}(t_{\rm A})| = d_{\rm AB}(t_{\rm B})+{\cal O}(c^{-1})$, we estimate the magnitudes of both terms:
{}
\begin{eqnarray}
\Delta d_{J_2}&\approx&\frac{GM_{\rm E}}{c^2}\frac{d_{\rm AB}}{a}\frac{R^2_\oplus J_{2\oplus}}{a^2}\Big\{\big(\gamma_{\epsilon\lambda}+3n_{\rm A\epsilon} n_{\rm A\lambda}\big)-{\textstyle\frac{3}{2}}\big[
(\gamma_{\epsilon\lambda}+5n_{\rm A\epsilon} n_{\rm A\lambda}\big)({\vec k}\cdot{\vec n}_A)-n_{\rm A\epsilon} k_{\lambda}-n_{\rm A\lambda} k_{\epsilon}\big]\frac{d_{\rm AB}}{a}\Big\}j^{\epsilon\lambda}_{\rm E}\approx\nonumber~~~\\
&\approx& 1.66\times10^{-7}~{\rm m}\cdot\big(\gamma_{\epsilon\lambda}+3n_{\rm A\epsilon} n_{\rm A\lambda}\big)j^{\epsilon\lambda}_{\rm E}+9.85\times10^{-9}~{\rm m}\cdot
\big(n_{\rm A\epsilon} k_{\lambda}+n_{\rm A\lambda} k_{\epsilon}\big)j^{\epsilon\lambda}_{\rm E},~~~
\label{eq:J2-val}
\end{eqnarray}
where we used (\ref{eq:rA.nAB}) and (\ref{eq:alpha}) to estimate $({\vec k}\cdot{\vec n}_{\rm A})=({\vec n}_{\rm AB}\cdot{\vec n}_{\rm A})=d_{\rm AB}/2a+{\cal O}(e)\approx 0.02$. Thus, even the term of the second order in $d_{\rm AB}/a$ in the quadrupole contribution to the delay is large enough to be observable by GRACE-FO.

Finally, we evaluate the contribution of the external gravity given by the last term in Eq.~(\ref{eq:phase_AB}):
\begin{eqnarray}
&&\sum_{b\not={\rm E}}\frac{GM_b}{2c^2}\frac{\gamma_{\epsilon\lambda}+3n^{}_{b{\rm E}\epsilon}n^{}_{b{\rm E}\lambda}}{r^3_{b{\rm E}}}R_{\rm AB}\Big(x^\epsilon_{\rm B} x^\lambda_{\rm A}+x^\lambda_{\rm B} x^\epsilon_{\rm A}+{\textstyle\frac{2}{3}}R^\epsilon_{\rm AB} R^\lambda_{\rm AB}\Big)\approx
\sum_{b\not={\rm E}}\frac{GM_b}{c^2}\Big(3({\vec n}^{}_{b{\rm E}}{\vec n}_{\rm A})^2-1\Big)\frac{d_{\rm AB}r_{\rm A}^2}{r^3_{b{\rm E}}}\leq
\nonumber\\
&&\hskip 80pt
\leq
\frac{2GM_m}{c^2}\frac{d_{\rm AB}a^2}{r^3_{m{\rm E}}}+\frac{2GM_\odot}{c^2}\frac{d_{\rm AB}a^2}{r^3_{\odot{\rm E}}}=
~2.41\times 10^{-11}~{\rm m}+1.11\times 10^{-11}~{\rm m}.
\label{eq:tides}
\end{eqnarray}
Clearly, the tidal contributions to the delay due to the Moon and the Sun are very small; contributions from other bodies of the solar system (i.e., Mars, Jupiter) are even smaller. Therefore, the entire contribution to light propagation due to the gravity of external bodies may be omitted.

\subsection{General relativistic phase model for GRACE-FO}
\label{sec:GRACE-FO-phase}

The evaluations conducted in the preceding section allowed us to neglect the contributions due to the spin moment and tidal gravity in Eq.~(\ref{eq:phase_AB}). As a result, the phase of a plane electromagnetic wave in the vicinity of the extended Earth can be expressed, at the level of accuracy required for the LRI experiment on GRACE-FO (see Sec.~\ref{sec:phasemag}), as
{}
\begin{eqnarray}
\varphi(t,{\vec x}) &=& \varphi_0+k_0\Big(c(t-t_{\rm 0})-
{\vec k}\cdot ({\vec x}-{\vec x}_{\rm 0})- \frac{2GM_{\rm E}}{c^2}\Big\{\ln\Big[\frac{r+{\vec k}\cdot {\vec x}}{r_{\rm 0}+{\vec k}\cdot {\vec x}_0}\Big]-\nonumber\\
&-&{\textstyle\frac{1}{6}}\Big[\frac{(n_\epsilon+k_\epsilon)(n_\lambda+k_\lambda)}{(r+{\vec k}\cdot{\vec x})^2}
+\frac{1}{r}\frac{\gamma_{\epsilon\lambda}+n_\epsilon n_\lambda}{(r+{\vec k}\cdot{\vec x})}-
\frac{(n_{\rm 0\epsilon}+k_\epsilon)(n_{\rm 0\lambda}+k_\lambda)}{(r_{0}+{\vec k}\cdot{\vec x}_{0})^2}
-\frac{1}{r_{\rm 0}}\frac{\gamma_{\epsilon\lambda}+n_{\rm 0\epsilon} n_{\rm 0\lambda}}{(r_{\rm 0}+{\vec k}\cdot{\vec x}_{\rm 0})}\Big]J_{\rm E}^{\epsilon\lambda}\Big\}\Big),
\label{eq:eik-phi-lamb-EJ}
\end{eqnarray}
which is accurate up to ${\cal O}$(0.2 nm). By dropping subscripts A and B and reinstating $(t_0,{\vec x}_0\equiv {\vec x}_0(t_0))$ and $(t,{\vec x})$ in (\ref{eq:eik-phi-lamb-EJ}) we returned to a generic form of the expression for $\varphi(t,{\vec x})$. This form is more convenient for the purpose of investigating the physical properties of the eikonal, aiming at a formulation of the LRI observables of the GRACE-FO.

It is instructional to present Eq.~(\ref{eq:eik-phi-lamb-EJ}) in the following equivalent from:
\begin{eqnarray}
\varphi(t,{\vec x}) &=& \varphi_0+k_0\Big(c(t-t_0)-{\cal R}\big({\vec x}_0(t_0),{\vec x}(t)\big)\Big),~~~~~
\label{eq:phase0}
\end{eqnarray}
where we introduced ${\cal R}(\vec{x}_0,\vec{x})$ which is the total distance traveled by light between the instant of emission $t_0$ and arbitrary instant $t$. At the level of accuracy appropriate for GRACE-FO, Eq.~(\ref{eq:eik-phi-lamb-EJ}) yields the following form of this function:
{}
\begin{eqnarray}
{\cal R}({\vec x}_0,{\vec x}) &=& {\vec k}\cdot({\vec x}-{\vec x}_0)+\frac{2GM_{\rm E}}{c^2}\Big\{\ln\Big[\frac{r+{\vec k}\cdot {\vec x}}{r_0+{\vec k}\cdot {\vec x}_0}\Big]-\nonumber\\
&-&
{\textstyle\frac{1}{6}}\Big[\frac{(n_\epsilon+k_\epsilon)(n_\lambda+k_\lambda)}{(r+{\vec k}\cdot{\vec x})^2}+\frac{1}{r}\frac{\gamma_{\epsilon\lambda}+n_\epsilon n_\lambda}{(r+{\vec k}\cdot{\vec x})}-
\frac{(n_{0\epsilon}+k_\epsilon)(n_{0\lambda}+k_\lambda)}{(r_0+{\vec k}\cdot{\vec x}_0)^2}
-\frac{1}{r_0}\frac{\gamma_{\epsilon\lambda}+n_{0\epsilon} n_{0\lambda}}{(r_0+{\vec k}\cdot{\vec x}_0)} \Big]J_{\rm E}^{\epsilon\lambda}\Big\}
+{\cal O}(G^2).~~~
\label{eq:R-total}
\end{eqnarray}
It does not depend on the wave's frequency and is determined solely by the geometry of the problem. Clearly, ${\cal R}\big({\vec x}_0,{\vec x}_0\big)=0$. The complete form of this quantity is easily recovered from Eq.~(\ref{eq:phase_t}). In the general case, in addition to the Euclidean distance $R={\vec k}\cdot({\vec x}-{\vec x}_0)$ traversed by the signal between the two points, the total path ${\cal R}$ includes several important general relativistic contributions, namely those due to the monopole, quadrupole and spin induced gravitational fields of the extended Earth and also tidal gravity terms induced by the external bodies.

Along the 4-dimensional path of a ray of light (a null geodesic in empty space), the phase stays always constant and equal to its initial value at the time of emission. By equating the phase (\ref{eq:phase0}) (or, in  a more general case, (\ref{eq:phase_t})) at two events --  the signal's emission at the point $(t_0,{\vec x}_0)$ and at an arbitrary event on the light part with coordinates $(t,{\vec x})$, we can write:
$\varphi(t_0,{\vec x}_0)\equiv\varphi_0=\varphi(t,{\vec x}) = \varphi_0+k_0\big(c(t-t_0)-{\cal R}({\vec x}_0(t_0),{\vec x}(t))\big),$
and recover the light-cone equation  synchronizing the events for the signal moving through a stationary spacetime:
\begin{equation}
c(t-t_0)={\cal R}\big({\vec x}_0(t_0),{\vec x}(t)\big).
\label{eq:light-cone}
\end{equation}
Note that in the post-Newtonian approximation of the general theory of relativity, Eq.~(\ref{eq:light-cone}) is exact and, as such, it is valid to all orders of the gravitational constant $G$. Eqs.~(\ref{eq:R-total}) and (\ref{eq:light-cone}) are the post-Minkowskian representation of the light cone \cite{Turyshev:2012nw} corresponding to the Green's function solution of the linearized homogeneous equations of the general theory of relativity for light propagation in the appropriate order. Any dependence on $G$ comes only via the geodesic distance ${\cal R}$ traveled by a ray of light, which, with the accuracy sufficient to analyze GRACE-FO, is given by (\ref{eq:R-total}).

\section{Laser ranging interferometric observables for GRACE-FO}
\label{sec:LRI-obs}

There are two types of data analysis that may be realized on GRACE-FO. The preferred LRI operating mode relies on a two-way configuration, in which the original signal sent by the first spacecraft is retransmitted by an active transponder on board the second spacecraft, to be ultimately received again on the first spacecraft. An alternative, which will be considered especially if coherent retransmission cannot be achieved reliably, may be to utilize dual one-way range (DOWR; see discussion in \cite{Turyshev-etal:2012} in the context of the GRAIL mission). The present section discusses the LRI operating mode in detail; the DOWR mode, which relies on precision timing and post-processing, is discussed in Sec.~\ref{sec:dowr} below.

\subsection{Relativistic clock synchronization and the geodesic signal path}
\label{sec:pseudo-ranges}

In the LRI operating mode, the interferometer on board of the first spacecraft compares the phase of the on-board laser oscillator to that of a signal received from the transponder on the second spacecraft. That transponder retransmits coherently a signal that was originally sent by the first spacecraft. The interferometer produces the phase difference and frequency observables from which the range and range rate between the two spacecraft are deduced.  These time series of phase and frequency values constitute the set of LRI observables of GRACE-FO \cite{Pierce-etal:2008,Sheard_etal:2012}.

To formulate a model for LRI observables, we consider a situation when a laser transponder system on spacecraft ${A}$, moving on a worldline $\vec{x}_{\rm A}(t)$, sends a continuous laser signal towards spacecraft ${B}$, which is then retransmitted by spacecraft ${\rm B}$ to be received by spacecraft $A$ (as shown in Fig.~\ref{fig:GRACE-FO-timing}). In the rest of this section, we shall use the shorthand $\vec{x}_{\rm A1}=\vec{x}_{\rm A}(t_{1})$, $\vec{x}_{\rm B2}=\vec{x}_{\rm B}(t_{\rm 2})$, and $\vec{x}_{\rm A3}=\vec{x}_{\rm A}(t_{\rm 3})$ to indicate the events of original transmission, retransmission by the $B$ transponder, and final reception by the $A$ spacecraft, with corresponding subscript notation for the quantities $r$, $\vec{n}$ and $n^\alpha$.

At the instant of reception on spacecraft $B$ during forward trip, from Eq.~(\ref{eq:phase0}) the signal's phase is characterized:
\begin{eqnarray}
\varphi(t_{\rm 2},{\vec x}_{\rm B2}) &=& \varphi(t_{\rm 1},{\vec x}_{\rm A1})+\frac{2\pi}{c}f_{\rm A0}\Big(\frac{d\tau_{\rm A}}{dt}\Big)_{t_1}\Big(c(t_{\rm 2}-t_{\rm 1})-{\cal R}_{\rm AB}({\vec x}_{\rm A1},{\vec x}_{\rm B2})\Big),
\label{eq:eik-GRACE-FO_A}
\end{eqnarray}
where $f_{\rm A0}$ is the proper frequency of the transmitter on spacecraft $A$, while  $f_{\rm A}=f_{\rm A0}(d\tau_{\rm A}/dt)_{t_1}$ being its coordinate frequency, as measured at the time of signal's emission $t_{\rm 1}$. We consider $\varphi(t_{\rm 1},\vec{x}_{\rm A1})$ to be the phase of the original transmission. ${\cal R}_{\rm AB}$ is the total geodesic distance traveled by the signal, which from (\ref{eq:R-total}) is determined as
{}
\begin{eqnarray}
\hskip -10pt
{\cal R}_{\rm AB}({\vec x}_{\rm A1},{\vec x}_{\rm B2})&=&
|{\vec x}_{\rm B}(t_{\rm 2}) - {\vec x}_{\rm A}(t_{\rm 1})|+\frac{2GM_{\rm E}}{c^2}\Big\{\ln\Big[\frac{r_{\rm A1}+r_{\rm B2}+R_{\rm A1B2}}{r_{\rm A1}+r_{\rm B2}-R_{\rm A1B2}}\Big]-\nonumber\\
&&\hskip -60pt -\,
{\textstyle\frac{1}{6}}\Big[\frac{(n_{\rm B2\epsilon}+k_\epsilon)(n_{\rm B2\lambda}+k_\lambda)}{(r_{\rm B2}+{\vec k}\cdot{\vec x}_{\rm B2})^2}
+\frac{1}{r_{\rm B2}}\frac{\gamma_{\epsilon\lambda}+n_{\rm B2\epsilon} n_{\rm B2\lambda}}{(r_{\rm B2}+{\vec k}\cdot{\vec x}_{\rm B2})}-
\frac{(n_{\rm A1\epsilon}+k_\epsilon)(n_{\rm A1\lambda}+k_\lambda)}{(r_{\rm A1}+{\vec k}\cdot{\vec x}_{\rm A1})^2}
-\frac{1}{r_{\rm A1}}\frac{\gamma_{\epsilon\lambda}+n_{\rm A1\epsilon} n_{\rm A1\lambda}}{(r_{\rm A1}+{\vec k}\cdot{\vec x}_{\rm A1})}\Big]J_{\rm E}^{\epsilon\lambda}\Big\}.~~~
\label{eq:t-AB}
\end{eqnarray}

The phase of the signal does not change along the signal's worldline: $\varphi(t_{\rm 2},\vec{x}_{\rm B2})=\varphi(t_{\rm 1},\vec{x}_{\rm A1})$. From (\ref{eq:eik-GRACE-FO_A}) we get:
\begin{equation}
t_{\rm 2} - t_{\rm 1}=c^{-1}{\cal R}_{\rm AB}\big({\vec x}_{\rm A}(t_1),{\vec x}_{\rm B}(t_2)\big),
\label{eq:tB-tA}
\end{equation}
which is the equation for coordinate time transfer between the time of emission $t_{\rm 1}$ and time of reception $t_{\rm 2}$.

The transponder on spacecraft $B$ responds with a phase coherent retransmission of a signal, which is then received on board spacecraft $A$ at $t_3$. At this point, using Eq.~(\ref{eq:phase0}), the retransmitted signal is characterized:
\begin{eqnarray}
\varphi(t_{\rm 3},{\vec x}_{\rm A3}) &=& \varphi(t_{\rm 2},{\vec x}_{\rm B2})+\frac{2\pi}{c}f_{\rm B0}\Big(\frac{d\tau_{\rm B}}{dt}\Big)_{t_2}\Big(c(t_{\rm 3}-t_{\rm 2})-{\cal R}_{\rm BA}({\vec x}_{\rm B2},{\vec x}_{\rm A3})\Big),
\label{eq:phase-BA}
\end{eqnarray}
where $f_{\rm B}=f_{\rm B0}(d\tau_{\rm B}/dt)_{t_2}$ being its coordinate frequency at the moment $t_{\rm 2}$ of the signal's coherent retransmission and $\varphi(t_{\rm 2},\vec{x}_{\rm B2})$ being the phase at this moment.
 ${\cal R}_{\rm BA}$ is with the total one-way distance of the return path, given by
{}
\begin{eqnarray}
\hskip -10pt
{\cal R}_{\rm BA}({\vec x}_{\rm B2},{\vec x}_{\rm A3})&=&
|{\vec x}_{\rm A}(t_{\rm 3}) - {\vec x}_{\rm B}(t_{\rm 2})|+\frac{2GM_{\rm E}}{c^2}\Big\{\ln\Big[\frac{r_{\rm A3}+r_{\rm B2}+R_{\rm B2A3}}{r_{\rm A3}+r_{\rm B2}-R_{\rm B2A3}}\Big]-\nonumber\\
&&\hskip-60pt -\,
{\textstyle\frac{1}{6}}\Big[\frac{(n_{\rm A3\epsilon}-k_\epsilon)(n_{\rm A3\lambda}-k_\lambda)}{(r_{\rm A3}-{\vec k}\cdot{\vec x}_{\rm A3})^2}
+\frac{1}{r_{\rm A3}}\frac{\gamma_{\epsilon\lambda}+n_{\rm A3\epsilon} n_{\rm A3\lambda}}{(r_{\rm A3}-{\vec k}\cdot{\vec x}_{\rm A3})}-
\frac{(n_{\rm B2\epsilon}-k_\epsilon)(n_{\rm B2\lambda}-k_\lambda)}{(r_{\rm B2}-{\vec k}\cdot{\vec x}_{\rm B2})^2}
-\frac{1}{r_{\rm B2}}\frac{\gamma_{\epsilon\lambda}+n_{\rm B2\epsilon} n_{\rm B2\lambda}}{(r_{\rm B2}-{\vec k}\cdot{\vec x}_{\rm B2})}\Big]J_{\rm E}^{\epsilon\lambda}\Big\},~~~
\label{eq:t-BA2}
\end{eqnarray}
where we accounted for the fact that during the return trip the wave vector points in the opposite direction, ${\vec k}\rightarrow -{\vec k}$. Using (\ref{eq:phase-BA}) and the fact that $\varphi(t_2,\vec{x}_{\rm B2})=\varphi(t_{\rm 3},\vec{x}_{\rm A3})$ the coordinate time transfer between $t_{\rm 2}$ and $t_{\rm 3}$ is given by
{}
\begin{equation}
t_{\rm 3} - t_{\rm 2}=c^{-1}{\cal R}_{\rm BA}\big({\vec x}_{\rm B}(t_2),{\vec x}_{\rm A}(t_3)\big).
\label{eq:transm-time:BA}
\end{equation}

We observe that, although events of the original signal emission at $t_{\rm 1}$ and its ultimate reception at $t_{\rm 3}$ are not directly connected by a light cone, we nevertheless may compute the total time elapsed between the two events (as was first observed in \cite{Fock-book:1959}). Indeed, with the help of Eqs.~(\ref{eq:tB-tA}) and (\ref{eq:transm-time:BA}) we have:
{}
\begin{equation}
t_{\rm 1}=t_{\rm 3} - c^{-1}\Big({\cal R}_{\rm AB}\big({\vec x}_{\rm A}(t_1),{\vec x}_{\rm B}(t_2)\big)+{\cal R}_{\rm BA}\big({\vec x}_{\rm B}(t_2),{\vec x}_{\rm A}(t_3)\big)\Big).
\label{eq:transm-time:ABA}
\end{equation}
Thus, the total coordinate time elapsed between the two events is fully determined by the geocentric positions of the two spacecraft at various specific instances along the two-way light path.

\subsection{The inter-spacecraft interferometric observables}
\label{sec:lri}

The GRACE-FO LRI observable is formed on spacecraft $A$ after a coherent retransmission of a signal originally transmitted by $A$ is received on board the $A$ spacecraft and compared to the on-board laser oscillator.
A coherent retransmission of the signal at $B_2$ implies a return signal originating at $\vec{x}_{\rm B2}$ with phase $\varphi(t_2,\vec{x}_{\rm B2})=\varphi(t_1,\vec{x}_{\rm A1})$, to be received at $\vec{x}_{\rm A3}$ with phase $\varphi(t_3,\vec{x}_{\rm A3})=\varphi(t_2,\vec{x}_{\rm B2})$. A phase observable is formulated by taking the difference between the received signal at $\vec{x}_{\rm A3}$ and the phase of the local oscillator on board spacecraft $A$ at $t_3$, that is, $\delta\varphi=\varphi(t_3,\vec{x}_{\rm A3})-\varphi(t_1,\vec{x}_{\rm A1})$.

To describe the LRI observables, first consider an oscillator with proper frequency $f_{\rm A0}$, located at a moving point $A_1$, that generates a signal with frequency $f_{\rm A0}(\tau_{\rm A1})$ at proper time $\tau_{\rm A1}$ measured along the worldline ${\vec x}_{\rm A}$ of the oscillator (see Fig.~\ref{fig:GRACE-FO-timing}). This signal is transmitted from point $A_1$ and received at point $B_2$, at proper time $\tau_{\rm B2}$ taken along the worldline ${\vec x}_{\rm B}(t)$. The instantaneous phase of the received signal is compared with the phase of the local oscillator located at point $B_2$ whose proper frequency at that instant is $f_{\rm B0}(\tau_{\rm B2})$.

At reception, the measurable quantity is the difference between the instantaneous phases of the two signals compared at ${\vec x}_{\rm B2}$. Instrumentally, at $\vec{x}_{\rm B2}$ one measures the infinitesimal difference $d n^{\rm rx}_{\rm AB}$ in the received number of cycles $dn_{\rm A}^{\rm B}(\tau_{\rm B2})$ originally transmitted at $\vec{x}_{\rm A1}$, and the number of the locally generated cycles $dn_{\rm B}(\tau_{\rm B2})$. This quantity may be expressed using proper frequencies and the infinitesimal proper time interval $d\tau_{\rm B2}$ as:
{}
\begin{equation}
d n^{\rm rx}_{\rm AB}(\tau_{\rm B2})=dn_{\rm B}(\tau_{\rm B2})-dn_{\rm A}^{\rm B}(\tau_{\rm B2})= f_{\rm B0}(\tau_{\rm B2})d\tau_{\rm B2}-f_{\rm A}^{\rm B}(\tau_{\rm B2})d\tau_{\rm B2},
\label{eq:dn-ab}
\end{equation}
where $f_{\rm A}^{\rm B}$ is the proper frequency of the oscillator $A$ as measured at $B$.

\begin{figure}[t]
\includegraphics[scale=0.67]{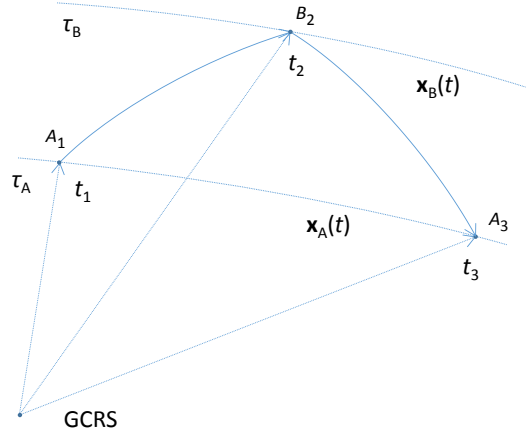}
\vskip -6pt
\caption{\label{fig:GRACE-FO-timing} Timing events on GRACE-FO: Depicted (not to scale) are the trajectories of the GRACE-FO-A and GRACE-FO-B spacecraft with corresponding proper times $\tau_{\rm A}$ and $\tau_{\rm B}$ and with three events in the GCRS, corresponding to signal transmission at $\vec{x}_{\rm A}(t_1)$, coherent retransmission by the $\rm B$ spacecraft transponder at $\vec{x}_{\rm B}(t_2)$, and final reception at $\vec{x}_{\rm A}(t_3)$.}
\end{figure}

Assuming that the fractional number of cycles sent from spacecraft $A$ at proper time $\tau_{\rm A1}$, denoted here as $n_{\rm A1}=n_{\rm A}(\tau_{\rm A1})$, and received on spacecraft $B$ at proper time $\tau_{\rm B2}$ and denoted as $n_{\rm A}^{\rm B2}=n_{\rm A}^{\rm B}(\tau_{\rm B2})$, are the same, or in infinitesimal form, $dn_{\rm A1}=dn_{\rm A}^{\rm B2}$, we can express the frequency $f_{\rm A}^{\rm B}(\tau_{\rm B2})$ via its value $f_{\rm A0}(\tau_{\rm A1})$ at the proper time $\tau_{\rm A1}$ of emission on spacecraft $A$:
{}
\begin{equation}
\frac{f_{\rm A}^{\rm B}(\tau_{\rm B2})}{f_{\rm A0}(\tau_{\rm A1})}=\frac{dn_{\rm A}^{\rm B2}}{d\tau_{\rm B2}}\frac{d\tau_{\rm A1}}{dn_{\rm A1}}=\frac{d\tau_{\rm A1}}{d\tau_{\rm B2}}.
\label{eq:dn-ab+}
\end{equation}
The infinitesimal difference between the number of cycles generated locally on spacecraft $B$ and those received from the spacecraft $A$, as given by Eq.~(\ref{eq:dn-ab}), takes the form:
{}
\begin{equation}
d n^{\rm rx}_{\rm AB}(\tau_{\rm B2})=
dn_{\rm B}(\tau_{\rm B2})-dn_{\rm A}(\tau_{\rm A1})=
f_{\rm B0}(\tau_{\rm B2})d\tau_{\rm B2}-f_{\rm A0}(\tau_{\rm A1})d\tau_{\rm A1}=\Big(f_{\rm B0}(\tau_{\rm B2})-f_{\rm A0}(\tau_{\rm A1})\frac{d\tau_{\rm A1}}{d\tau_{\rm B2}}\Big)d\tau_{\rm B2}.
\label{eq:dn-ab++-0}
\end{equation}

The laser transponder system at spacecraft $B$ is a phase-locked transponder capable of locking onto the incoming signal. It will be able to respond to frequency fluctuations of the received signal for its subsequent retransmission. Coherency between the received signal and the local oscillator implies $n^{\rm rx}_{\rm AB}=$~const., and, from (\ref{eq:dn-ab++-0}), the frequency of the signal and the infinitesimal number of cycles received at spacecraft $B$ are given as
\begin{equation}
f_{\rm B0}(\tau_{\rm B2})=f_{\rm A0}(\tau_{\rm A1})\frac{d\tau_{\rm A1}}{d\tau_{\rm B2}}
~~~~~{\rm and}~~~~~
dn_{\rm B}(\tau_{\rm B2})=dn_{\rm A}(\tau_{\rm A1}).
\label{eq:dn-ab2-coh}
\end{equation}
However, the transponder on GRACE-FO-B is an offset phase-locked transponder in which a fixed frequency offset, $f^{\rm off}_{\rm B}(\tau_{\rm B2})$, is added to (\ref{eq:dn-ab2-coh}) \cite{Pierce-etal:2008,Sheard_etal:2012}, so the retransmission will be done at the shifted frequency
{}
\begin{equation}
f^{\rm tx}_{\rm B0}(\tau_{\rm B2})=f_{\rm B0}(\tau_{\rm B2})+f^{\rm off}_{\rm B}(\tau_{\rm B2}).
\label{eq:dn-ab2}
\end{equation}
The introduction of the offset frequency will also affect the transmitted phase, given here by the number of transmitted cycles, $n^{\rm tx}_{\rm B}(\tau_{\rm B2})$. Thus, compared to a coherent transmission where the properties of the transmitted signal are identical to those of the received signal (as summarized by (\ref{eq:dn-ab2-coh})), the presence of the offset frequency results in adding to the infinitesimal number of transmitted cycles a linear ramp of $f^{\rm off}_{\rm B}d\tau_{\rm B2}$:
{}
\begin{equation}
dn^{\rm tx}_{\rm B}(\tau_{\rm B2})=
dn_{\rm B}(\tau_{\rm B2})+f^{\rm off}_{\rm B}(\tau_{\rm B2})d\tau_{\rm B2}.
\label{eq:dn-ab22'}
\end{equation}

The LRI observable on GRACE-FO is formed by comparing the properties of the signal generated by the local oscillator (that is already shifted by the offset frequency) to those of the incoming signal \cite{Sheard_etal:2012}. As before, the measurable quantity is the infinitesimal difference in the number of cycles, $d n^{\rm tx}_{\rm AB}(\tau_{\rm B2})$, given as
{}
\begin{equation}
d n^{\rm tx}_{\rm AB}(\tau_{\rm B2})=
dn^{\rm tx}_{\rm B}(\tau_{\rm B2})-dn_{\rm A}(\tau_{\rm A1})=
\Big(f^{\rm tx}_{\rm B0}(\tau_{\rm B2})-f_{\rm A0}(\tau_{\rm A1})\frac{d\tau_{\rm A1}}{d\tau_{\rm B2}}\Big)d\tau_{\rm B2}.
\label{eq:dn-ab++}
\end{equation}
This expression represents the beatnote signal measured by the phasemeter on the transponder spacecraft $B$ \cite{Ware-etal:2006}.

Similarly, we develop an expression describing light propagation from spacecraft $B$ back to spacecraft $A$, where it is received at proper time $\tau_{\rm A3}$ and measured with respect to the local oscillator. In this case the quantity of interest is the infinitesimal number of cycles $dn_{\rm BA}$, which is given as
{}
\begin{equation}
d n_{\rm BA}(\tau_{\rm A3})=dn_{\rm A}(\tau_{A3})-dn^{\rm tx}_{\rm B}(\tau_{\rm B2})=
\Big(f_{\rm A0}(\tau_{\rm A3})-f^{\rm tx}_{\rm B0}(\tau_{\rm B2})\frac{d\tau_{\rm B2}}{d\tau_{\rm A3}}\Big)d\tau_{\rm A3}.
\label{eq:dn-ba00-0}
\end{equation}

Substituting the transmitted frequency $f^{\rm tx}_{\rm B0}(\tau_{\rm B2})$, given by (\ref{eq:dn-ab2}), and the number of cycles $dn^{\rm tx}_{\rm B}(\tau_{\rm B2})$, given by (\ref{eq:dn-ab22'}),  in Eqs.~(\ref{eq:dn-ab++})--(\ref{eq:dn-ba00-0}), we obtain models for the beatnote signals measured by the phasemeters:
{}
\begin{eqnarray}
d n^{\rm tx}_{\rm AB}(\tau_{\rm B2})&=&
d\Big(n_{\rm B}(\tau_{\rm B2})-n_{\rm A}(\tau_{\rm A1})\Big)+f^{\rm off}_{\rm B}(\tau_{\rm B2})d\tau_{\rm B2}=
\Big(f_{\rm B0}(\tau_{\rm B2})+f^{\rm off}_{\rm B}(\tau_{\rm B2})-f_{\rm A0}(\tau_{\rm A1})\frac{d\tau_{\rm A1}}{d\tau_{\rm B2}}\Big)d\tau_{\rm B2},
\label{eq:dn-AB*}\\
d n_{\rm BA}(\tau_{\rm A3})&=&d\Big(n_{\rm A}(\tau_{A3})-n_{\rm B2}(\tau_{\rm B2})\Big)-f^{\rm off}_{\rm B}(\tau_{\rm B2})d\tau_{\rm B2}=
\Big(f_{\rm A0}(\tau_{\rm A3})-\big(f_{\rm B0}(\tau_{\rm B2})+f^{\rm off}_{\rm B}(\tau_{\rm B2})\big)\frac{d\tau_{\rm B2}}{d\tau_{\rm A3}}\Big)d\tau_{\rm A3}.
\label{eq:dn-ba00}
\end{eqnarray}

Equations (\ref{eq:dn-AB*}) and (\ref{eq:dn-ba00}) represent an important starting point in derivation of observational equations needed to process interferometric data on GRACE-FO. If one assumes coherent reception and retransmission on spacecraft $B$, this mode of operation leads to the LRI-type of observables (which we discuss in Sec.~\ref{sec:phase-LRI}). If no coherent link between the spacecraft is assumed, the process will depend on the stability of local oscillators on board the spacecraft and a phase reconstruction based, for example, on a high-precision interpolation of phase measurements \cite{Shaddock:2004ua}. This process leads to the DOWR-type of observables (discussed in Sec.~\ref{sec:dowr} below).

\subsection{Formulating the phase for LRI}
\label{sec:phase-LRI}

For LRI measurements, which are based on the coherent reception of the signal from spacecraft $A$ and its retransmission back to the same spacecraft, the quantity of interest is $n_{\rm BA}$, which will be provided by a phasemeter on spacecraft $A$. Coherent operation of the transponder on board spacecraft $B$ is captured by Eq.~(\ref{eq:dn-ab2-coh}). Substituting these expressions into (\ref{eq:dn-AB*}), we obtain
{}
\begin{eqnarray}
d n^{\rm tx}_{\rm AB}(\tau_{\rm B2})&=&f^{\rm off}_{\rm B}(\tau_{\rm B2})d\tau_{\rm B2}.
\label{eq:dn-ab+=}
\end{eqnarray}
Equation~(\ref{eq:dn-ab+=}) describes the phasemeter's signal of the offset phase-locked transponder \cite{Sheard_etal:2012} on the spacecraft $B$, for which the laser phase will be controlled by feeding back the detected signal, so that the beatnote phase on the transponder spacecraft, $n^{\rm tx}_{\rm AB}(\tau_{\rm B2})$, is driven to follow a linear ramp, i.e. $f^{\rm off}_{\rm B}(\tau_{\rm B2})\tau_{\rm B2}$.

The right-hand side of (\ref{eq:dn-ba00}) together with the conditions (\ref{eq:dn-ab2-coh}) of coherent reception yields
{}
\begin{eqnarray}
d n_{\rm BA}(\tau_{\rm A3})
&=&
\Big(f_{\rm A0}(\tau_{\rm A3})-f_{\rm A0}(\tau_{\rm A1})\frac{d\tau_{\rm A1}}{d\tau_{\rm A3}}-f^{\rm off}_{\rm B}(\tau_{\rm B2})\frac{d\tau_{\rm B2}}{d\tau_{\rm A3}}\Big)d\tau_{\rm A3},~~~~
\label{eq:dn-ba42}
\end{eqnarray}
which describes the phase of the beatnote on the phasemeter on the spacecraft $A$ -- our signal of interest for LRI.

To develop Eq.~(\ref{eq:dn-ba42}) further, we use the differential equation that relates the rate of the spacecraft proper  times, $\tau_{\rm A}$ and $\tau_{\rm B}$, as measured by an on-board clock in Earth's orbit, to the time in GCRS, denoted here as $t$ (see Ref.~\cite{Turyshev-Toth:2013}) as
{}
\begin{eqnarray}
\frac{d\tau_{\rm A}}{dt}&=& 1-\frac{1}{c^2}\Big[\frac{{\vec v}^2_{\rm A}}{2}+U_{\rm E}({\vec y}_{\rm A})\Big]+{\cal O}(c^{-4}) ~~~~{\rm and}~~~~
\frac{d\tau_{\rm B}}{dt}= 1-\frac{1}{c^2}\Big[\frac{{\vec v}^2_{\rm B}}{2}+U_{\rm E}({\vec y}_{\rm B})\Big]+{\cal O}(c^{-4}).
\label{eq:proper-coord-t}
\end{eqnarray}
Using $({d\tau_{\rm A}}/{dt})_{t_i}$ to mean the value of the expression (\ref{eq:proper-coord-t}) at $t=t_i~(i=1,3)$, we have
{}
\begin{eqnarray}
d\tau_{\rm A1}= \Big(\frac{d\tau_{\rm A}}{dt}\Big)_{t_1}dt_1,
~~~~~~
d\tau_{\rm A3}= \Big(\frac{d\tau_{\rm A}}{dt}\Big)_{t_3}dt_3,
~~~~~~{\rm and}~~~~~~
d\tau_{\rm B2}= \Big(\frac{d\tau_{\rm B}}{dt}\Big)_{t_2}
dt_2,
\label{eq:proper-coord-t_A}
\end{eqnarray}
where the instances of coordinate time $t_1$ and $t_3$ corresponding to the events of the signal's emission and its ultimate reception by the same spacecraft, are not independent and are related by (\ref{eq:transm-time:ABA}). Similarly, $t_2$ and $t_3$ are related by (\ref{eq:transm-time:BA}).
Therefore, the ratio of proper times in (\ref{eq:dn-ba42}) may be expressed via the ratio of their coordinate counterparts as
{}
\begin{equation}
\frac{d\tau_{\rm A1}}{d\tau_{\rm A3}}
=\Big(\frac{d\tau_{\rm A}}{dt}\Big)_{t_1}\Big(\frac{d\tau_{\rm A}}{dt}\Big)^{-1}_{t_3}\frac{dt_{\rm 1}}{dt_{\rm 3}}
~~~~~~ {\rm and} ~~~~~~
\frac{d\tau_{\rm B2}}{d\tau_{\rm A3}}
=\Big(\frac{d\tau_{\rm B}}{dt}\Big)_{t_2}\Big(\frac{d\tau_{\rm A}}{dt}\Big)^{-1}_{t_3}\frac{dt_{\rm 2}}{dt_{\rm 3}}.
\label{eq:tau-comb}
\end{equation}

As a result, Eq.~(\ref{eq:dn-ba42}) takes the form
{}
\begin{eqnarray}
d n_{\rm BA}(\tau_{\rm A3})&=&
\Big(f_{\rm A0}(\tau_{\rm A3})-f_{\rm A0}(\tau_{\rm A1})\Big(\frac{d\tau_{\rm A}}{dt}\Big)_{t_1}\Big(\frac{d\tau_{\rm A}}{dt}\Big)^{-1}_{t_3}\frac{dt_{\rm 1}}{dt_{\rm 3}}-
f^{\rm off}_{\rm B}(\tau_{\rm B2})
\Big(\frac{d\tau_{\rm B}}{dt}\Big)_{t_2}\Big(\frac{d\tau_{\rm A}}{dt}\Big)^{-1}_{t_3}\frac{dt_{\rm 2}}{dt_{\rm 3}}\Big)d\tau_{\rm A3}.~~~~
\label{eq:dn-ba42+}
\end{eqnarray}

Using  (\ref{eq:transm-time:BA}) and (\ref{eq:transm-time:ABA}) we have the following exact expression for the ratio of coordinate times present in this equation:
{}
\begin{eqnarray}
\frac{dt_{\rm 1}}{dt_{\rm 3}}&=&1-\frac{1}{c}\frac{d}{dt_{\rm 3}}\Big[{\cal R}_{\rm AB}({\vec x}_{\rm A1},{\vec x}_{\rm B2})+{\cal R}_{\rm BA}({\vec x}_{\rm B2},{\vec x}_{\rm A3})
\Big],
\label{eq:dt_A/dt_B0}\\
\frac{dt_{\rm 2}}{dt_{\rm 3}}&=&1-\frac{1}{c}\frac{d}{dt_{\rm 3}}\Big[{\cal R}_{\rm BA}({\vec x}_{\rm B2},{\vec x}_{\rm A3})\Big].
\label{eq:dt_A/dt_B0*}
\end{eqnarray}

The results given by Eqs.~(\ref{eq:dt_A/dt_B0}) and (\ref{eq:dt_A/dt_B0*}) allow us to present (\ref{eq:dn-ba42+}), describing the infinitesimal difference between the number of cycles coherently retransmitted from $B$ and received at spacecraft $A$, and the number of cycles generated by the oscillator on board spacecraft $A$, as
{}
\begin{eqnarray}
d n_{\rm BA}(\tau_{\rm A3})&=&\Big(
f_{\rm A0}(\tau_{\rm A3})-f_{\rm A0}(\tau_{\rm A1})\Big(\frac{d\tau_{\rm A}}{dt}\Big)_{t_1}\Big(\frac{d\tau_{\rm A}}{dt}\Big)^{-1}_{t_3}-
f^{\rm off}_{\rm B}(\tau_{\rm B2})
\Big(\frac{d\tau_{\rm B}}{dt}\Big)_{t_2}\Big(\frac{d\tau_{\rm A}}{dt}\Big)^{-1}_{t_3}+\nonumber\\
&&\hskip 46pt +\,
\frac{1}{c}f_{\rm A0}(\tau_{\rm A1})\Big(\frac{d\tau_{\rm A}}{dt}\Big)_{t_1}\Big(\frac{d\tau_{\rm A}}{dt}\Big)^{-1}_{t_3}\frac{d}{dt_{\rm 3}}\Big[{\cal R}_{\rm AB}({\vec x}_{\rm A1},{\vec x}_{\rm B2})+{\cal R}_{\rm BA}({\vec x}_{\rm B2},{\vec x}_{\rm A3})\Big]+\nonumber\\
&&\hskip 46pt +\,
\frac{1}{c}f^{\rm off}_{\rm B}(\tau_{\rm B2})
\Big(\frac{d\tau_{\rm B}}{dt}\Big)_{t_2}\Big(\frac{d\tau_{\rm A}}{dt}\Big)^{-1}_{t_3}\frac{d}{dt_{\rm 3}}\Big[{\cal R}_{\rm BA}({\vec x}_{\rm B2},{\vec x}_{\rm A3})\Big]
\Big)d\tau_{\rm A3}.
\label{eq:dn_ABA0=}
\end{eqnarray}
The resulting expression is valid for arbitrary trajectories of spacecraft $A$ and $B$. Although (\ref{eq:dn_ABA0=}) contains all three values of time, $t_{\rm 1}, t_{\rm 2}, t_{\rm 3}$, any two of these values are determined by the third. The logic of the LRI  measurements dictates that it is $t_{\rm 3}$, the instant of signal reception back at spacecraft $A$, that should be treated as an independent variable. The values of $t_{\rm 1}$ and $t_{\rm 2}$ may be explicitly expressed via $t_{\rm 3}$ as $t_{\rm 1}(t_3)$ and $t_{\rm 2}(t_3)$ by applying the transformations (\ref{eq:t-AB}), (\ref{eq:tB-tA}), and (\ref{eq:t-BA2}), (\ref{eq:transm-time:BA}). Furthermore, (\ref{eq:proper-coord-t})  and (\ref{eq:proper-coord-t_A}) relate the coordinate time $t_3$ to the proper time $\tau_{\rm A3}$ and can be integrated to determine $t_3=t_3(\tau_{\rm A3})$ and vice versa.

Substituting (\ref{eq:dn-ab2-coh}) in the left-hand side of (\ref{eq:dn-ba00}), this quantity may be expressed in infinitesimal form as
{}
\begin{equation}
d n_{\rm BA}(\tau_{\rm A3})=
d\Big(n_{\rm A}(\tau_{\rm A3})-n_{\rm A}(\tau_{\rm A1})\Big)-f^{\rm off}_{\rm B}(\tau_{\rm B2})d\tau_{\rm B2}.
\label{eq:dn_aba31}
\end{equation}
The first two terms in (\ref{eq:dn_aba31}) represent the infinitesimal difference between the number of cycles generated locally at spacecraft $A$, $n_{\rm A}(\tau_{\rm A3})$ at proper time $\tau_{\rm A3}$, and received cycles, $n_{\rm A}(\tau_{\rm A1})$, that were originally generated by the same oscillator at proper time $\tau_{\rm A1}$.

The GRACE-FO optical transponder experiment uses a continuous signal.
The on-board optical interferometer is designed to track the instantaneous phase of the received signal. Information on the phase of the received signal and its rate of change (and the associated range and range rate), are used to address the GRACE-FO science objectives.

To capture this logic of the LRI measurements, we now treat $\tau_{\rm A3}$, $\tau_{\rm B2}$ and $\tau_{\rm A1}$ as continuous variables, allowing us to formally integrate Eq.~(\ref{eq:dn_aba31}). This yields, up to an arbitrary integration constant that represents the combined phases of the transmitter, receiver, and offset generator at the beginning of the integration interval, the expression
{}
\begin{equation}
\Delta\varphi(\tau_{\rm A3})=2\pi\int dn_{\rm BA}=\varphi(\tau_{\rm A3})-\varphi(\tau_{\rm A1})-2\pi f^{\rm off}_{\rm B}(\tau_{\rm B2})\tau_{\rm B2}.
\label{eq:D-phi}
\end{equation}
The quantity $\Delta \varphi(\tau_{\rm A3})$ is one of the  LRI observables on GRACE-FO formed at spacecraft $A$. It compares the phase of the local oscillator on spacecraft $A$, denoted here by $\varphi_{\rm A}(\tau_{\rm A3})$, with itself but taken a round-trip light time earlier, $\varphi_{\rm A}(\tau_{\rm A1})=\varphi_{\rm A}(\tau_{\rm A3}-2c^{-1}\rho_{\rm AB}(\tau_{\rm A3}))$, while applying a linear phase ramp $2\pi f^{\rm off}_{\rm B}(\tau_{\rm B2})\tau_{\rm B2}$ due to the offset frequency.

To investigate how $\Delta\varphi(\tau_{\rm A3})$ evolves with time, we (\ref{eq:dn_ABA0=}). This integral expresses the cumulative difference over an arbitrary interval of time in the cycle count between the oscillator on board spacecraft $A$ and the two-way return signal received from spacecraft B, yielding $\Delta\varphi/2\pi$:
{}
\begin{eqnarray}
\frac{\Delta\varphi(\tau_{\rm A3})}{2\pi}&=&
\int \Big[
f_{\rm A0}(\tau_{\rm A3})-f_{\rm A0}(\tau_{\rm A1})\Big(\frac{d\tau_{\rm A}}{dt}\Big)_{t_1}\Big(\frac{d\tau_{\rm A}}{dt}\Big)_{t_3}^{-1}
-f^{\rm off}_{\rm B}(\tau_{\rm B2})
\Big(\frac{d\tau_{\rm B}}{dt}\Big)_{t_2}\Big(\frac{d\tau_{\rm A}}{dt}\Big)^{-1}_{t_3}
\Big]d\tau_{\rm A3}+\nonumber\\
&&\hskip -40pt +\,
\frac{1}{c} \int \Big[f_{\rm A0}(\tau_{\rm A1})\Big(\frac{d\tau_{\rm A}}{dt}\Big)_{t_1}\Big(\frac{d\tau_{\rm A}}{dt}\Big)_{t_3}^{-1}\Big(\dot{\cal R}_{\rm AB}(t'_{\rm 3})+\dot{\cal R}_{\rm BA}(t'_{\rm 3})\Big)+
f^{\rm off}_{\rm B}(\tau_{\rm B2})
\Big(\frac{d\tau_{\rm B}}{dt}\Big)_{t_2}\Big(\frac{d\tau_{\rm A}}{dt}\Big)^{-1}_{t_3}\dot{\cal R}_{\rm BA}(t'_{\rm 3})\Big]d\tau_{\rm A3},~~~~~
\label{eq:dn_ABA2}
\end{eqnarray}
where the geodesic distances ${\cal R}_{\rm AB}(t_3)={\cal R}_{\rm AB}({\vec x}_{\rm A1}(t_3),{\vec x}_{\rm B2}(t_3))$ and ${\cal R}_{\rm BA}(t_3)={\cal R}_{\rm BA}({\vec x}_{\rm B2}(t_3),{\vec x}_{\rm A3}(t_3))$ are expressed via $t_3$ and are presented by (\ref{eq:shap-corr}) and (\ref{eq:nt-BA+**}) for ${\cal R}_{\rm AB}(t_3)$ and (\ref{eq:nt-BA2}) for ${\cal R}_{\rm BA}(t_3)$, correspondingly. The overdot in $\dot{\cal R}_{\rm AB}(t_3)$ and $\dot{\cal R}_{\rm BA}(t_3)$ denotes differentiation with respect to $t_3$.

To evaluate the first two terms in square brackets of the first integral of (\ref{eq:dn_ABA2}), we present them equivalently:
{}
\begin{equation}
f_{\rm A0}(\tau_{\rm A3})-f_{\rm A0}(\tau_{\rm A1})\Big(\frac{d\tau_{\rm A}}{dt}\Big)_{t_1}\Big(\frac{d\tau_{\rm A}}{dt}\Big)_{t_3}^{-1}=f_{\rm A0}(\tau_{\rm A3})-f_{\rm A0}(\tau_{\rm A1})+f_{\rm A0}(\tau_{\rm A1})\Big[1-\Big(\frac{d\tau_{\rm A}}{dt}\Big)_{t_1}\Big(\frac{d\tau_{\rm A}}{dt}\Big)_{t_3}^{-1}\Big].
\label{eq:1-term}
\end{equation}

The difference $f_{\rm A0}(\tau_{\rm A3})-f_{\rm A0}(\tau_{\rm A1})$ on the right hand side of (\ref{eq:1-term}) depends on the laser frequency stabilization on spacecraft $A$. The anticipated frequency fluctuations $\delta f_{\rm A0}=f_{\rm A0}(\tau_{\rm A3})-f_{\rm A0}(\tau_{\rm A1})={\dot f_{\rm A0}}\Delta\tau_{\rm A}+{\cal O}(\Delta\tau_{\rm A}^2)$ are of the order of $\delta f_{\rm A0}/f_{\rm A0}={\dot f_{\rm A0}}\Delta\tau_{\rm A}/f_{\rm A0}\leq 2\times 10^{-15}$. Therefore, during a round-trip transit time of $2d_{\rm AB}/c=1.8$~ms, this term will contribute less then 1 nm to the range error. Expecting that the frequency stabilization goal will be achieved, the first two terms on the right-hand side of (\ref{eq:1-term}) may be omitted.

We evaluate the third term on the right-hand side of (\ref{eq:1-term}) using the orbital configuration chosen for GRACE-FO. To do this, we can estimate the first ratio in (\ref{eq:tau-comb}). Thus, using  (\ref{eq:proper-coord-t}) we have:
\begin{equation}
\Big(\frac{d\tau_{\rm A}}{dt}\Big)_{t_1}\Big(\frac{d\tau_{\rm A}}{dt}\Big)_{t_3}^{-1}= 1+\frac{1}{c^2}\frac{d}{dt}\Big[\frac{{\vec v}^2_{\rm A}}{2}+U_{\rm E}({\vec y}_{\rm A})\Big]\Delta t_{13}+{\cal O}(\Delta t^2_{13}, c^{-4}),
\label{eq:tau-comb-A}
\end{equation}
where $\Delta t_{13}=t_{\rm 3}-t_{\rm 1}$.
The magnitude of the $1/c^2$ term here can be easily evaluated using Eq.~(\ref{eq:tau_A}). Thus, for a round-trip time of transmission between the two spacecraft $\Delta t_{13}\sim 2d_{\rm AB}/c$, this term has the magnitude
{}
\begin{eqnarray}
\frac{d}{dt}\Big[{\frac{{\vec v}^2_{\rm A}}{2}}+\frac{GM}{r_{\rm A}}\Big]\cdot\frac{2d_{\rm AB}}{c^3}&\simeq&-\frac{4GM}{a^2}\sqrt{\frac{GM}{a}}\frac{d_{\rm AB}}{c^3}
e\sin\omega_{\rm G} t= 2.6\times 10^{-18}\cdot\sin\omega_{\rm G} t,
\label{eq:tau_Ad}
\end{eqnarray}
which is equivalent to a contribution of $\sim 7.1\times 10^{-13}$~m to the round-trip travel and is clearly negligible for GRACE-FO.

Similarly we evaluate the third term in the square brackets of the first integral of (\ref{eq:dn_ABA2}). Using the anticipated similarity of the orbits of the two GRACE-FO spacecraft, we evaluate the ratio involving the proper times as
\begin{eqnarray}
\Big(\frac{d\tau_{\rm B}}{dt}\Big)_{t_2}\Big(\frac{d\tau_{\rm A}}{dt}\Big)^{-1}_{t_3}-1&=&
\frac{1}{c^2}\Big[{\textstyle\frac{1}{2}}\big({{\vec v}^2_{\rm B}-{\vec v}^2_{\rm A}}\big)+U_{\rm E}({\vec y}_{\rm B})-U_{\rm E}({\vec y}_{\rm A})\Big]+{\cal O}(c^{-4})\approx \nonumber\\
&\approx &
\frac{2GM}{c^2}\frac{d_{\rm AB}}{a^2}e\sin\omega_{\rm G} t= 5.2\times 10^{-14}\cdot\sin\omega_{\rm G} t,
\label{eq:tau-comb-AB}
\end{eqnarray}
As we shall see below, in equations that model observables this term will always appear multiplied by the small factor $f^{\rm off}_{\rm B}/f_{\rm A0}=2.13\times 10^{-8}$, where $f^{\rm off}_{\rm B}=6\times 10^{6}$~Hz is the offset frequency \cite{Pierce-etal:2008} and $f_{\rm A0}=2.82\times 10^{14}$~Hz, which is set by the operating wavelength $\lambda_{\rm A0}=1064$~nm. Consequently we will omit terms containing the left side of Eq.~(\ref{eq:tau-comb-AB}).

Therefore, accounting for (\ref{eq:1-term})--(\ref{eq:tau-comb-AB}) and treating the frequencies $f_{\rm A0}(\tau_{\rm A3})=f_{\rm A0}$ and $f^{\rm off}_{\rm B}(\tau_{\rm B2})=f^{\rm off}_{\rm B}$ to be constant, the phase difference (\ref{eq:dn_ABA2}), formed between the phase of the coherently retransmitted light completing its two-way round-trip and the phase of the local oscillator at the moment of reception, may be given by the following expression:
{}
\begin{eqnarray}
\frac{\Delta\varphi(\tau_{\rm A3})}{2\pi}&=&
\frac{1}{c} \int_{\tau^0_{\rm A3}}^{\tau_{\rm A3}}
 \Big[f_{\rm A0}\Big(\dot{\cal R}_{\rm AB}(t'_{\rm 3})+\dot{\cal R}_{\rm BA}(t'_{\rm 3})\Big)+
f^{\rm off}_{\rm B}\dot{\cal R}_{\rm BA}(t'_{\rm 3})\Big]d\tau_{\rm A3}-
f^{\rm off}_{\rm B}
\,\big(\tau_{\rm A3}-\tau^0_{\rm A3}\big).~~~~~
\label{eq:dn_ABA2=+*}
\end{eqnarray}

We can now introduce the inter-satellite range $\rho_{\rm AB}(t)$:
\begin{eqnarray}
\rho_{\rm AB}(t_3)=\frac{1}{2}\Big({\cal R}_{\rm AB}(t_3)+{\cal R}_{\rm BA}(t_3)\Big)+
\frac{f^{\rm off}_{\rm B}}{2f_{\rm A0}+f^{\rm off}_{\rm B}}\frac{1}{2}\Big({\cal R}_{\rm BA}(t_3)-{\cal R}_{\rm AB}(t_3)\Big),
\label{eq:rho_0}
\end{eqnarray}
where ${\cal R}_{\rm AB}(t_3)\equiv {\cal R}_{\rm AB}(t^*_3)$ is taken at a delayed time given as $t_2\equiv t^*_3=t_3-c^{-1}{\cal R}_{\rm BA}(t_3)$, in accord with (\ref{eq:T_BA2}). The first term in (\ref{eq:rho_0}) is the geometric range between the two spacecraft, $\rho^{\,0}_{\rm AB}(t_3)$, which is defined using (\ref{eq:transm-time:ABA}) as:
\begin{eqnarray}
\rho^{\,0}_{\rm AB}(t_3)&\equiv&{\textstyle\frac{1}{2}}c(t_3-t_1)=
{\textstyle\frac{1}{2}}
\Big[{\cal R}_{\rm AB}({\vec x}_{\rm A1},{\vec x}_{\rm B2})+{\cal R}_{\rm BA}({\vec x}_{\rm B2},{\vec x}_{\rm A3})\Big],
\label{eq:range=}
\end{eqnarray}
where ${\cal R}_{\rm AB}({\vec x}_{\rm A1},{\vec x}_{\rm B2})$ and ${\cal R}_{\rm BA}({\vec x}_{\rm B2},{\vec x}_{\rm A3})$ expressed as functions of time $t_3$.
The second term in (\ref{eq:rho_0}) is a correction to the physical range between the two spacecraft due to the fact that the frequency of the signal that will be used to measure the one-way range ${\cal R}_{\rm BA}$ is higher by $f^{\rm off}_{\rm B}$ compared to the frequency that will be used to measure ${\cal R}_{\rm AB}$. It is also convenient to introduce the
inter-satellite range rate $\dot\rho_{\rm AB}(t_3)=d\rho_{\rm AB}(t_3)/dt_3$:
{}
\begin{eqnarray}
\dot\rho_{\rm AB}(t_3)=\frac{1}{2}\Big(\dot{\cal R}_{\rm AB}(t_3)+\dot{\cal R}_{\rm BA}(t_3)\Big)+
\frac{f^{\rm off}_{\rm B}}{2f_{\rm A0}+f^{\rm off}_{\rm B}}\frac{1}{2}\Big(\dot{\cal R}_{\rm BA}(t_3)-\dot{\cal R}_{\rm AB}(t_3)\Big).
\label{eq:range-rate=}
\end{eqnarray}

Substituting (\ref{eq:range-rate=}) into (\ref{eq:dn_ABA2=+*}), we can present one of the LRI observables, the phase difference, as
{}
\begin{eqnarray}
\frac{\Delta\varphi(\tau_{\rm A3})}{2\pi}&=&
\frac{1}{c} \big(2f_{\rm A0}+f^{\rm off}_{\rm B}\big)
\int_{\tau^0_{\rm A3}}^{\tau_{\rm A3}}
\dot\rho_{\rm AB}(t'_3)d\tau'_{\rm A3}
-f^{\rm off}_{\rm B}
\,\big(\tau_{\rm A3}-\tau^0_{\rm A3}\big).
\label{eq:dn_ABA2=+}
\end{eqnarray}

The primary LRI observable on GRACE-FO is the fractional number of cycles received at spacecraft $A$ per unit of proper time $\tau_{\rm A3}$ (as opposed to the absolute number of cycles \cite{Sheard_etal:2012}). This quantity can be developed by differentiating (\ref{eq:dn_ABA2=+}) with respect to proper time $ d\tau_{\rm A3}$, which results in the following:
{}
\begin{eqnarray}
\frac{d\Delta\varphi(\tau_{\rm A3})}{2\pi \, d\tau_{\rm A3}}&=&
\frac{1}{c} \big(2f_{\rm A0}+f^{\rm off}_{\rm B}\big)
\dot\rho_{\rm AB}(t_3)
-f^{\rm off}_{\rm B},
\label{eq:dn_ABA0+}
\end{eqnarray}
where the range rate $\dot\rho_{\rm AB}(t_{\rm 3})\equiv\dot\rho_{\rm AB}\big(t_3(\tau_{\rm A3})\big)$ is given by  (\ref{eq:range-rate=}), with $t_3=t_3(\tau_{\rm A3})$ determined with (\ref{eq:proper-coord-t}).

In addition to phase and phase rate observables, LRI data will be used to numerically  compute phase rate fluctuations. A model for this quantity may be developed
by differentiating (\ref{eq:dn_ABA0+}) with respect to the proper time $\tau_{\rm A3}$:
{}
\begin{eqnarray}
\frac{d^2\Delta\varphi(\tau_{\rm A3})}{2\pi \, d\tau^2_{\rm A3}}&=&
\frac{1}{c} \big(2f_{\rm A0}+f^{\rm off}_{\rm B}\big)
\Big(\frac{d\tau_{\rm A}}{dt}\Big)^{-1}_{t_3}\ddot\rho_{\rm AB}(t_3),
\label{eq:Dopp-dot}
\end{eqnarray}
where, similarly to Eqs.~(\ref{eq:range=})--(\ref{eq:range-rate=}), the range acceleration $\ddot\rho_{\rm AB}(t_3)=d^2\rho_{\rm AB}(t_3)/dt_3^2$ may be computed from (\ref{eq:rho_0}).

\subsection{Phase difference observable and associated range}
\label{sec:range}

As we saw in the preceding sections, the instances of time corresponding to the events of emission, retransmission and reception on GRACE-FO are not independent and are linked by the light-cone equations. Given an instant of reception $t_3$, we can reconstruct the corresponding instances of retransmission ($t_2$) and original transmission ($t_1$). This is done by using light-cone equations discussed in Sec.~\ref{sec:pseudo-ranges} and with the help of instantaneous positions for both spacecraft, ${\vec x}_{\rm A}(t_3)$ and ${\vec x}_{\rm B}(t_3)$. In other words, we can express  $t_2$ and $t_1$ as functions of the final time of reception $t_3$: $t_2=t_2(t_3)$ and $t_1=t_1(t_3)$.

As a result, using (\ref{eq:dn_ABA2=+}), we present the observable phase difference $\Delta\varphi(\tau_{\rm A3})$, which is the difference between the phase of the coherently retransmitted signal completing its two-way round-trip and the phase of the local oscillator, relating this differential quantity to the inter-satellite range, $\rho_{\rm AB}$ as
{}
\begin{eqnarray}
\frac{c}{2f_{\rm A0}+f^{\rm off}_{\rm B}}\Big[\frac{\Delta\varphi(\tau_{\rm A3})}{2\pi}+f^{\rm off}_{\rm B}\,\big(\tau_{\rm A3}-\tau^0_{\rm A3}\big)\Big]&=&
\int_{\tau^0_{\rm A3}}^{\tau_{\rm A3}}
\dot\rho_{\rm AB}(t'_3)d\tau'_{\rm A3},
\label{eq:dn_ABA2=+*=6}
\end{eqnarray}
with the proper time $\tau_{\rm A}$ is related to GCRS time $t$ via (\ref{eq:proper-coord-t}).

The integral in (\ref{eq:dn_ABA2=+*=6}) depends on the coordinate-to-proper time conversion and may be evaluated as
{}
\begin{eqnarray}
\int_{\tau^0_{\rm A3}}^{\tau_{\rm A3}}
\dot\rho_{\rm AB}(t'_3)d\tau'_{\rm A3}=
\int_{t^0_{\rm 3}}^{t_{\rm 3}}
\dot\rho_{\rm AB}(t'_3)\Big(\frac{d\tau'_{\rm A3}}{dt'_3}\Big)dt'_3&=&
\rho_{\rm AB}(t_3)-\rho_{\rm AB}(t^0_3)+
\int_{t_3^0}^{t_3}
\dot\rho_{\rm AB}(t'_3)\Big[\Big(\frac{d\tau'_{\rm A3}}{dt'_3}\Big)-1\Big]dt'_3,
\label{eq:dn_ABA2=+*=*}
\end{eqnarray}
where the instantaneous range $\rho_{\rm AB}(t)$ is given by (\ref{eq:t-AB-lri2}).
Using (\ref{eq:kepler-r}), (\ref{eq:kepler-va}), and (\ref{eq:vAB}), we evaluate the integrand in the expression on the right-hand side of (\ref{eq:dn_ABA2=+*=*}) as below
{}
\begin{eqnarray}
\dot\rho_{\rm AB}(t)\Big[\Big(\frac{d\tau_{\rm A}}{dt}\Big)-1\Big]&=&-\frac{1}{c^2}({\vec n}_{\rm AB}\cdot{\vec v}_{\rm AB})\Big[\frac{{\vec v}^2_{\rm A}}{2}+U_{\rm E}({\vec y}_{\rm A})\Big]=\nonumber\\
&&\hskip -120pt
=-\frac{1}{c^2}\Big(\frac{G M_{\rm E}}{a}\Big)^{3/2}\frac{3d_{\rm AB}}{2a}\Big(e\cos\beta+\frac{17}{14}e^2\sin2\beta+{\cal O}(e^2)\Big)
=0.3~{\rm nm/s}\cdot\cos\omega_{\rm G}t+0.8~{\rm pm/s}\cdot\sin2\omega_{\rm G}t+{\cal O}(e^2).
\label{eq:int**}
\end{eqnarray}
The magnitude of the largest term here is almost 0.3 nm/s, it comes at the orbital frequency and, thus, after a sufficiently long integration it may contribute to the science measurements of GRACE-FO. Therefore, we advocate to keep the term on the right-hand side of (\ref{eq:dn_ABA2=+*=6})  in the integral form.

To develop an analytical expression for the physical inter-satellite range (\ref{eq:rho_0}), we start from the geometric range which is given by (\ref{eq:range=}). The optical path length ${\cal R}_{\rm AB}(t_3)$ as a function of $t_3$ is presented by (\ref{eq:nt-BA+**}) and (\ref{eq:shap-corr}), whereas ${\cal R}_{\rm BA}(t_3)$ is given by (\ref{eq:nt-BA2}).
Assuming that the quadrupole moment does not change during the round-trip light transit time, the geometric inter-satellite range (\ref{eq:range=}) has the form:
{}
\begin{eqnarray}
\hskip -10pt
\rho^{\,0}_{\rm AB}(t_{\rm 3})&=&
\frac{1}{2}\Big(|{\vec x}_{\rm A}(t_{\rm 3}) - {\vec x}_{\rm B}(t_{\rm 2})|+|{\vec x}_{\rm B}(t_{\rm 2}) - {\vec x}_{\rm A}(t_{\rm 1})|\Big)+\frac{GM_{\rm E}}{c^2}\ln\Big[\Big(\frac{r_{\rm A3}+r_{\rm B2}+R_{\rm B2A3}}{r_{\rm A3}+r_{\rm B2}-R_{\rm B2A3}}\Big)\Big(\frac{r_{\rm A1}+r_{\rm B2}+R_{\rm A1B2}}{r_{\rm A1}+r_{\rm B2}-R_{\rm A1B2}}\Big)\Big]-\nonumber\\
&&\hskip-50pt -\,
\frac{GM_{\rm E}}{6c^2}\Big\{\Big[\frac{(n_{\rm A3\epsilon}-k_\epsilon)(n_{\rm A3\lambda}-k_\lambda)}{(r_{\rm A3}-{\vec k}\cdot{\vec x}_{\rm A3})^2}
+\frac{1}{r_{\rm A3}}\frac{\gamma_{\epsilon\lambda}+n_{\rm A3\epsilon} n_{\rm A3\lambda}}{(r_{\rm A3}-{\vec k}\cdot{\vec x}_{\rm A3})}-
\frac{(n_{\rm B2\epsilon}-k_\epsilon)(n_{\rm B2\lambda}-k_\lambda)}{(r_{\rm B2}-{\vec k}\cdot{\vec x}_{\rm B2})^2}
-\frac{1}{r_{\rm B2}}\frac{\gamma_{\epsilon\lambda}+n_{\rm B2\epsilon} n_{\rm B2\lambda}}{(r_{\rm B2}-{\vec k}\cdot{\vec x}_{\rm B2})}+\nonumber\\
&&\hskip-13pt +\,
\frac{(n_{\rm B2\epsilon}+k_\epsilon)(n_{\rm B2\lambda}+k_\lambda)}{(r_{\rm B2}+{\vec k}\cdot{\vec x}_{\rm B2})^2}
+\frac{1}{r_{\rm B2}}\frac{\gamma_{\epsilon\lambda}+n_{\rm B2\epsilon} n_{\rm B2\lambda}}{(r_{\rm B2}+{\vec k}\cdot{\vec x}_{\rm B2})}-
\frac{(n_{\rm A1\epsilon}+k_\epsilon)(n_{\rm A1\lambda}+k_\lambda)}{(r_{\rm A1}+{\vec k}\cdot{\vec x}_{\rm A1})^2}
-\frac{1}{r_{\rm A1}}\frac{\gamma_{\epsilon\lambda}+n_{\rm A1\epsilon} n_{\rm A\lambda}}{(r_{\rm A1}+{\vec k}\cdot{\vec x}_{\rm A1})}\Big]J_{\rm E}^{\epsilon\lambda}\Big\},~~
\label{eq:t-AB-lri2}
\end{eqnarray}
were we neglected the terms that are ${\cal O}(c^{-3})$ and also those that are below 0.1 nm in range for GRACE-FO.

We use (\ref{eq:T_ABb})--(\ref{eq:nt-BA2}) to express the Euclidean distances traveled by light as functions of  geocentric time with $t_3\equiv t$, while $t_1=t_1(t)$ and $t_2=t_2(t)$, as given below:
{}
\begin{eqnarray}
\frac{1}{2}\Big(|{\vec x}_{\rm A}(t_{\rm 3}) - {\vec x}_{\rm B}(t_{\rm 2})|+|{\vec x}_{\rm B}(t_{\rm 2}) - {\vec x}_{\rm A}(t_{\rm 1})|\Big)&=&\nonumber\\
&&\hskip -160 pt =
d_{\rm AB}-\frac{1}{c}({\vec d}_{\rm AB}\cdot {\vec v}_{\rm AB})
+\frac{d_{\rm AB}}{2c^2}\Big({\vec v}_{\rm AB}^2+{\vec v}_{\rm A}^2+({{\vec n}}_{\rm AB}\cdot {\vec v}_{\rm B})^2+({\vec d}_{\rm AB}\cdot ({\vec a}_{\rm AB}-{\vec a}_{\rm A}))\Big)
+{\cal O}(c^{-3}, G),~~~
\label{eq:t-AB-lri3}
\end{eqnarray}
where all the terms on the right-hand side are taken at time $t_{\rm 3}\equiv t$. Expressing the remaining $GM_{\rm E}/c^2$ terms in Eq.~(\ref{eq:t-AB-lri2}) also as functions of $t_3$, and neglecting ${\cal O}(1/c^3)$ contributions, with the help of result (\ref{eq:t-AB-lri3}) we can present an expression for the range $\rho_{\rm AB}$ given by (\ref{eq:t-AB-lri2}) in the following form:
{}
\begin{eqnarray}
\rho^{\,0}_{\rm AB}(t_3)&=&
d_{\rm AB}-\frac{1}{c}({\vec d}_{\rm AB}\cdot {\vec v}_{\rm AB})
+\frac{d_{\rm AB}}{2c^2}\Big({\vec v}_{\rm AB}^2+{\vec v}_{\rm A}^2+({{\vec n}}_{\rm AB}\cdot {\vec v}_{\rm B})^2+({\vec d}_{\rm AB}\cdot ({\vec a}_{\rm AB}-{\vec a}_{\rm A}))\Big)+
\nonumber\\
&+&
\frac{2GM_{\rm E}}{c^2}\ln\Big[\frac{r_{\rm A}+r_{\rm B}+d_{\rm AB}}{r_{\rm A}+r_{\rm B}-d_{\rm AB}}\Big]-
\frac{GM_{\rm E}}{3c^2}\Big\{\big(\frac{n_{\rm B\epsilon}}{r^2_{\rm B}}-\frac{n_{\rm A\epsilon}}{r^2_{\rm A}}\big)k_\lambda +\big(\frac{n_{\rm B\lambda}}{r^2_{\rm B}}-\frac{n_{\rm A\lambda}}{r^2_{\rm A}}\big)k_\epsilon\Big\}J_{\rm E}^{\epsilon\lambda}+
\nonumber\\
&+&\frac{GM_{\rm E}}{6c^2}d_{\rm AB}\Big\{
\big(\gamma_{\epsilon\lambda}+2k_\epsilon k_\lambda\big)\big(\frac{1}{r^3_{\rm B}}+\frac{1}{r^3_{\rm A}}\big)
+\frac{3n_{\rm B\epsilon} n_{\rm B\lambda}}{r^3_{\rm B}}+\frac{3n_{\rm A\epsilon} n_{\rm A\lambda}}{r^3_{\rm A}}
\Big\}J_{\rm E}^{\epsilon\lambda}+
{\cal O}(0.5~{\rm nm}).~~~
\label{eq:t-AB-lri3=}
\end{eqnarray}
To further simplify the expression for the quadrupole term in (\ref{eq:t-AB-lri3=}), we took into account Eq.~(\ref{eq:rA.nAB}), which allowed us to introduce the following approximation holds: $({\vec k}\cdot{\vec x}_{\rm A})=({\vec n}_{\rm AB}\cdot{\vec x}_{\rm A})=-{\textstyle\frac{1}{2}}d_{\rm AB}+{\cal O}(e)$. The size of the error term of ${\cal O}(0.5~{\rm nm})$ in (\ref{eq:t-AB-lri3=}) is determined by the omitted quadrupole terms as a result of this approximation.

We can now estimate the sizes of the terms involved. Using (\ref{eq:dAB.vAB}),  the second term in (\ref{eq:t-AB-lri3=}) can be estimated:
{}
\begin{eqnarray}
\frac{1}{c}({\vec d}_{\rm AB}\cdot{\vec v}_{\rm AB})&=&
-\frac{1}{c}\sqrt{\frac{GM}{a}}\frac{d_{\rm AB}^2}{a}e\sin\omega_{\rm G}t\approx -272.6~\mu{\rm m}\cdot \sin\omega_{\rm G}t.
\label{eq:dAB.vAB/c}
\end{eqnarray}
In addition to the once per orbit (1/rev) periodic term, Eq.~(\ref{eq:dAB.vAB/c}) also contributes periodic terms up to the $\sim e^3$ order which will bring 2/rev and 3/rev terms that are important to the GRACE-FO science data analysis.

The combination of the terms in Eq.~(\ref{eq:t-AB-lri3=}),
\begin{eqnarray}
\frac{d_{\rm AB}}{2c^2}\Big({\vec v}_{\rm AB}^2+({\vec d}_{\rm AB}\cdot {\vec a}_{\rm AB})\Big)=\frac{d_{\rm AB}}{2c^2}\,\frac{d}{dt}({\vec d}_{\rm AB}\cdot {\vec v}_{\rm AB})=-\frac{GM}{2c^2}\frac{d^3_{\rm AB}}{a^3}e\cos\omega_{\rm G}t=1.38\times 10^{-10}~{\rm m}\cdot\cos\omega_{\rm G}t,
\label{eq:1/c^2}
\end{eqnarray}
can be evaluated using (\ref{eq:dAB.vAB/c}) and (\ref{eq:E}). This combination of terms is too small to be accounted for in the range model.

Next, we look at the remaining $1/c^2$ terms in (\ref{eq:t-AB-lri3=}) and evaluate them:
{}
\begin{eqnarray}
\frac{d_{\rm AB}}{2c^2}\Big({\vec v}_{\rm A}^2+({{\vec n}}_{\rm AB}\cdot {\vec v}_{\rm B})^2-({\vec d}_{\rm AB}\cdot {\vec a}_{\rm A})\Big)&=&\frac{GM_{\rm E}}{c^2}\frac{d_{\rm AB}}{a}(1+2e\cos\omega_{\rm G}t)+\frac{GM_{\rm E}}{c^2}\frac{d^3_{\rm AB}}{4a^3}=\nonumber\\
&=&1.76\times 10^{-4}~{\rm m}+3.51\times 10^{-7}~{\rm m}\cdot\cos\omega_{\rm G}t+6.88\times 10^{-8}~{\rm m}.
\label{eq:1/c^2a}
\end{eqnarray}
Therefore, all of these terms should be included in the range model that is accurate to 1~nm.

The Shapiro term in (\ref{eq:t-AB-lri3=}) was already evaluated in (\ref{eq:Shapiro}) and, with a magnitude of 376.5~$\mu$m, it must be accounted for. A periodic contribution from the Shapiro term appears because of the eccentricity. As such, it comes at the orbital frequency and was estimated to be $0.377~\mu{\rm m}\cdot\cos\omega_{\rm G}t$, which is significant.

The first among the quadrupole terms in (\ref{eq:t-AB-lri3=}), which is antisymmetric under the exchange of A and B, can be estimated as
{}
\begin{eqnarray}
-\frac{GM_{\rm E}}{3c^2}\Big\{\big(\frac{n_{\rm B\epsilon}}{r^2_{\rm B}}-\frac{n_{\rm A\epsilon}}{r^2_{\rm A}}\big)k_\lambda +\big(\frac{n_{\rm B\lambda}}{r^2_{\rm B}}-\frac{n_{\rm A\lambda}}{r^2_{\rm A}}\big)k_\epsilon\Big\}J_{\rm E}^{\epsilon\lambda}
\approx \frac{2}{c^2}a_{\rm AB} R_\oplus^2 J_{2\oplus}||j_{\rm E}^{\epsilon\lambda}||=\frac{2GM_{\rm E}}{c^2}\frac{d_{\rm AB}}{a^3}R_\oplus^2 J_{2\oplus}=3.32\times 10^{-7}~{\rm m}.
\label{eq:1/c^2J}
\end{eqnarray}
The second quadrupole term, symmetric under the exchange of $A$ and $B$, is evaluated to be nearly of the same size as the first one $\sim4.98\times 10^{-7}~{\rm m}$.  The periodic terms in the quadrupole contribution are further reduced by the eccentricity down to $\sim 5\times 10^{-10}$~m, have 1/rev periodicity and are omitted here. Thus, both of these terms should be present in the 1~nm range model.

To develop the second term in (\ref{eq:dn_ABA2=+*}), we observe that its contribution will be reduced by the small factor of $({f^{\rm off}_{\rm B}}/{4f_{\rm A0}})=5.3\times 10^{-9}$, which determines the size of the terms that we would need to keep in ${\cal R}_{\rm BA}(t)-{\cal R}_{\rm AB}(t)$.
Using (\ref{eq:shap-corr}), (\ref{eq:nt-BA+**}) and (\ref{eq:nt-BA2}), to sufficient accuracy we have
that
\begin{equation}
\frac{f^{\rm off}_{\rm B}}{2f_{\rm A0}+f^{\rm off}_{\rm B}}\frac{1}{2}\Big({\cal R}_{\rm BA}(t)-{\cal R}_{\rm AB}(t)\Big)=-
\frac{f^{\rm off}_{\rm B}}{2f_{\rm A0}}\frac{1}{c}({\vec d}_{\rm AB}\cdot {\vec v}_{\rm A})=\frac{f^{\rm off}_{\rm B}}{2f_{\rm A0}}\frac{1}{c}\sqrt{\frac{GM}{a}}d_{\rm AB}=73~{\rm nm}.
\end{equation}
Thus, this constant term deserves to be in the range model.

Substituting the results obtained in Eq.~(\ref{eq:rho_0}), we can present the  inter-satellite range in the form
{}
\begin{eqnarray}
\rho_{\rm AB}(t)&=&
d_{\rm AB}-\frac{1}{c}({\vec d}_{\rm AB}\cdot {\vec v}_{\rm AB})+\frac{d_{\rm AB}}{2c^2}\Big({\vec v}_{\rm A}^2+({{\vec n}}_{\rm AB}\cdot {\vec v}_{\rm B})^2-({\vec d}_{\rm AB}\cdot {\vec a}_{\rm A})\Big)+
\frac{2GM_{\rm E}}{c^2}\ln\Big[\frac{r_{\rm A}+r_{\rm B}+d_{\rm AB}}{r_{\rm A}+r_{\rm B}-d_{\rm AB}}\Big]-
\nonumber\\
&&\hskip -48pt -
\frac{GM_{\rm E}}{3c^2}\Big\{\big(\frac{n_{\rm B\epsilon}}{r^2_{\rm B}}-\frac{n_{\rm A\epsilon}}{r^2_{\rm A}}\big)k_\lambda +\big(\frac{n_{\rm B\lambda}}{r^2_{\rm B}}-\frac{n_{\rm A\lambda}}{r^2_{\rm A}}\big)k_\epsilon
-\frac{d_{\rm AB}}{2}\Big(
\big(\gamma_{\epsilon\lambda}+2k_\epsilon k_\lambda\big)\big(\frac{1}{r^3_{\rm B}}+\frac{1}{r^3_{\rm A}}\big)
+\frac{3n_{\rm B\epsilon} n_{\rm B\lambda}}{r^3_{\rm B}}+\frac{3n_{\rm A\epsilon} n_{\rm A\lambda}}{r^3_{\rm A}}\Big)
\Big\}J_{\rm E}^{\epsilon\lambda}-\nonumber\\
&-&
\frac{f^{\rm off}_{\rm B}}{2f_{\rm A0}+f^{\rm off}_{\rm B}}\frac{1}{c}({\vec d}_{\rm AB}\cdot {\vec v}_{\rm A})+
{\cal O}(0.5~{\rm nm}),~~~
\label{eq:range-ABA0}
\end{eqnarray}
where all the quantities involved are functions of the GCRS time $t$ (were we set $t_{\rm 3}\equiv t$).
With an ultimate precision of 0.5~nm, the model for LRI-enabled range given in Eq.~(\ref{eq:range-ABA0}) accounts for all the terms that one needs to include in order to develop a model for the range observable for GRACE-FO measurements accurate to 1 nm.

Equation~(\ref{eq:dn_ABA2=+*=6}) together with the expression for the modeled range (\ref{eq:range-ABA0}), are the equations needed for processing the LRI-enabled phase difference observable on GRACE-FO.

\subsection{Phase rate observable and associated range rate}
\label{sec:range-rate}

The phase rate observable at spacecraft $A$ is obtained by differentiating the interferometric phase difference observable with respect to the proper time at the moment of signal reception, as given in Eq.~(\ref{eq:dn_ABA0+}). We can use this equation to solve for the range rate, $\dot\rho$, allowing us to express this quantity in terms of the proper time derivative of the phase difference $\Delta\varphi$:
{}
\begin{eqnarray}
\frac{c}{2f_{\rm A0}+f^{\rm off}_{\rm B}}\Big[\frac{d\Delta\varphi(\tau_{\rm A})}{2\pi \, d\tau_{\rm A}}+f^{\rm off}_{\rm B}\Big]&=&\dot\rho_{\rm AB}(t).
\label{eq:dn-ba50}
\end{eqnarray}
To derive an explicit expression for the range rate, $\dot\rho_{\rm AB}$, we differentiate the range, $\rho_{\rm AB}$, (\ref{eq:range-ABA0}) with respect to time. To the required $1/c^3$ order, the derivative $d\rho_{\rm AB}(t)/dt$ is obtained as:
{}
\begin{eqnarray}
{\dot \rho_{\rm AB}}(t)&=&({\vec n}_{\rm AB}\cdot {\vec v}_{\rm AB})-\frac{1}{c}\Big({\vec v}_{\rm AB}^2+({\vec a}_{\rm AB}\cdot {\vec d}_{\rm AB})\Big)+\frac{1}{2c^2}\Big\{\big({\vec v}^2_{\rm A}-({\vec n}_{\rm AB}\cdot {\vec v}_{\rm B})^2-({\vec d}_{\rm AB}\cdot {\vec a}_{\rm A})\big)({\vec n}_{\rm AB}\cdot {\vec v}_{\rm AB})
+\nonumber\\
&+&
2({\vec v}_{\rm AB}\cdot{\vec v}_{\rm B})({\vec n}_{\rm AB}\cdot {\vec v}_{\rm B})+
d_{\rm AB}\Big(2({\vec v}_{\rm A}\cdot {\vec a}_{\rm A})+2({\vec n}_{\rm AB}\cdot {\vec v}_{\rm B})({\vec n}_{\rm AB}\cdot{\vec a}_{\rm B})-({\vec v}_{\rm AB}\cdot {\vec a}_{\rm A})-({\vec d}_{\rm AB}\cdot\dot{\vec a}_{\rm A})\Big)\Big\}+\nonumber\\
&+&
\frac{4GM}{c^2}\Big[
\frac{({\vec n}_{\rm AB}\cdot {\vec v}_{\rm AB})}{(r_{\rm A}+r_{\rm B})}-
\frac{d^{}_{\rm AB}\big(({\vec n}_{\rm B}\cdot {\vec v}_{\rm B})+({\vec n}_{\rm A}\cdot {\vec v}_{\rm A})\big)}{(r_{\rm A}+r_{\rm B})^2}
\Big]-\frac{f^{\rm off}_{\rm B}}{2f_{\rm A0}+f^{\rm off}_{\rm B}}\frac{1}{c}\Big(({\vec v}_{\rm AB}\cdot {\vec v}_{\rm A})+({\vec d}_{\rm AB}\cdot {\vec a}_{\rm A})\Big)+
\nonumber\\
&&~~\hskip -55pt +\frac{GM_{\rm E}}{3c^2}\Big\{
-\Big(\frac{v_{\rm B}^\mu}{r_{\rm B}^3}-\frac{v_{\rm A}^\mu}{r_{\rm A}^3}\Big)\Big(\gamma_{\epsilon\mu}k_\lambda+\gamma_{\lambda\mu}k_\epsilon\Big)+
\frac{d_{\rm AB}}{2}\Big(
\frac{v_{\rm B}^\mu}{r_{\rm B}^4}\big(
\gamma_{\epsilon\mu}n_{\rm B\lambda}+
\gamma_{\lambda\mu}n_{\rm B\epsilon}\big)+
\frac{v_{\rm A}^\mu}{r_{\rm A}^4}\big(
\gamma_{\epsilon\mu}n_{\rm A\lambda}+
\gamma_{\lambda\mu}n_{\rm A\epsilon}\big)\Big)
\Big\}J^{\epsilon\lambda}_{\rm E}.~~~
\label{eq:rat-AB0}
\end{eqnarray}
Note that while evaluating the quadrupole terms in (\ref{eq:rat-AB0}), we omitted the terms that contain a dot product between a positional unit vector, ${\vec n}$, and a velocity vector, ${\vec v}$. According to (\ref{eq:rva0}), the overall contribution of a term containing such a product is multiplied by the orbital eccentricity, $e$, as $({\vec n}\cdot {\vec v})\simeq|{\vec v}|e$, and, thus, it will be $e$ times smaller than the other terms in that expression. Recognizing the fact that the quadrupole terms in (\ref{eq:rat-AB0}) are already very small, we therefore omitted those quadrupole terms that contain $\dot r_{\rm A}=({\vec n}_{\rm A}\cdot {\vec v}_{\rm A}),$ $\dot r_{\rm B}=({\vec n}_{\rm B}\cdot {\vec v}_{\rm B})$, and $\dot d_{\rm AB}=({\vec n}_{\rm AB}\cdot {\vec v}_{\rm AB})$.

As we discussed above, while developing the range model (\ref{eq:range-ABA0}), we omitted several 1/rev terms in the quadrupole contribution. The cumulative effect of these terms behaves as $\sim 0.5~{\rm nm}\cdot \sin\omega_{\rm G}t$, setting the accuracy limit for the range model (\ref{eq:range-ABA0}). The accuracy of the range rate model (\ref{eq:rat-AB0}) may be determined directly by using this quadrupole contribution. To do that, we differentiate this term with respect to time and observe that the range rate model (\ref{eq:rat-AB0}) is accurate to $\sim1~{\rm pm/s}\cdot \cos\omega_{\rm G}t$, which is sufficient to study the long-term tends and seasonal variations in the GRACE-FO data. At the same time, in order to detect gravity variations on short spatial scales, GRACE-FO will use short data arcs integrating interferometric phase over periods of time
of 30--100 sec. Clearly, in this case one does not need the full precision available in the range rate model (\ref{eq:rat-AB0}); a simplified version is sufficient. We will develop such a model below.

We can now evaluate the magnitude of each of the terms in Eq.~(\ref{eq:rat-AB0}) using the basic orbital formation parameters for GRACE-FO. Defining the semi-major axis of GRACE-FO spacecraft as $a=R_\oplus+h_G$, with $h_\oplus$ being spacecraft altitude, we see that the first term in (\ref{eq:rat-AB0}) can be evaluated using Eqs.~(\ref{eq:alpha}), (\ref{eq:nAB.vAB}) as
{}
\begin{eqnarray}
({\vec n}_{\rm AB}\cdot{\vec v}_{\rm AB})&\approx&-
\sqrt{\frac{GM}{a}}\frac{d_{\rm AB}}{a}e\sin\beta_{\rm AB}=30.3~{\rm cm/s}\cdot\sin\omega_{\rm G}t +{\cal O}(e^2).
\label{eq:nAB.vAB-ap}
\end{eqnarray}
The size of this term motivates including all terms up to the $\sim e^4$ order. As a result, the term (\ref{eq:nAB.vAB-ap}) will contribute at several different frequencies ranging from one to four times the orbital frequency (or from 1/rev to 4/rev), thereby affecting the GRACE-FO science program.

Next, to estimate the size of the $1/c$ term, we observe that this combination represents a full time derivative of $({\vec v}_{\rm AB}\cdot {\vec d}_{\rm AB})/c$, the magnitude of which was estimated in (\ref{eq:dAB.vAB/c}). Using this result, we have
{}
\begin{eqnarray}
-\frac{1}{c}\Big({\vec v}^2_{\rm AB}+({\vec a}_{\rm AB}\cdot {\vec d}_{\rm AB})\Big)=-\frac{1}{c}\frac{d}{dt}({\vec v}_{\rm AB}\cdot {\vec d}_{\rm AB})&=& 272.6~\mu{\rm m}~\sqrt{\frac{GM}{a^3}}\cos\omega_{\rm G}t= 306~{\rm nm/s} \cdot\cos\omega_{\rm G}t.
\label{eq:dAB.vAB-rr}
\end{eqnarray}
Correspondingly, both of these terms must be accounted for in the range rate model for GRACE-FO. Also, this term will likely to contribute at 2/rev and, may be, even at 3/rev frequencies (especially, if there will be a small orbit mismatch between the two spacecraft). This possibility needs further investigation with realistic orbits for GRACE-FO constellation.

Among the $1/c^2$ terms, there are several that cancel each other at the level of 4 nm/s, ultimately resulting in the following estimate for the magnitude of this set of terms:
\begin{eqnarray}
&&\hskip -10pt
\frac{1}{2c^2}\Big\{\big({\vec v}^2_{\rm A}-({\vec n}_{\rm AB}\cdot {\vec v}_{\rm B})^2-({\vec d}_{\rm AB}\cdot {\vec a}_{\rm A})\big)({\vec n}_{\rm AB}\cdot {\vec v}_{\rm AB})+2({\vec v}_{\rm AB}\cdot{\vec v}_{\rm B})({\vec n}_{\rm AB}\cdot {\vec v}_{\rm B})+\nonumber\\
&&\hskip 43pt +\,
d_{\rm AB}\Big(2({\vec v}_{\rm A}\cdot {\vec a}_{\rm A})+2({\vec n}_{\rm AB}\cdot {\vec v}_{\rm B})({\vec n}_{\rm AB}\cdot{\vec a}_{\rm B})-({\vec v}_{\rm AB}\cdot {\vec a}_{\rm A})-({\vec d}_{\rm AB}\cdot\dot{\vec a}_{\rm A})\Big)\Big\}\approx
\nonumber\\
&&\hskip 60pt \approx-\,
\frac{1}{c^2}\Big(\frac{GM}{a}\Big)^\frac{3}{2}\frac{3d_{\rm AB}}{a}e\sin\beta_{\rm AB}=0.6~{\rm nm/s}\cdot\sin\omega_{\rm G}t.
\label{eq:term-c-3.1}
\end{eqnarray}

The Shapiro term in Eq.~(\ref{eq:rat-AB0}) evaluates as
\begin{eqnarray}
\frac{4GM}{c^2}\Big[\frac{({\vec n}_{\rm AB}\cdot {\vec v}_{\rm AB})}{(r_{\rm A}+r_{\rm B})}-
\frac{d^{}_{\rm AB}\big(({\vec n}_{\rm A}\cdot {\vec v}_{\rm A})+({\vec n}_{\rm B}\cdot {\vec v}_{\rm B})\big)}{(r_{\rm A}+r_{\rm B})^2}\Big]&\approx& \frac{1}{c^2}\Big(\frac{GM}{a}\Big)^\frac{3}{2}\frac{4d_{\rm AB}}{a}\,e\sin\beta_{\rm AB}=
0.8~{\rm nm/s}\cdot \sin\omega_{\rm G}t.
\label{eq:shap1}
\end{eqnarray}

Using $J_{2\oplus}\sim 1\times 10^{-3}$ for the Earth's quadrupole coefficient and introducing, for convenience, the quantity $j^{\epsilon\lambda}_{\rm E}=J^{\epsilon\lambda}_{\rm E}/(3J_{2\oplus}R^2_\oplus)\lesssim1$, the entire quadrupole contribution in Eq.~(\ref{eq:rat-AB0}) can be evaluated as
{}
\begin{eqnarray}
&&\frac{GM_{\rm E}}{3c^2}\Big\{-
\Big(\frac{v_{\rm B}^\mu}{r_{\rm B}^3}-\frac{v_{\rm A}^\mu}{r_{\rm A}^3}\Big)\big(\gamma_{\epsilon\mu}k_\lambda+
\gamma_{\lambda\mu}k_\epsilon\big)+
\frac{3d_{\rm AB}}{2}\Big(
\frac{v_{\rm B}^\mu}{r_{\rm B}^4}\big(
\gamma_{\epsilon\mu}n_{\rm B\lambda}+
\gamma_{\lambda\mu}n_{\rm B\epsilon}\big)+
\frac{v_{\rm A}^\mu}{r_{\rm A}^4}\big(
\gamma_{\epsilon\mu}n_{\rm A\lambda}+
\gamma_{\lambda\mu}n_{\rm A\epsilon}\big)\Big)
\Big\}J^{\epsilon\lambda}_{\rm E}\approx
\nonumber\\
&&\hskip +65pt
\approx\frac{1}{c^2}\Big(\frac{GM}{a}\Big)^\frac{3}{2}\frac{6d_{\rm AB}}{a}\frac{R^2_\oplus}{a^2}J_{2\oplus}\mu_{\rm G}\sin\beta_{\rm AB}\lesssim
1.1\mu_{\rm G}~{\rm nm/s}\cdot\sin\omega_{\rm G}t.
\label{eq:j2-eval}
\end{eqnarray}
The factor $\mu_{\rm G}$ introduced in (\ref{eq:j2-eval}) is defined as $\mu_{\rm G}={\textstyle\frac{1}{3}}\big(\hat v_{\rm AB\epsilon}k_{\lambda}\big)j^{\epsilon\lambda}_{\rm E}+\big(\hat v_{\rm A\epsilon}n_{\rm A\lambda}\big)j^{\epsilon\lambda}_{\rm E}$, with $\hat v^\epsilon_{\rm AB}$ and $\hat v^\epsilon_{\rm A}$ being the velocity unit vectors $\hat v^\epsilon_{\rm AB}=v^\epsilon_{\rm AB}/v_{\rm AB}$ and $\hat v^\epsilon_{\rm A}=v^\epsilon_{\rm A}/v_{\rm A}$, correspondingly. For the GRACE-FO configuration, it is estimated that $|\mu_{\rm G}|\lesssim {\textstyle{\frac{1}{3}}}$, suggesting that, at the level of precision expected in the GRACE-FO LRI observable, none of the terms in (\ref{eq:j2-eval}) would contribute to the range rate.

Finally, the frequency offset term in (\ref{eq:rat-AB0}) was evaluated to make a negligible contribution to range rate:
{}
\begin{eqnarray}
\frac{f^{\rm off}_{\rm B}}{2f_{\rm A0}+f^{\rm off}_{\rm B}}\frac{1}{c}\Big(({\vec v}_{\rm AB}\cdot {\vec v}_{\rm A})+({\vec d}_{\rm AB}\cdot {\vec a}_{\rm A})\Big)\approx
\frac{f^{\rm off}_{\rm B}}{2f_{\rm A0}}\frac{1}{c}
\Big(\frac{GM}{a}\Big)\frac{d_{\rm AB}}{a}e\,\sin\beta_{\rm AB}=
8.2\times 10^{-14}~{\rm m/s}\cdot\sin\omega_{\rm G}t.
\label{eq:rat-AB0-eval}
\end{eqnarray}

The contribution of Eq.~(\ref{eq:rat-AB0-eval}) is clearly insignificant. However, the contribution of the terms represented by Eqs.~(\ref{eq:term-c-3.1})--(\ref{eq:j2-eval}) must be treated with care. All these terms contribute to the range rate at a frequency of 1/rev, and such contributions are likely absorbed into estimates of non-gravitational noise, such as solar heating that has similar periodicity. However, the difference between orbital and solar heating frequencies (due to the drift of the day-night terminator line) can introduce a slowly varying unmodeled residual with an annual frequency; this term can either mask or enhance estimates of seasonal variability of components of the Earth's gravitational field.

Putting these considerations aside, small spatial scale resolution can be achieved even if the terms in (\ref{eq:term-c-3.1})--(\ref{eq:j2-eval}) are omitted. As a result, (\ref{eq:rat-AB0}), the model for LRI-enabled range rate, may be presented in a simplified form:
{}
\begin{equation}
\dot\rho_{\rm AB}(t)=({\vec n}_{\rm AB}\cdot {\vec v}_{\rm AB})-\frac{1}{c}\Big({\vec v}^2_{\rm AB}+({\vec a}_{\rm AB}\cdot {\vec d}_{\rm AB})\Big)+
{\cal O}(0.8~{\rm nm/s}\cdot\sin\omega_{\rm G}t).
\label{eq:range-rate}
\end{equation}
As before, the GCRS time $t$ expressed via proper time $\tau_{\rm A}$ as $t=t(\tau_{\rm A})$ using (\ref{eq:proper-coord-t}). Equation (\ref{eq:range-rate}) generalizes the usual Lorentz frequency transformation to the case of accelerated motion  \cite{Turyshev-etal:2012}. The first term in the equation above, as shown by (\ref{eq:nAB.vAB-ap}), contributes $\sim$30.3 cm/s to range rate, while the $1/c$ term was evaluated in (\ref{eq:dAB.vAB-rr}) to be $\sim306~{\rm nm/s}$. Correspondingly, both of these terms must be accounted for in the range rate model for GRACE-FO.

Finally, with the help of Eq.~(\ref{eq:range-rate}), the GRACE-FO LRI phase rate observable given in Eq.~(\ref{eq:dn-ba50}) takes the following form:
{}
\begin{eqnarray}
\frac{c}{2f_{\rm A0}+f^{\rm off}_{\rm B}}\Big[\frac{d\Delta\varphi(\tau_{\rm A})}{2\pi \, d\tau_{\rm A}}+f^{\rm off}_{\rm B}\Big]&=&
({\vec n}_{\rm AB}\cdot {\vec v}_{\rm AB})-\frac{1}{c}\Big({\vec v}^2_{\rm AB}+({\vec a}_{\rm AB}\cdot {\vec d}_{\rm AB})\Big)+
{\cal O}(0.8~{\rm nm/s}\cdot\sin\omega_{\rm G}t).
\label{eq:range-rate-0}
\end{eqnarray}
Note that because of the orbital design chosen for the GRACE-FO constellation, featuring nearly identical spacecraft orbits with a very small orbital eccentricity of $e=0.001$,
the range rate model (\ref{eq:range-rate-0}) does not include contributions from general relativistic terms even as it accounts, in the form as presented, all terms larger than $0.8~{\rm nm/s}\cdot\sin\omega_{\rm G}t$. 
At the same time, the GRACE-FO mission objectives include investigation of long-term phenomena in the Earth's gravity field. Any study of long term trends in the GRACE-FO data will require using long data arcs over several years. For these investigations, one would have to use the complete range rate model given by (\ref{eq:rat-AB0}), which is accurate to $\sim1~{\rm pm/s}\cdot \cos\omega_{\rm G}t$.

\subsection{Phase rate fluctuations and range acceleration}
\label{sec:Doppler-fluct}

Equation (\ref{eq:Dopp-dot}) defines fluctuations in the phase rate by connecting they with the the range acceleration $\ddot\rho_{\rm AB}(t_3)$ as
{}
\begin{eqnarray}
\frac{c}{2f_{\rm A0}+f^{\rm off}_{\rm B}}
\frac{d^2\Delta\varphi(\tau_{\rm A})}{2\pi \, d\tau^2_{\rm A}}\Big(\frac{d\tau_{\rm A}}{dt}\Big)&=&
\ddot\rho_{\rm AB}(t),
\label{eq:Dopp-dot+}
\end{eqnarray}
where $\ddot\rho_{\rm AB}(t_3)=d^2\rho_{\rm AB}(t_3)/dt_3^2$ is computed from (\ref{eq:rho_0}). The analytical form of range acceleration may be computed directly from (\ref{eq:range-rate}) (that came from the first two terms in (\ref{eq:range-ABA0})) by differentiating it with respect to time $t$:
{}
\begin{align}
\ddot\rho_{\rm AB}(t)&=\frac{1}{d_{\rm AB}}\Big({\vec v}^2_{\rm AB}-({\vec n}_{\rm AB}\cdot {\vec v}_{\rm AB})^2+({\vec d}_{\rm AB}\cdot {\vec a}_{\rm AB})\Big)-\nonumber\\
&-\frac{1}{c}\Big(2({\vec v}_{\rm AB}\cdot{\vec a}_{\rm AB})+(\dot{\vec a}_{\rm AB}\cdot {\vec d}_{\rm AB})+({\vec a}_{\rm AB}\cdot {\vec v}_{\rm AB})\Big)+
{\cal O}(0.7~{\rm pm/s}^2\cdot\sin\omega_{\rm G}t).
\label{eq:range-accel+!}
\end{align}
The first term in (\ref{eq:range-accel+!}) was evaluated with the help of (\ref{eq:nAB.vAB-ap}) to be $340~\mu{\rm m/s}^2$. Similarly, the second term was evaluated with the help of (\ref{eq:dAB.vAB-rr}) to be $343~{\rm pm/s}^2$. Thus, both of these terms must be included in the acceleration model.

As discussed in the preceding subsection \ref{sec:range-rate}, while the magnitude of the remaining error term is small, it is a systematic term with 1/rev periodicity. As such, while it may be absorbed into estimates for non-gravitational forces, it may be necessary to account for it fully when long-duration studies of GRACE-FO LRI data are used to estimate seasonal or annual variabilities of the Earth's gravitational field.

\section{Processing interferometric data using a dual one-way approach}
\label{sec:dowr}

To provide initial estimates for the GRACE-FO orbit and inter-spacecraft separation vector, the mission will rely on a Ka-band microwave system called the Ka-band ranging (KBR) system. This system records instances of signal reception times independently on both spacecraft using on-board clocks that are synchronized using standard algorithms \cite{Turyshev:2012nw}. When processed together, these observations can be used to estimate the double one-way range (DOWR) and double one-way range rate (DOWRR).

Motivated by the possibility that both GRACE-FO spacecraft will be able to capture high-resolution interferometric data with sufficient precision, we may also consider utilizing the interferometric data to extract DOWR-style observables. Such an approach is currently being developed as a technology demonstration for the Time Delay Interferometric Ranging (TDIR), proposed for the LISA mission \cite{Tinto:2004yw,Shaddock:2004ua}. The difference between the standard two-way LRI and the DOWR-style range and range-rate observables must be captured in the corresponding model. The model may also be used to develop appropriate instrumentation requirements to enable these observables for the GRACE-FO mission.

\begin{figure}[t]
\includegraphics[scale=0.67]{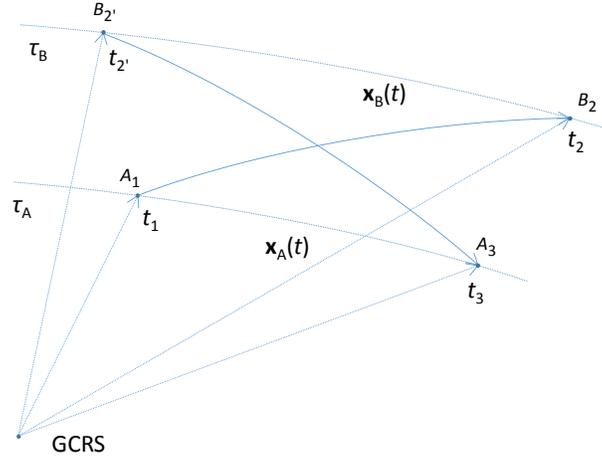}
\vskip -6pt
\caption{\label{fig:DOWR-events}
Timing events on GRACE-FO for a DOWR scenario: Depicted (not to scale) are the trajectories of the GRACE-FO-A and GRACE-FO-B spacecraft with corresponding proper times $\tau_{\rm A}$ and $\tau_{\rm B}$ and with four events in the GCRS, corresponding to a one-way signal transmission at $\vec{x}_{\rm A}(t_1)$ and its reception by the $B$ spacecraft at $\vec{x}_{\rm B}(t_2)$, and, similarly, another one-way signal transmission at $\vec{x}_{\rm B}(t'_2)$ by spacecraft $B$ and reception of this signal at $\vec{x}_{\rm A}(t_3)$.}
\end{figure}

Similarly to the standard LRI operations discussed in Sec.~\ref{sec:lri}, we model a signal that is transmitted by spacecraft $A$ at proper time $\tau_{\rm A1}$ and received by spacecraft $B$ at proper time $\tau_{\rm B2}$, as described by (\ref{eq:dn-ab++}) and depicted in Fig.~\ref{fig:DOWR-events}. In this case, the infinitesimal difference between the number of cycles initially emitted at $A$ and generated locally at the spacecraft $B$
(taking into account that the local oscillator is ramped by the offset frequency) is given by (\ref{eq:dn-AB*}) as:
{}
\begin{equation}
dn^{\rm tx}_{\rm AB}(\tau_{\rm B2})=\Big(f_{\rm B0}(\tau_{\rm B2})
+f_{\rm B}^{\rm off}(\tau_{\rm B2})
-f_{\rm A0}(\tau_{\rm A1})\frac{d\tau_{\rm A1}}{d\tau_{\rm B2}}\Big)d\tau_{\rm B2}.
\label{eq:n_AB-d}
\end{equation}
We no longer assume coherency in the form of Eq.~(\ref{eq:dn-ab2-coh}): $f_{\rm B0}(\tau_{\rm B2})$ and $f_{\rm A0}(\tau_{\rm A1})$ are now independent.

To develop the second observable, we assume that another signal was generated at spacecraft $B$ at proper time $\tau_{{\rm B2}'}$. This signal, consisting of a fractional number of cycles, $dn_{\rm B}(\tau_{{\rm B2}'})$ and combined with the offset frequency, will be received  and recorded at the spacecraft $A$ at proper time $\tau_{\rm A3}$. Similarly to (\ref{eq:dn-ba00}), the quantity of interest in the infinitesimal number of cycles $dn_{\rm BA}$, given as:
{}
\begin{equation}
dn_{\rm BA}(\tau_{\rm A3})=\Big(f_{\rm A0}(\tau_{\rm A3})-\big(f_{\rm B2}(\tau_{{\rm B2}'})+f_{\rm B}^{\rm off}(\tau_{{\rm B2}'})\big)\frac{d\tau_{{\rm B2}'}}{d\tau_{\rm A3}}\Big)d\tau_{\rm A3}.
\label{eq:n_BA-d}
\end{equation}

In the case of (\ref{eq:n_AB-d}), the quantity that will be recorded on spacecraft $B$ is a high-resolution time-series  of the infinitesimal difference in the fractional number of cycles between cycles generated by the local oscillator at the moment of reception, $n_{\rm B}(\tau_{\rm B2})$, and those originally emitted, $n_{\rm A}(\tau_{\rm A1})$.
Multiplied by $2\pi$, this provides the one-way phase difference (referenced to a local oscillator on board the receiving spacecraft) for the signals traveling from spacecraft $A$ to $B$:
{}
\begin{equation}
n^{\rm tx}_{\rm AB}(\tau_{\rm B2})=\frac{1}{2\pi}\Big(\varphi_{\rm B}(\tau_{\rm B2})
+2\pi f_{\rm B}^{\rm off}(\tau_{\rm B2})\tau_{\rm B2}
-\varphi_{\rm A}(\tau_{\rm A1})\Big)=\frac{\Delta\varphi^{\rm dowr}_{\rm AB}(\tau_{\rm B2})}{2\pi}.
\label{eq:n_AB-d-ph}
\end{equation}
Similarly, (\ref{eq:n_BA-d}) provides the second observable as:
\begin{equation}
n_{\rm BA}(\tau_{\rm A3})=\frac{1}{2\pi}\Big(\varphi_{\rm A}(\tau_{\rm A3})-\varphi_{\rm B}(\tau_{{\rm B2}'})-2\pi f_{\rm B}^{\rm off}(\tau_{{\rm B2}'})\tau_{{\rm B2}'}\Big)=\frac{\Delta\varphi^{\rm dowr}_{\rm BA}(\tau_{\rm A3})}{2\pi}.
\label{eq:n_BA-d-ph}
\end{equation}

The phase differences $\Delta\varphi_{\rm AB}(\tau_{\rm B2})$ and $\Delta\varphi_{\rm BA}(\tau_{\rm A3})$ will be recorded at spacecraft $B$ and $A$, respectively, and will be made available as a high-resolution time-series. Laser frequency stabilization on the order of $\delta f_{\rm A0}/f_{\rm A0}\leq 2\times 10^{-15}$, combined with
milli-cycle level phase interpolation (as opposed to a precision timing implemented for KBR on GRACE and GRAIL missions \cite{Turyshev:2012nw,Konopliv-etal:2013} and a
phase interpolation to the level of a micro-cycle that is being developed for LISA \cite{Shaddock:2004ua}),
may provide the conditions necessary to process these phase differences in a DOWR configuration.

Similarly to the development performed in Sec.~\ref{sec:lri} for LRI observables, we can now develop the DOWR observational model. We follow the approach presented in \cite{Turyshev-etal:2012}. Adding (\ref{eq:n_AB-d}) and (\ref{eq:n_BA-d}), while using the definitions for the phase difference observables $\Delta\varphi_{\rm AB}(\tau_{\rm B2})$ and $\Delta\varphi_{\rm BA}(\tau_{\rm A3})$ given by (\ref{eq:n_AB-d-ph}) and (\ref{eq:n_BA-d-ph}), we obtain the following expression:
{}
\begin{eqnarray}
d\Big(\frac{\Delta\varphi^{\rm dowr}_{\rm AB}(\tau_{\rm B2})}{2\pi}+\frac{\Delta\varphi^{\rm dowr}_{\rm BA}(\tau_{\rm A3})}{2\pi}\Big)
&=&
\Big(f_{\rm A0}(\tau_{\rm A3})-f_{\rm A0}(\tau_{\rm A1})
\Big(\frac{d\tau_{\rm A}}{dt}\Big)_{\rm 1}
\Big(\frac{d\tau_{\rm B}}{dt}\Big)^{-1}_{\rm 2}
\frac{d\tau_{\rm B2}}{d\tau_{\rm A3}}\Big)d\tau_{\rm A3}+
\nonumber\\
&&\hskip -118pt+\,
\Big(f_{\rm B0}(\tau_{\rm B2})+f_{\rm B}^{\rm off}(\tau_{\rm B2})-
\big[f_{\rm B0}(\tau_{{\rm B2}'})+
f_{\rm B}^{\rm off}(\tau_{{\rm B2}'})\big]
\Big(\frac{d\tau_{\rm B}}{dt}\Big)_{2'}
\Big(\frac{d\tau_{\rm A}}{dt}\Big)^{-1}_{\rm 3}
\frac{d\tau_{\rm A3}}{d\tau_{\rm B2}}\Big)d\tau_{\rm B2}+
\nonumber\\
&&\hskip -170pt+\,
\frac{1}{c}
f_{\rm A0}(\tau_{\rm A1})
\Big(\frac{d\tau_{\rm A}}{dt}\Big)_1
\Big(\frac{d\tau_{\rm B}}{dt}\Big)^{-1}_{\rm 2}
\frac{d{\cal R}_{\rm AB}(t_2)}{dt_2}
d\tau_{\rm B2}+
\frac{1}{c}
\Big(f_{\rm B0}(\tau_{{\rm B2}'})+f_{\rm B}^{\rm off}(\tau_{{\rm B2}'})\Big)
\Big(\frac{d\tau_{\rm B}}{dt}\Big)_{2'}
\Big(\frac{d\tau_{\rm A}}{dt}\Big)^{-1}_{\rm 3}
\frac{d{\cal R}_{\rm BA}(t_3)}{dt_3}
d\tau_{\rm A3}.
\label{eq:n_ABA-dowr}
\end{eqnarray}
We used the fact that coordinate times $t_1$ and $t_2$ are connected by the light cone equations (\ref{eq:T_ABb}), (\ref{eq:nt-BA+**}). To connect coordinate times $t_{2'}$ and $t_3$ we used the second pair of light cone equations, (\ref{eq:T_BA2}) and  (\ref{eq:nt-BA2}), replacing $t_{2}$ with $t_{2'}$.

Expression (\ref{eq:n_ABA-dowr}) describes a generic combination of the observable phase differences.
A model for DOWR may be obtained simply by assuming in this expression that the clocks on both spacecraft are synchronized. We can capture this assumption by equating the appropriate proper and coordinate times, namely $\tau_{\rm B2}=\tau_{\rm A3}=\tau$ and $t_{\rm 2}=t_{\rm 3}=t$. In addition, the fact that GRACE-FO will rely on nearly identical Keplerian orbits with very small eccentricities also allows for a simplification. Thus, relying on the results (\ref{eq:tau-comb-A})--(\ref{eq:tau-comb-AB}) and also on the numerical values of the GRACE-FO mission parameters from Table~\ref{tb:params}, and assuming that frequency stabilization and phase matching
goals are achieved, we can treat the frequencies $f_{\rm A0}$, $f_{\rm B0}$ and $f_{\rm B}^{\rm off}$ as constants and
present (\ref{eq:n_ABA-dowr}) in the following form:
{}
\begin{eqnarray}
d\Big(\frac{\Delta\varphi^{\rm dowr}_{\rm AB}(\tau_{\rm})}{2\pi}+\frac{\Delta\varphi^{\rm dowr}_{\rm BA}(\tau_{\rm})}{2\pi}\Big)
&=&
\frac{1}{c}\Big(f_{\rm A0}\dot{\cal R}_{\rm AB}(t)+
\big(f_{\rm B0}+f_{\rm B}^{\rm off}\big)
\dot{\cal R}_{\rm BA}(t)\Big)d\tau.
\label{eq:DOWR}
\end{eqnarray}

The inter-spacecraft range, $\rho^{\rm dowr}_{\rm AB}$, is defined by combining the one-way ranges in the following manner:
{}
\begin{eqnarray}
\rho^{\rm dowr}_{\rm AB}(t)&=&
\frac{1}{2}\Big({\cal R}_{\rm AB}(t)+{\cal R}_{\rm BA}(t)\Big)+
\frac{f_{\rm B0}+f^{\rm off}_{\rm B}-f_{\rm A0}}{f_{\rm A0}+f_{\rm B0}+f^{\rm off}_{\rm B}}\frac{1}{2}
\Big({\cal R}_{\rm BA}(t)-{\cal R}_{\rm AB}(t)\Big).~~~~
\label{eq:DOWR-def}
\end{eqnarray}
Although the functional form of (\ref{eq:DOWR-def}) is similar to that of (\ref{eq:dn_ABA2=+*}), the one-way light travel distances ${\cal R}_{\rm AB}(t)$ and ${\cal R}_{\rm BA}(t)$ are taken at the same time, $t$. This is contrary to (\ref{eq:dn_ABA2=+*}), where ${\cal R}_{\rm AB}(t^*)$ is taken at a delayed time of $t^*=t-c^{-1}{\cal R}_{\rm BA}(t)$.

Similarly to the (\ref{eq:rho_0}), the first term in (\ref{eq:DOWR-def}) is the geometric DOWR range between the two spacecraft, defined as $\rho^{\rm dowr\,0}_{\rm AB}(t)={\textstyle\frac{1}{2}}\big({\cal R}_{\rm AB}(t)+{\cal R}_{\rm BA}(t)\big)$. This quantify has the form:
{}
\begin{eqnarray}
\rho^{\rm dowr\,0}_{\rm AB}(t) &=& \frac{1}{2}\Big(|{\vec x}_{\rm A}(t_{\rm}) - {\vec x}_{\rm B}(t'_{\rm 2})|+|{\vec x}_{\rm B}(t_{\rm}) - {\vec x}_{\rm A}(t_{\rm 1})|\Big)+
\frac{2GM_{\rm E}}{c^2}\ln\Big[\frac{r_{\rm A}+r_{\rm B}+d_{\rm AB}}{r_{\rm A}+r_{\rm B}-d_{\rm AB}}
\Big]-
\nonumber\\
&&\hskip -48pt -
\frac{GM_{\rm E}}{3c^2}\Big\{\big(\frac{n_{\rm B\epsilon}}{r^2_{\rm B}}-\frac{n_{\rm A\epsilon}}{r^2_{\rm A}}\big)k_\lambda +\big(\frac{n_{\rm B\lambda}}{r^2_{\rm B}}-\frac{n_{\rm A\lambda}}{r^2_{\rm A}}\big)k_\epsilon
-\frac{d_{\rm AB}}{2}\Big(
\big(\gamma_{\epsilon\lambda}+2k_\epsilon k_\lambda\big)\big(\frac{1}{r^3_{\rm B}}+\frac{1}{r^3_{\rm A}}\big)
+\frac{3n_{\rm B\epsilon} n_{\rm B\lambda}}{r^3_{\rm B}}+\frac{3n_{\rm A\epsilon} n_{\rm A\lambda}}{r^3_{\rm A}}\Big)
\Big\}J_{\rm E}^{\epsilon\lambda}.~~~~
\label{eq:range-DOWR+}
\end{eqnarray}

To develop $\rho^{\rm dowr}_{\rm AB}$, we first apply light-cone equations from Appendix~\ref{sec:ranges}, to estimate the following quantity:
{}
\begin{eqnarray}
\frac{1}{2}\Big(|{\vec x}_{\rm A}(t_{\rm}) - {\vec x}_{\rm B}(t'_{\rm 2})|+|{\vec x}_{\rm B}(t_{\rm}) - {\vec x}_{\rm A}(t_{\rm 1})|\Big) &=&
d_{\rm AB}-\frac{1}{2c}({\vec d}_{\rm AB}\cdot {\vec v}_{\rm AB})
+\frac{d_{\rm AB}}{2c^2}\Big(({\vec v}_{\rm A}\cdot{\vec v}_{\rm B})+({{\vec n}}_{\rm AB}\cdot {\vec v}_{\rm A})({{\vec n}}_{\rm AB}\cdot {\vec v}_{\rm B})\Big),~~~~
\label{eq:range-DOWR-Euc}
\end{eqnarray}
which is accurate to ${\cal O}(c^{-3}, G).$
Note that the gravitational contributions to the DOWR and LRI ranges, given respectively by Eqs.~(\ref{eq:range-ABA0}) and (\ref{eq:range-DOWR}), are identical. Any difference would appear at the next level of approximation, $\propto G/c^3$, which is negligible for GRACE-FO.

Similarly to (\ref{eq:dn_ABA2=+*}), the contribution of the second term in (\ref{eq:DOWR-def}) will be reduced by the small factor of $({f^{\rm off}_{\rm B}}/{4f_{\rm A0}})=5.3\times 10^{-9}$, which determines the size of the terms that we would need to keep in ${\cal R}_{\rm BA}(t)-{\cal R}_{\rm AB}(t)$. Using (\ref{eq:nt-BA+**}) and (\ref{eq:nt-BA2}) and accounting for the fact that $f_{\rm A0}\approx f_{\rm B0}$, to sufficient accuracy we have
that
\begin{eqnarray}
\frac{f_{\rm B0}+f^{\rm off}_{\rm B}-f_{\rm A0}}{f_{\rm A0}+f_{\rm B0}+f^{\rm off}_{\rm B}}\frac{1}{2}
\Big({\cal R}_{\rm BA}(t)-{\cal R}_{\rm AB}(t)\Big)&=&-
\frac{f^{\rm off}_{\rm B}}{2f_{\rm A0}}\frac{1}{2c}\big({\vec d}_{\rm AB}\cdot({\vec v}_{\rm A}+{\vec v}_{\rm B})\big)=
\nonumber\\
&=&
\frac{f^{\rm off}_{\rm B}}{2f_{\rm A0}}\frac{1}{2c}\sqrt{\frac{GM}{a}}\frac{d^2_{\rm AB}}{a}e\sin\beta_{\rm AB}=1.45~{\rm pm}\cdot\sin\omega_{\rm G}t.
\end{eqnarray}
Therefore, contribution of the frequency offset to the DOWR range is negligible and, thus, it may be omitted.

Finally, after considering all the simplifying assumptions above, the observational equation to process the DOWR data with interferometric phase differences, $\Delta\varphi^{\rm dowr}_{\rm AB}$ and $\Delta\varphi^{\rm dowr}_{\rm AB}$, recorded at the GRACE-FO spacecraft takes the form:
{}
\begin{eqnarray}
\frac{c}{f_{\rm A0}+f_{\rm B0}+f_{\rm B}^{\rm off}}
\Big(
\frac{\Delta\varphi^{\rm dowr}_{\rm AB}(\tau)}{2\pi}+
\frac{\Delta\varphi^{\rm dowr}_{\rm BA}(\tau)}{2\pi}
\Big)
=\int_{\tau_0}^\tau
\dot{\rho}^{\rm dowr}_{\rm BA}(t)d\tau,~~~~
\label{eq:DOWR+}
\end{eqnarray}
which is accurate to the order of ${\cal O}(0.3~\mu{\rm m})$. The inter-satellite range $\rho^{\rm dowr}_{\rm AB}(t)$, as expressed via instantaneous quantities, was obtained by substituting (\ref{eq:range-DOWR-Euc}) in (\ref{eq:range-DOWR+}) and was found to have the form:
{}
\begin{eqnarray}
\rho^{\rm dowr}_{\rm AB}(t) &=& d_{\rm AB}-\frac{1}{2c}({\vec d}_{\rm AB}\cdot {\vec v}_{\rm AB})
+\frac{d_{\rm AB}}{2c^2}\Big(({\vec v}_{\rm A}\cdot{\vec v}_{\rm B})+({{\vec n}}_{\rm AB}\cdot {\vec v}_{\rm A})({{\vec n}}_{\rm AB}\cdot {\vec v}_{\rm B})\Big)+
\frac{2GM_{\rm E}}{c^2}\ln\Big[\frac{r_{\rm A}+r_{\rm B}+d_{\rm AB}}{r_{\rm A}+r_{\rm B}-d_{\rm AB}}
\Big]-
\nonumber\\
&&\hskip -48pt -
\frac{GM_{\rm E}}{3c^2}\Big\{\big(\frac{n_{\rm B\epsilon}}{r^2_{\rm B}}-\frac{n_{\rm A\epsilon}}{r^2_{\rm A}}\big)k_\lambda +\big(\frac{n_{\rm B\lambda}}{r^2_{\rm B}}-\frac{n_{\rm A\lambda}}{r^2_{\rm A}}\big)k_\epsilon
-\frac{d_{\rm AB}}{2}\Big(
\big(\gamma_{\epsilon\lambda}+2k_\epsilon k_\lambda\big)\big(\frac{1}{r^3_{\rm B}}+\frac{1}{r^3_{\rm A}}\big)
+\frac{3n_{\rm B\epsilon} n_{\rm B\lambda}}{r^3_{\rm B}}+\frac{3n_{\rm A\epsilon} n_{\rm A\lambda}}{r^3_{\rm A}}\Big)
\Big\}J_{\rm E}^{\epsilon\lambda}.
~~~~
\label{eq:range-DOWR}
\end{eqnarray}
This expression is accurate to 0.5 nm, which is adequate for DOWR measurements. The magnitudes of the terms in (\ref{eq:range-DOWR}) were estimated to be: the $\sim 1/c$ term is $136.3~\mu$m (in part, because of the small eccentricity), the $\sim 1/c^2$ Sagnac term is $175.6~\mu$m,  the Shapiro term is 376.5~$\mu$m, and the entire quadrupole term  contributes up to 0.3~$\mu$m. Note the difference in this equation and its LRI counterpart, given by (\ref{eq:t-AB-lri3}), where the one of the legs of the round-trip travel was delayed by a half of the round-trip light-travel time.

Similarly, at the appropriate level of accuracy, the DOWR-enabled range rate model for GRACE-FO
will have the form:
{}
\begin{eqnarray}
\frac{c}{f_{\rm A0}+f_{\rm B0}+f_{\rm B}^{\rm off}}\Big(
\frac{d\Delta\varphi^{\rm dowr}_{\rm AB}(\tau)}{2\pi\,d\tau}&+&
\frac{d\Delta\varphi^{\rm dowr}_{\rm BA}(\tau)}{2\pi\,d\tau}
\Big)=\dot\rho^{\rm dowr}_{\rm AB}(t),~~~~~~
\label{eq:DOWRR}
\end{eqnarray}
where with accuracy sufficient for GRACE-FO, the model for range rate, $\dot\rho^{\rm dowr}_{\rm AB}$, is given as below
{}
\begin{equation}
\dot\rho^{\rm \,dowr}_{\rm AB}(t)=({\vec n}_{\rm AB}\cdot {\vec v}_{\rm AB})-\frac{1}{2c}\Big({\vec v}^2_{\rm AB}+({\vec a}_{\rm AB}\cdot {\vec d}_{\rm AB})\Big)+
{\cal O}(0.8~{\rm nm/s}\cdot\sin\omega_{\rm G}t).
\label{eq:DOWRR-rho}
\end{equation}
The second term on the right-hand side would contribute only one half of (\ref{eq:range-rate}) or $0.153~\mu{\rm m/s} \cdot\cos\omega_{\rm G}t$.

Finally, we can develop the equation determining fluctuations in phase rate. Similarly to (\ref{eq:Dopp-dot+}), we differentiate (\ref{eq:DOWRR}) with respect to proper time, to obtain
{}
\begin{eqnarray}
\frac{c}{f_{\rm A0}+f_{\rm B0}+f_{\rm B}^{\rm off}}\Big[
\frac{d^2\Delta\varphi^{\rm dowr}_{\rm AB}(\tau)}{2\pi\,d\tau^2}\Big(\frac{d\tau_{\rm B}}{dt}\Big)+
\frac{d^2\Delta\varphi^{\rm dowr}_{\rm BA}(\tau)}{2\pi\,d\tau^2}\Big(\frac{d\tau_{\rm A}}{dt}\Big)\Big]=
\ddot\rho^{\rm \,dowr}_{\rm AB}(t),~~~~~~
\label{eq:DOWRA}
\end{eqnarray}
where the range acceleration $\ddot\rho^{\rm \,dowr}_{\rm AB}(t)$ may be computed directly from (\ref{eq:DOWRR-rho}) (that by itself came from the first two terms in (\ref{eq:range-DOWR})) by differentiating it with respect to time $t$:
{}
\begin{align}
\ddot\rho^{\rm \,dowr}_{\rm AB}(t)&=\frac{1}{d_{\rm AB}}\Big({\vec v}^2_{\rm AB}-({\vec n}_{\rm AB}\cdot {\vec v}_{\rm AB})^2+({\vec d}_{\rm AB}\cdot {\vec a}_{\rm AB})\Big)-\nonumber\\
&-\frac{1}{2c}\Big(2({\vec v}_{\rm AB}\cdot{\vec a}_{\rm AB})+(\dot{\vec a}_{\rm AB}\cdot {\vec d}_{\rm AB})+({\vec a}_{\rm AB}\cdot {\vec v}_{\rm AB})\Big)+
{\cal O}(0.7~{\rm pm/s}^2\cdot\sin\omega_{\rm G}t).
\label{eq:DOWR-accel}
\end{align}

Although Eqs.~(\ref{eq:range-DOWR}) and (\ref{eq:DOWRR-rho}) are consistent with the results that we obtained for the GRAIL mission \cite{Turyshev:2012nw}, they are different from the LRI-enabled measurements of range and range rate presented here in the form of Eqs.~(\ref{eq:range-ABA0})  and (\ref{eq:range-rate}). One clear difference between Eqs.~(\ref{eq:range-ABA0}) and (\ref{eq:range-DOWR}) and between Eqs.~(\ref{eq:range-rate}) and (\ref{eq:DOWRR-rho}) is the factor $\frac{1}{2}$ present in the $1/c$ terms for the DOWR-enabled measurements. This difference is due the fact that the LRI relies on a two-way experimental configuration. The signal traverses the inter-spacecraft distance twice before it is received by the interferometer and compared against the on-board laser oscillator. In contrast, the dual one-way DOWR-enabled range and
range rate observables are formed using signals that only traversed the inter-spacecraft distance once.

As GRACE-FO navigation and data analysis will also rely on a KBR system (similar to \cite{Turyshev:2012nw,Konopliv-etal:2013}) for initial estimation of the spacecraft's orbital parameters and, especially, for the inter-spacecraft separation vector, one needs to make sure that the appropriate range and range rate models (LRI vs. DOWR) are used for different types of observables.

Concluding, we emphasize that, although there is a possibility to digitally post-process the downlinked phase differences recorded at each spacecraft in order to simulate the conditions that may allow processing LRI data in a manner similar to DOWR processing, technical details of this procedure (i.e., details relevant to the on-board instrumentation and data flow algorithms) and their implementation on GRACE-FO are still in development. Nevertheless, we hope that the model we presented here will be useful to develop these capabilities.

\section{Conclusions and recommendations}
\label{sec:sonc}

We studied, in the post-Newtonian approximation of the general theory of relativity, the propagation of a plane electromagnetic wave traveling through the gravitational field in the vicinity of an extended body. An arbitrarily shaped and rotating distribution of matter is represented by an arbitrary energy-momentum tensor. We derived a compact closed form general relativistic solution describing the phase of a plane wave that accounts for contributions of all the mass and current multipoles of the body itself (Eqs.~(\ref{eq:eik-sol-all}) and (\ref{eq:eik-sol-all-S})), as well as the contribution due to the tidal gravity field (given by Eq.~(\ref{eq:eq_eik-phi-lamb-T1})) produced by the external bodies forming a system of $N$ astronomical bodies. As such, the solution that we obtained significantly extends previous similar derivations found in the literature.

We evaluated a solution for relativistic phase (\ref{eq:phase_t}), which, in addition to the usual Shapiro term, accounts for the quadrupole term of the Earth's mass distribution, the Earth's spin, and the tidal gravity introduced by the potentials of external bodies in the GCRS.  At the level of accuracy anticipated from GRACE-FO, the recommended solution for relativistic phase is given by (\ref{eq:eik-phi-lamb-EJ}). This formulation allows one to achieve a self-consistent analytical result that can be evaluated for the GRACE-FO LRI experiment. All the necessary information regarding the geometry of the experiment and the background gravitational field is captured in the total signal path ${\cal R}$, a quantity we introduced in Sec.~\ref{sec:GRACE-FO-phase}.

Based on the established solution, we presented a new formulation for the relativistic phase transformation that describes a coherent signal transmission between the two spacecraft of the GRACE-FO constellation. We developed LRI observables for GRACE-FO including both the phase and phase rate of the signal received at the master spacecraft, including relativistic treatment of the transponder offset frequency. Eq.~(\ref{eq:dn_ABA2=+*=6}), together with  the relativistic range model (\ref{eq:range-ABA0}), provide a high-precision formulation for the LRI phase observable on board GRACE-FO. Similarly, Eq.~(\ref{eq:range-rate-0}) provides a high-precision model for the phase rate observable on board GRACE-FO. These results will allow GRACE-FO to reach the desired resolution of 1 nm in range and 1~nm/s in range rate.

Our formulation justifies the basic assumptions behind the design of the GRACE-FO mission. In particular, our analysis demonstrates the importance of achieving nearly circular and nearly identical orbits (with eccentricity of $e\sim 0.001$) for the twin GRACE-FO spacecraft. If these requirements  are satisfied, the observables can be represented by a set of very simple models, allowing the project to streamline data conditioning and ultimate science data analysis. Conversely, should the orbital parameters be less ideal than assumed (as presented in Table~\ref{tb:params}; see also Appendix~\ref{sec:useful-rels}), the range and range rate models would have to be updated to include terms that we were able to omit, as their magnitudes were sufficiently small for the planned, ideal orbits. Of particular concern are several omitted terms in the range rate model (\ref{eq:range-rate}), namely the $1/c^2$ Sagnac term given by (\ref{eq:term-c-3.1}), the Shapiro term discussed in (\ref{eq:shap1}), and the quadrupole term, evaluated in (\ref{eq:j2-eval}). If needed, the omitted terms can be easily identified and reinstated by retracing the computational steps presented in this paper.

The analysis of relativistic effects presented here dealt mainly with the magnitudes of the largest terms of these contributions. These are either constant or vary at the orbital frequency (i.e., 1/rev). Our analysis can be used to validate the choices made in mission design, including decisions concerning terms that are assumed to be removable in the form of an empirically introduced bias. Moreover, our results may be easily extended to analyze the contributions of other periodic terms that come at twice the orbital frequency, i.e., 2/rev, and higher. The analytical models that we present for the range and range rate are, therefore, also applicable in situations that take the mission outside the empirical domain, for instance, when optimal orbital parameters are not achieved and the aforementioned empirical bias can no longer be reliably used.

In addition to developing a model for relativistic contributions to the two-way GRACE-FO LRI observable, we also considered the possibility that the laser-ranging instrument will be used in a DOWR configuration. In this operating mode, the transmitters on the two spacecraft are no longer operating in coherent mode, and the quality of the observable relies on the frequency stability of the two independent laser oscillators on board the twin spacecraft.

As a part of our analysis, we identified an intriguing possibility that may lead to an improvement in the accuracy of the Eddington parameter $\gamma$. As we discussed in Sec.~\ref{sec:phasemag}, if the GRACE-FO laser interferometer instrument achieves a 1~nm range resolution, this experiment could improve the accuracy of the estimate of the Eddington's relativistic parameter $\gamma$ to $5.3\times 10^{-6}$, which exceeds by a factor of 5 or more the accuracy of the current best estimate for $\gamma$, provided by the Cassini mission to Saturn \cite{Bertotti-Iess-Tortora:1999}. This possibility is truly remarkable; it is indicative  of the high level of accuracy achieved by modern geodesy when an engineering team needs to account for a number of general relativistic effects in order to reach the stated science objectives of an Earth-orbiting mission.

The analysis presented in this paper was conducted using an idealized set of conditions without considering noise. It can nonetheless be used to study the propagation of various forms of noise through the GRACE-FO architecture and investigate the impact of various noise sources on future science investigations with GRACE-FO. The formulation for the GRACE-FO LRI observables presented here allows for the direct introduction of noise terms (such as the frequency stability of the on-board laser oscillator or the accuracy of orbit determination), as part of an investigation of the noise contribution to the ultimate accuracy of the experiment.

One of the principal objectives of GRACE-FO is to monitor seasonal changes in the Earth's gravitational field using long-duration data spans. In this analysis, the difference between the orbital frequency and the frequency of thermal noise due to solar heating can be significant, as it can introduce an unmodeled annual term that can mask or enhance estimates of annual variability. Therefore, it is important to model the corresponding effects correctly and disentangle relativistic, thermal, and instrumental effects.

Concluding, we emphasize that the eikonal-based approach and the corresponding solution for the phase of a plane electromagnetic wave developed in Sec.~\ref{sec:em-phase} provides an efficient way to account for general relativistic effects on the propagation of light in the solar system, as needed, for instance, for the high-precision astrometric campaign recently initiated by the ESA's Gaia mission \cite{Klioner:2003} (see details at {\tt http://sci.esa.int/gaia/}). Also, the new approach we presented in Sec.~\ref{sec:LRI-obs} makes it possible to develop a highly accurate description of signal (re)transmission in the post-Newtonian approximation of a gravitational theory. This approach is based on the generally covariant notion of the phase of a plane wave and allows one to formulate observables in the proper reference frames of the transmitting and receiving spacecraft. The results are applicable to other past and planned missions with similar observables, notably to GRACE and GRAIL missions \cite{Turyshev:2012nw,Konopliv-etal:2013} and new space gravimetry missions like Satellite-to-Satellite Interferometry (SSI) mission \cite{Cesare-etal:2006} that is currently studied. Furthermore, our approach can be readily applied in other contexts, including experiments conducted away from the Earth, that involve high-precision measurements between multiple spacecraft relying on precision phase measurements done with microwave signals \cite{Asmar2005} or laser ranging interferometry (or both): for instance, the formulation of the TDIR observables for the LISA mission \cite{Tinto:2004yw,Shaddock:2004ua}. Such possibilities will be investigated and the results will be reported elsewhere.

\begin{acknowledgments}
We thank William F. Folkner, Gerhard L. Kruizinga, Robert Spero, Michael M. Watkins, and Dah-Ning Yuan of JPL for their interest and many useful comments provided during the work and preparation of this manuscript. This work was performed at the Jet Propulsion Laboratory, California Institute of Technology, under a contract with the National Aeronautics and Space Administration.
\end{acknowledgments}


\appendix

\section{The post-Newtonian approximation of the general relativity}
\label{sec:GR-background}

General relativity represents gravitation as a tensor field with universal coupling to the particles and fields of the Standard Model. It describes gravity as a universal deformation of the flat spacetime Minkowski metric, $\gamma_{mn}$:
\begin{equation}
g_{mn}(x^k)=\gamma_{mn}+h_{mn}(x^k).
\label{eq:gmn}
\end{equation}
The theory can be defined by postulating the action describing the gravitational field and its coupling to matter fields.
The propagation and self-interaction of the gravitational field are described by the action
\begin{eqnarray}
\label{eq:hilb-ens-action}
{\cal S}_{\rm G}[g_{mn}]
&=&\frac{c^4}{16\pi G}\int d^4x
\sqrt{-g} R,
\end{eqnarray}
where $G$ is Newton's gravitational constant, $g^{mn}$ is the inverse of $g_{mn}$, $g=\det g_{mn}$ and $R$ is the trace of the Ricci tensor.

The universal, minimal coupling of $g_{mn}$ to all matter fields of the Standard Model of particle physics is accomplished by using it to replace the Minkowski metric everywhere \cite{Turyshev-2008ufn}. By varying the total action ${\cal S}_{\rm tot}[\psi,A_m,H;g_{mn}]={\cal S}_{\rm G}[g_{mn}]+{\cal S}_{\rm SM}[\psi, A_m, H; g_{mn}],$ with respect to $g_{mn}$, one obtains Einstein's gravitational field equations:
\begin{equation}
R^{mn}= \frac{8\pi G}{c^4}\Big(T^{mn}-\frac{1}{2} g^{mn} T\Big).
\label{eq:GR-eq}
\end{equation}

Equation (\ref{eq:GR-eq}) connects the geometry of a 4-dimensional Riemannian manifold (spacetime), represented here by the Ricci-tensor $R^{mn}$, to the matter content of that spacetime, represented by a symmetric and conserved energy-momentum tensor, $T^{mn}=(2/\sqrt{-g})\delta {\cal L}_{\rm SM}/\delta g_{mn}$, which obeys the covariant conservation equation
\begin{equation}
\nabla_k\big(\sqrt{-g}T^{mk}\big)=0.
\label{eq:T-conserv}
\end{equation}

The theory is invariant under arbitrary coordinate transformations: $x^{\prime m}=f^m(x^n)$. This freedom to choose coordinates allows for the introduction of gauge conditions that may offer some technical convenience in solving the field equations~(\ref{eq:GR-eq}). For instance, in analogy with the Lorenz gauge of electromagnetism, the harmonic gauge  \cite{Fock-book:1959} is often used. It corresponds to imposing the condition,
\begin{equation}
\partial_n \big(\sqrt{-g}g^{mn}\big) = 0.
\label{eq:gauge-c}
\end{equation}

To solve the equations of the general theory of relativity, one assumes that the spacetime is asymptotically flat and there is no gravitational radiation coming from outside the body.
In terms of perturbations of the Minkowski metric, $h_{mn}$,
introduced in Eq.~(\ref{eq:gmn}), this amounts to introducing the following two boundary conditions \cite{Fock-book:1959}:
\begin{equation}
\lim_{\substack{r\rightarrow\infty\\t+r/c={\rm const}}} h_{mn}=0
~~~~~~{\rm and}~~~\lim_{\substack{r\rightarrow\infty\\t+r/c={\rm const}}}
\big[(rh_{mn})_{,r}+(rh_{mn})_{,0}\big]=0,
\label{eq:ass-flat}
\end{equation}
where $r$ represents the spatial distance, $t$ the time relative to the origin of a coordinate system.

In the weak gravitational field limit, these solutions can be expressed as perturbations of the flat spacetime Minkowski metric, in the post-Newtonian approximation.
The solution to Einstein's field equations in the post-Newtonian approximation that is sufficient to describe the gravitational field in the solar system has the form (see also \cite{Turyshev-Toth:2013} and references therein):
{}
\begin{eqnarray}
g_{00}&=&1-\frac{2w}{c^2}+\frac{2w^2}{c^4} + {\cal O}(c^{-6}),
\hskip 20pt
g_{0\alpha}=-\gamma_{\alpha\lambda}\frac{4w^\lambda}{c^3}+ {\cal O}(c^{-5}),
\hskip 20pt
g_{\alpha\beta}=\gamma_{\alpha\beta}+\gamma_{\alpha\beta}\frac{2w}{c^2} + {\cal O}(c^{-4}),\label{eq:gab1B*}
\end{eqnarray}
{}
where the scalar and vector gravitational potentials $w$ and $w^\alpha$ are determined from the following harmonic equations:
{}

\begin{eqnarray}
\Box w&=&4\pi G \sigma + {\cal O}(c^{-4}),
\hskip 20pt
\Delta w^\alpha= 4\pi G\sigma^\alpha + {\cal O}(c^{-2}),
\label{eq:eq-w^a*}
\end{eqnarray}
where we have introduced the scalar, $\sigma$, and vector, $\sigma^\alpha$, densities connected to the energy-momentum tensor:
{}
\begin{eqnarray}
\sigma &=& c^{-2}(T^{00} -
\gamma_{\mu\lambda}T^{\mu\lambda}) + {\cal O}(c^{-4}),
\hskip 20pt \sigma^\alpha =  c^{-1}T^{0\alpha} + {\cal O}(c^{-3}).
\label{eq:(sig_a)=}
\end{eqnarray}
The energy-momentum conservation $\nabla_k\big(\sqrt{-g}T^{mk}\big)=0$ together with the harmonic gauge conditions (\ref{eq:gauge-c}) lead to the existence of the Newtonian continuity equations first for the matter densities
$c\partial_0 \sigma+\partial_\epsilon \sigma^\epsilon=  {\cal O}(c^{-2})$ and then for the gravitational potentials
$c\partial_0 w+\partial_\epsilon w^\epsilon=  {\cal O}(c^{-2})$; see details in \cite{Turyshev-Toth:2013}.

A general solution for $w$ and $w^\alpha$ of Eq.~(\ref{eq:eq-w^a*}) satisfying the asymptotic flatness condition (\ref{eq:ass-flat}) can be written in terms of advanced and retarded potentials. The recommended solution \cite{Soffel:2003cr}, half advanced and half retarded,  reads:
{}
\begin{eqnarray}
w(t,\vec{x})&=&w_0(t,\vec{x})+G\int \frac{\sigma(t,\vec{x}')d^3x'}{|\vec{x}-\vec{x}'|} + \frac{1}{2c^2}G\frac{\partial^2}{\partial t^2}\int d^3x'
{\sigma(t,\vec{x}')}{|\vec{x}-\vec{x}'|}+{\cal O}(c^{-3}),
\label{eq:w*}\\
{}
w^\alpha(t,\vec{x})&=& w^\alpha_0(t,\vec{x})+ G\int \frac{\sigma^\alpha(t,\vec{x}')d^3x'}{|\vec{x}-\vec{x}'|}+{\cal O}(c^{-2}),
\label{eq:w^a*}
\end{eqnarray}
where $w_0$ and $w^\alpha_0$ are the solutions of the homogeneous equations: $\Box w_0={\cal O}(c^{-3})$ and $\Delta w^\alpha_0={\cal O}(c^{-2})$.

Equations (\ref{eq:gab1B*}) and  (\ref{eq:w*})--(\ref{eq:w^a*}) represent a well-known solution to the gravitational field equations of the general theory of relativity in the post-Newtonian approximation  \cite{MTW,Will_book93,Soffel:2003cr,Turyshev-2008ufn,Turyshev-Toth:2013}. We use these expressions to study light propagation in the vicinity of an extended gravitating body.

\section{Instantaneous distances between the spacecraft}
\label{sec:ranges}

In Section \ref{sec:pseudo-ranges} we obtained the light-cone equations that depend on two instants of time -- the time of a signal's emission and the time of its reception. Clearly, one may use either one of the two instants, as the second one is determined by the light cone. In particular, we observe that Eq.~(\ref{eq:t-AB}) can be used to express either $t_{\rm 1}$ as a function of $t_{\rm 2}$ or vice versa. Similar to KBR observables that are  timestamped using the time of reception (that is, $t_{\rm 2}$), a one-way LRI observable would be formed at the time of signal reception (in this case, on spacecraft $B$). To reflect this fact, the Euclidean range, $R_{\rm AB}\big({\vec x}_{\rm A}(t_1),{\vec x}_{\rm B}(t_{\rm 2})\big)=|{\vec x}_{\rm B}(t_{\rm 2}) - {\vec x}_{\rm A}(t_{\rm 1})|$, gets modified by Sagnac correction terms (as observed in Ref.~\cite{Blanchet-etal:2001} and also developed in Ref.~\cite{Turyshev:2012nw}) consistently to the order $1/c^3$:
{}
\begin{equation}
R_{\rm AB}\big({\vec x}_{\rm A}\big(t_1(t_2)\big),{\vec x}_{\rm B}(t_{\rm 2})\big) = d_{\rm AB}+\frac{1}{c}({\vec d}_{\rm AB}\cdot {\vec v}_{\rm A})
+\frac{d_{\rm AB}}{2c^2}\Big({\vec v}_{\rm A}^2+({{\vec n}}_{\rm AB}\cdot {\vec v}_{\rm A})^2-({\vec d}_{\rm AB}\cdot {\vec a}_{\rm A})\Big)+{\cal O}(c^{-3}),
\label{eq:nt-AB+**}
\end{equation}
where ${\vec d}_{\rm AB} = {\vec x}_{\rm B}(t_{\rm 2})-{\vec x}_{\rm A}(t_{\rm 2})$ is the ``instantaneous'' Euclidean coordinate distance between $A$ and $B$ at the instant of reception at $B$ (we have $d_{\rm AB} = |{\vec d}_{\rm AB}|$ and ${{\vec n}}_{\rm AB}={\vec d}_{\rm AB}/d_{\rm AB}$), where ${\vec v}_{\rm A}={\vec v}_{\rm A}(t_{\rm 2})$ and ${\vec a}_{\rm A}={\vec a}_{\rm A}(t_{\rm 2})$ denote the coordinate velocity and  acceleration of spacecraft $A$ correspondingly, both taken at $t_{\rm 2}$. With the help of Eq.~(\ref{eq:t-AB}), we determine the following expression for the instantaneous delay between the two spacecraft measured at time $t_{\rm 2}$:
{}
\begin{eqnarray}
t_{\rm 2} - t_{\rm 1}&=& c^{-1}{\cal R}_{\rm AB}\Big({\vec x}_{\rm A}\big(t_1(t_2)\big),{\vec x}_{\rm B}(t_{\rm 2})\Big)\equiv c^{-1}{\cal R}_{\rm AB}(t_{\rm 2}),
\label{eq:T_ABb}
\end{eqnarray}
with the instantaneous light travel distance ${\cal R}_{\rm AB}(t_{\rm 2})$ expressed as
{}
\begin{eqnarray}
{\cal R}_{\rm AB}(t_{\rm 2}) &=& d_{\rm AB}+\frac{1}{c}({\vec d}_{\rm AB}\cdot {\vec v}_{\rm A})
+\frac{d_{\rm AB}}{2c^2}\Big({\vec v}_{\rm A}^2+({{\vec n}}_{\rm AB}\cdot {\vec v}_{\rm A})^2-({\vec d}_{\rm AB}\cdot {\vec a}_{\rm A})\Big)+ \nonumber\\
&+& \frac{2GM_{\rm E}}{c^2}\Big\{\ln\Big[\frac{r_{\rm A}+r_{\rm B}+d_{\rm AB}}{r_{\rm A}+r_{\rm B}-d_{\rm AB}}
\Big(1+\frac{2}{c}\frac{({\vec d}_{\rm AB}\cdot {\vec v}_{\rm A})}{(r_{\rm A}+r_{\rm B})}\Big)\Big]-\nonumber\\
&-&
{\textstyle\frac{1}{6}}\Big[\frac{(n_{\rm B\epsilon}+k_\epsilon)(n_{\rm B\lambda}+k_\lambda)}{(r_{\rm B}+{\vec k}\cdot{\vec r}_{\rm B})^2}
+\frac{1}{r_{\rm B}}\frac{\gamma_{\epsilon\lambda}+n_{\rm B\epsilon} n_{\rm B\lambda}}{(r_{\rm B}+{\vec k}\cdot{\vec r}_{\rm B})}-
\frac{(n_{\rm A\epsilon}+k_\epsilon)(n_{\rm A\lambda}+k_\lambda)}{(r_{\rm A}+{\vec k}\cdot{\vec r}_{\rm A})^2}
-\frac{1}{r_{\rm A}}\frac{\gamma_{\epsilon\lambda}+n_{\rm A\epsilon} n_{\rm A\lambda}}{(r_{\rm A}+{\vec k}\cdot{\vec r}_{\rm A})}\Big]J_{\rm E}^{\epsilon\lambda}\Big\},~~~
\label{eq:nt-BA+**}
\end{eqnarray}
where all quantities here are taken at the instant of reception $t_{\rm B}$. The second term in Eq.~(\ref{eq:nt-BA+**}) represents the Sagnac term of order $1/c$. Taking $d_{\rm AB}=270$~km (which yields $d_{\rm AB}/c\sim900.6~\mu$s) and, using Eq.~(\ref{eq:nAB.vA}), it was estimated to be 6.89~m (or $\sim$22.97~ns). The third Sagnac term, of order $1/c^2$, is 175.5~$\mu$m (or $\sim$0.59~ps), comparable to the Earth's Shapiro delay term, which is 351.3~$\mu$m (or $\sim$1.17~ps). The Sagnac-type contribution term in the Shapiro delay $\sim \ln [1+2({\vec d}_{\rm AB}\cdot {\vec v}_{\rm A})/c(r_{\rm A}+r_{\rm B})]$  contributes 8.96 nm (or $\sim$0.0298~fs) and must be kept in the model.  Furthermore, the quadrupole term was evaluated, with the help of Eq.~(\ref{eq:J2-val}),  to be 167~nm (or $\sim$0.557~fs).

Similarly, for the return leg in case of a two-way transmission, we can express ${\cal R}_{\rm BA}$ in (\ref{eq:t-BA2}) as a function of the coordinate time of reception $t_{\rm 3}$. Thus, for a signal, emitted at spacecraft $B$ at time $t_{\rm 2}$ and received at spacecraft $A$ at $t_{\rm 3}$, we can write (\ref{eq:transm-time:BA}) as given below:
{}
\begin{eqnarray}
t_{\rm 3} - t_{\rm 2}&=& c^{-1}{\cal R}_{\rm BA}\Big({\vec x}_{\rm B}\big(t_{\rm 2}(t_3)\big),{\vec x}_{\rm A}(t_{\rm 3})\Big)\equiv c^{-1}{\cal R}_{\rm BA}(t_{\rm 3}),
\label{eq:T_BA2}
\end{eqnarray}
with the instantaneous one-way light travel distance ${\cal R}_{\rm BA}(t_{\rm 3})$ is given as
\begin{eqnarray}
{\cal R}_{\rm BA}(t_{\rm 3}) &=& d_{\rm AB}-\frac{1}{c}({\vec d}_{\rm AB}\cdot {\vec v}_{\rm B})
+\frac{d_{\rm AB}}{2c^2}\Big({\vec v}_{\rm B}^2+({{\vec n}}_{\rm AB}\cdot {\vec v}_{\rm B})^2+({\vec d}_{\rm AB}\cdot {\vec a}_{\rm B})\Big)+ \nonumber\\
&+&
\frac{2GM_{\rm E}}{c^2}\Big\{\ln\Big[\frac{r_{\rm A}+r_{\rm B}+d_{\rm AB}}{r_{\rm A}+r_{\rm B}-d_{\rm AB}}\Big(1-\frac{2}{c}\frac{({\vec d}_{\rm AB}\cdot {\vec v}_{\rm B})}{(r_{\rm A}+r_{\rm B})}\Big)\Big]-\nonumber\\
&-&
{\textstyle\frac{1}{6}}\Big[\frac{(n_{\rm A\epsilon}-k_\epsilon)(n_{\rm A\lambda}-k_\lambda)}{(r_{\rm A}-{\vec k}\cdot{\vec r}_{\rm A})^2}+\frac{1}{r_{\rm A}}\frac{\gamma_{\epsilon\lambda}+n_{\rm A\epsilon} n_{\rm A\lambda}}{(r_{\rm A}-{\vec k}\cdot{\vec r}_{\rm A})}-\frac{(n_{\rm B\epsilon}-k_\epsilon)(n_{\rm B\lambda}-k_\lambda)}{(r_{\rm B}-{\vec k}\cdot{\vec r}_{\rm B})^2}
-\frac{1}{r_{\rm B}}\frac{\gamma_{\epsilon\lambda}+n_{\rm B\epsilon} n_{\rm B\lambda}}{(r_{\rm B}-{\vec k}\cdot{\vec r}_{\rm B})}\Big]J_{\rm E}^{\epsilon\lambda}\Big\}.~~~
\label{eq:nt-BA2}
\end{eqnarray}

In addition, we may need an expression for the instantaneous light travel distance ${\cal R}_{\rm AB}(t_{\rm 2})$ expressed as a function of $t_{\rm 3}$ as ${\cal R}_{\rm AB}(t_{\rm 2})\equiv{\cal R}_{\rm AB}(t_{\rm 2}(t_3))$. Using (\ref{eq:T_BA2}) and (\ref{eq:nt-BA2}), from (\ref{eq:nt-BA+**}) we obtain
{}
\begin{eqnarray}
{\cal R}_{\rm AB}(t_{\rm 2}) &\equiv&{\cal R}_{\rm AB}\big(t_3-c^{-1}{\cal R}_{\rm BA}(t_{\rm 3})\big)={\cal R}_{\rm AB}(t_{\rm 3}) -\frac{1}{c}({\vec d}_{\rm AB}\cdot {\vec v}_{\rm AB})+
\frac{d_{\rm AB}}{2c^2}\Big({\vec v}_{\rm AB}^2+({\vec d}_{\rm AB}\cdot {\vec a}_{\rm AB})-
\nonumber\\
&-&
2({{\vec v}}_{\rm AB}\cdot {\vec v}_{\rm A})-
2({\vec d}_{\rm AB}\cdot {\vec a}_{\rm A})+
({{\vec n}}_{\rm AB}\cdot {\vec v}_{\rm AB})^2+
2({{\vec n}}_{\rm AB}\cdot {\vec v}_{\rm AB})({{\vec n}}_{\rm AB}\cdot {\vec v}_{\rm A})\Big)
+{\cal O}(c^{-3}),~~~~~
\label{eq:nt-BA+*=}
\end{eqnarray}
where all quantities here are taken at the instant of the reception time $t_{\rm 3}$. Note that this expression would formally have a kinematic Sagnac-type contribution in the Shapiro delay, that comes from transforming $t_{\rm 2}$ into $t_{\rm 3}$ via (\ref{eq:T_BA2})
\begin{eqnarray}
\frac{2GM_{\rm E}}{c^2}\ln\Big[1-\frac{2}{c}\frac{({\vec d}_{\rm AB}\cdot {\vec v}_{\rm AB})}{(r_{\rm A}+r_{\rm B})}\Big]\approx
\frac{2GM_{\rm E}}{c^3}\sqrt{\frac{GM_{\rm E}}{a}}\frac{d^2_{\rm AB}}{a^2}e=3.5\times 10^{-13}~{\rm m}\cdot\sin\omega_{\rm G}t,
\label{eq:shap-corr}
\end{eqnarray}
where we used (\ref{eq:dAB.vAB}). Clearly, this term is too small for our purposes and, thus, it was omitted.

\section{Contribution of the Earth's relativistic quadrupole moment}
\label{sec:need-for-J2}

GRACE-FO will rely on three standard coordinate systems: the GCRS, which is centered at the Earth's center of mass and is used to track orbits in the vicinity of the Earth (discussed in Sec~\ref{sec:GCRS}); the Topocentric Coordinate Reference System (TCRS), which is used to provide the positions of objects on the surface of the Earth, such as DSN ground stations; and the Satellite Coordinate Reference System (SCRS), which is needed for proper-to-coordinate time transformations. Definition and properties of TCRS together with useful details on relativistic time-keeping in the solar system are given in \cite{Turyshev-etal:2012}. The SCRS was discussed in \cite{Turyshev-etal:2012} in the context of the GRAIL mission. Here we investigate a need for an update for the standard general relativistic models for spacetime coordinates and equations of motion \cite{Petit-Luzum:2010}.

Refs.~\cite{Turyshev-Toth:2013,Turyshev-etal:2012}, show that transformations between the harmonic coordinates of the SCRS introduced on spacecraft $A$ and denoted here by $\{y^m_{\rm A}\}=\{c\tau_{\rm A}{\vec y}\}$, and coordinates of GCRS $\{x^m\}=\{ct, {\vec x}\}$, to sufficient accuracy, are  given by
{}
\begin{eqnarray}
\tau_{\rm A}&=& t-c^{-2}\Big\{\!\int_{t_{0}}^{t}\!\!\! \Big(
{\textstyle\frac{1}{2}}{\vec v}^2_{\rm A}+
U_{\rm E}+u^{\tt tidal}_{\rm E}\Big)dt'+({\vec v}^{}_{\rm A}\cdot {\vec r}^{}_{\rm A})\Big\}+{\cal O}({c^{-4}}),
\label{eq:trans++y0}\\[3pt]
{\vec y}_{\rm A}&=& {\vec r}_{\rm A}+
c^{-2}\Big\{{\textstyle\frac{1}{2}}{\vec v}_{\rm A} ({\vec v}_{\rm A}\cdot{\vec r}_{\rm A})+{\vec r}_{\rm A}U_{\rm E}+
{\vec r}^{}_{\rm A} ({\vec r}^{}_{\rm A}\cdot{\vec a}_{\rm A})-{\textstyle\frac{1}{2}}{\vec a}^{}_{\rm A}{r}^2_{\rm A}
\Big\}+{\cal O}(c^{-4}),
\label{eq:trans++ya}
\end{eqnarray}
where ${\vec r}_{\rm A}={\vec x} - {\vec x}_{\rm A0}$. The quantity $U_{\rm E}$ in (\ref{eq:trans++y0})--(\ref{eq:trans++ya}) is the Newtonian gravitational potential of the Earth (\ref{eq:pot_w_0}), evaluated at the location of the spacecraft. ${\vec v}_{\rm A}$ and ${\vec a}_{\rm A}$ are velocity and Newtonian acceleration of spacecraft $A$ in GCRS. Contribution of the tidal potential, $u^{\tt tidal}_{\rm E}$, is at most $4\times 10^{-20}$ and is negligible. The $c^{-4}$ terms in Eq.~(\ref{eq:trans++y0}) are of order $\sim v^4/c^4\simeq 10^{-19}$ and also negligible for GRACE-FO. The first two of the $1/c^2$ terms in (\ref{eq:trans++ya}) produce correction $\sim9.8\times 10^{-10}\cdot {\vec r}^{}_{\rm A}$, which, at $|r_{\rm A}|=d_{\rm AB}$, amounts to $265~\mu$m. The acceleration-dependent terms at $d_{\rm AB}$ amount to a correction on the order of $6~\mu$m. Thus, the terms in (\ref{eq:trans++ya}) are very small and are rarely used in a mission analysis; we present them here only for completeness. For a complete post-Newtonian form of these transformations, including the terms $c^{-4}$, and their explicit derivation, consult Ref.~\cite{Turyshev-Toth:2013}.

Eq.~(\ref{eq:trans++y0}) yields the differential equation (\ref{eq:proper-coord-t}) that relates the rate of the spacecraft proper $\tau_{\rm A}$ time, as measured by an on-board clock in Earth orbit, to the time in GCRS, $t$. Substituting in Eq.~(\ref{eq:proper-coord-t}) potential $U_{\rm E}$ from (\ref{eq:pot_w_0}), we have:
{}
\begin{eqnarray}
\frac{d\tau_{\rm A}}{dt}&=&
1-\frac{1}{c^2}\Big[\frac{{\vec v}^2_{\rm A}}{2}+
\frac{GM_{\rm E}}{r_{\rm A}}\Big(1-\sum_{\ell=2}^\infty\Big(\frac{R_{\rm E}}{r_{\rm A}}\Big)^\ell J_\ell P_{\ell0}(\cos\theta)-\sum_{\ell=2}^\infty\sum_{k=1}^{+\ell}\Big(\frac{R_{\rm E}}{r}\Big)^\ell P_{\ell k}(\cos\theta)(C^{{\rm E}}_{\ell k}\cos k\phi+S^{{\rm E}}_{\ell k}\sin k\phi)\Big)
\Big],~~~~~
\label{eq:prop-coord-time}
\end{eqnarray}
where $J_\ell$ are the zonal harmonics coefficients of the Earth mass distribution. Their contribution to (\ref{eq:prop-coord-time}) is expected to be the largest among all of the terms in the expression above.
Taking the values for these coefficients to be $J_2=1.08264\times 10^{-3}$, $J_3=-2.5326\times 10^{-6}$, $J_4=-1.61998\times 10^{-6}$, $J_5=2.1025\times 10^{-7}$, and $J_6=5.406878\times 10^{-7}$, we can estimate their contributions to (\ref{eq:prop-coord-time}). The anticipated range precision of $\Delta d=1$~nm implies a timing precision of the order of $\Delta d/d_{\rm AB}=3.7\times 10^{-15}$. We will use this number to evaluate the terms in (\ref{eq:prop-coord-time}). The largest contribution to $d\tau_{\rm A}/dt$, of course, comes from the velocity and mass monopole terms, which are estimated to produce an effect of the order of $c^{-2}({\textstyle\frac{1}{2}}{\vec v}^2_{\rm A}+G M_{\rm E}/r_{\rm A})\sim 9.80\times 10^{-10}$. The quadrupole term produces contribution of the order of $c^{-2}G M_{\rm E}/r_{\rm A}(R_{\rm E}/r_{\rm A})^2J_2\sim 6.14\times 10^{-13}$, which is large enough to be included in the model. Contributions of other zonal harmonics ranging from $\sim 1.33\times 10^{-15}$ (from $J_3$) to $\sim 2.33\times 10^{-16}$ (from $J_6$). Although their individual contributions are quite small to warrant their place in the model, their cumulative effect may be noticeable; this possibility will be further investigated. Therefore, for GRACE-FO in the (\ref{eq:prop-coord-time}) we must keep only the quadrupole term:
{}
\begin{eqnarray}
\frac{d\tau_{\rm A}}{dt}&=&
1-\frac{1}{c^2}\Big[\frac{{\vec v}^2_{\rm A}}{2}+
\frac{GM_{\rm E}}{r_{\rm A}}\Big(1+\frac{n_{\rm A\epsilon} n_{\rm A \lambda}}{2 r^3_{\rm A}}J^{\epsilon\lambda}_{\rm E}\Big)
\Big]+{\cal O}(1.33\times 10^{-15}),~~~~~
\label{eq:prop-coord-time-J2}
\end{eqnarray}
where the three-dimensional tensor of the quadrupole moment, $J^{\epsilon\lambda}_{\rm E}$, was introduced in (\ref{eq:mass-spin-moments}).

To determine the orbits of the spacecraft, one must also describe the propagation of electromagnetic signals between any two points in space.  The light-time equation corresponding to the metric tensor (\ref{eq:gab-E})--(\ref{eq:pot_loc-w_a+}) and written to the accuracy sufficient for GRACE-FO has the form given by (\ref{eq:tB-tA}) with ${\cal R}$ from (\ref{eq:t-AB}):

\begin{eqnarray}
t_2-t_1&=& \frac{1}{c}|{\vec x}(t_{\rm 2}) - {\vec x}(t_{\rm 1})|+(1+\gamma)\sum_b \frac{G M_b}{c^3}
\ln\left[\frac{r_1^b+r_2^b+r_{12}^b}{r_1^b+r_2^b-r_{12}^b}\right]-
\nonumber\\
&-&
\frac{GM_{\rm E}}{3c^2}\Big[\frac{(n_{\rm 2\epsilon}+k_\epsilon)(n_{\rm 2\lambda}+k_\lambda)}{(r_{\rm 2}+{\vec k}\cdot{\vec x}_{\rm 2})^2}+
\frac{1}{r_{\rm 2}}\frac{\gamma_{\epsilon\lambda}+n_{\rm 2\epsilon} n_{\rm 2\lambda}}{(r_{\rm 2}+{\vec k}\cdot{\vec x}_{\rm 2})}-
\frac{(n_{\rm 1\epsilon}+k_\epsilon)(n_{\rm 1\lambda}+k_\lambda)}{(r_{\rm 1}+{\vec k}\cdot{\vec x}_{\rm 1})^2}
-\frac{1}{r_{\rm 1}}\frac{\gamma_{\epsilon\lambda}+n_{\rm 1\epsilon} n_{\rm 1\lambda}}{(r_{\rm 1}+{\vec k}\cdot{\vec x}_{\rm 1})}\Big]J_{\rm E}^{\epsilon\lambda},
\label{eq:light-time}
\end{eqnarray}
where $t_1$ refers to the time instant of signal transmission and $t_2$ refers to the reception time, while  ${\vec x}(t_1)$ and ${\vec x}(t_2)$ are the geocentric positions of the transmitter and receiver. Also, $r^b_{1,2}$ are the distances of the transmitter and receiver from the body $b$ and $r^b_{12}$ is their spatial separation. The logarithmic contribution in (\ref{eq:light-time}) is the Shapiro gravitational time delay that, in the case of GRACE-FO, is mostly due to the Earth, the Moon, and the Sun. The last term is due to Earth's quadrupole whose presence extends the standard formulation given, for instance, in  \cite{Moyer:2003,Turyshev:2008}.

Finally, the relativistic geocentric equations of motion of a satellite that are recommended by IERS \cite{Petit-Luzum:2010} must be updated to include the contribution from the relativistic quadrupole moment of the Earth at the $1/c^2$ order. It is estimated that the corresponding $J^{\epsilon\lambda}_{\rm E}$-term in the equations of motion produces a contribution of the order of $\sim21~{\rm pm/s}^2$, which may still be noticeable in the orbits of the GRACE-FO spacecraft. However, the differential nature of the GRACE-FO architecture will further reduce the GRACE-FO sensitivity to such small accelerations by a factor of $d_{\rm AB}/a\approx 0.04$ thereby reducing the differential acceleration to less than 0.8 pm/s$^2$. Nevertheless, as there is still a controversy on the explicit form of the contribution of the Earth's oblateness to the $1/c^2$ terms in the relativistic equations of motion of an Earth-orbiting satellite, some additional work to unambiguously  identify this form is required.

\section{Useful relations for nearly identical Keplerian orbits}
\label{sec:useful-rels}

In this Appendix we introduce several useful relations that help in the evaluation of the magnitudes of various expressions that involve combinations of orbital quantities of the GRACE-FO spacecraft. We assume that both spacecraft orbit the Earth in the same planar orbit with identical semi-major axis, $a$, and  eccentricity, $e$, but different eccentric anomalies ${\cal E}_{\rm A}$ and ${\cal E}_{\rm B}$, correspondingly. Clearly, real mission analysis will involve navigational solutions based on the use of relativistic equations of motion for both spacecraft that are perturbed by the presence of  nongravitational forces acting on them which will result in slightly different orbits. Nevertheless, expressions below are useful for estimation purposes and, as such, they are used thought the paper.
Under these assumptions, expressions for the position vector ${\vec r}_{\rm A}$, velocity, ${\vec v}_{\rm A}=\dot{\vec r}_{\rm A}$, and acceleration ${\vec a}_{\rm A}=\ddot{\vec r}_{\rm A}$ of the spacecraft $A$, for example, are given as:
{}
\begin{eqnarray}
{\vec r}_{\rm A}&=&a\big(\cos {\cal E}_{\rm A}-e, \sqrt{1-e^2}\sin {\cal E}_{\rm A}\big), ~~~r_{\rm A}=a(1-e\cos {\cal E}_{\rm A}),
~~~
{\vec n}_{\rm A}=\Big(\frac{\cos {\cal E}_{\rm A}-e}{1-e\cos {\cal E}_{\rm A}}, \frac{\sqrt{1-e^2}\sin {\cal E}_{\rm A}}{1-e\cos {\cal E}_{\rm A}}\Big),~~~
\label{eq:kepler-r}\\
{\vec v}_{\rm A}&=&\frac{\sqrt{GMa}}{r_{\rm A}}\Big(-\sin {\cal E}_{\rm A}, \sqrt{1-e^2}\cos {\cal E}_{\rm A}\Big),~~~
{\vec a}_{\rm A}=-\frac{GM}{r_{\rm A}^2}\Big(\frac{\cos {\cal E}_{\rm A}-e}{1-e\cos {\cal E}_{\rm A}}, \frac{\sqrt{1-e^2}\sin {\cal E}_{\rm A}}{1-e\cos {\cal E}_{\rm A}}\Big),
\label{eq:kepler-va}
\end{eqnarray}
where the eccentric anomaly, ${\cal E}_{\rm A}$, is connected with the mean anomaly, ${\cal M}_{\rm A}$, as usual:
\begin{equation}
{\cal E}_{\rm A}(t)-e\sin {\cal E}_{\rm A}(t)={\cal M}_{\rm A}(t)={\cal M}_{\rm A0}+n(t-t_0), ~~~{\rm with}~~~ n=\sqrt{\frac{GM}{a^3}}~~~~{\rm and}~~~~a{\dot {\cal E}}_{\rm A}=\frac{\sqrt{GMa}}{r_{\rm A}}.
\label{eq:E}
\end{equation}
Similar expressions for spacecraft $B$ may be obtained simply by changing $A$ to $B$ in Eqs.~(\ref{eq:kepler-r})--(\ref{eq:E}).

Using Eqs.~(\ref{eq:kepler-r})--(\ref{eq:E}) we develop  a set of useful relations that will help determining the magnitudes of various effects in the range and range rate observables of GRACE-FO, namely:
{}
\begin{eqnarray}
({\vec r}_{\rm A}\cdot {\vec v}_{\rm A})&=&\sqrt{GMa}\,e\sin {\cal E}_{\rm A}, ~~~~
({\vec n}_{\rm A}\cdot {\vec v}_{\rm A})=
\sqrt{\frac{GM}{a}}\frac{e\sin {\cal E}_{\rm A}}{1-e\cos {\cal E}_{\rm A}}, \label{eq:rva0}\\
({\vec r}_{\rm A}\cdot {\vec a}_{\rm A})&=&-\frac{GM}{r_{\rm A}}, ~~~~
({\vec n}_{\rm A}\cdot {\vec a}_{\rm A})=-\frac{GM}{r^2_{\rm A}}, ~~~~
({\vec v}_{\rm A}\cdot {\vec a}_{\rm A})=-\frac{GM}{r^2_{\rm A}}\frac{\sqrt{GMa}}{r_{\rm A}}\,e\sin {\cal E}_{\rm A}.
\label{eq:rva}
\end{eqnarray}
We also have the following two relations:
{}
\begin{eqnarray}
\frac{d}{dt}\Big(\frac{{\vec v}^2_{\rm A}}{2}\Big)&=&\frac{d}{dt}\Big(\frac{GM}{r_{\rm A}}\Big)=-\frac{GM}{a^2}\sqrt{\frac{GM}{a}}\frac{e\sin{\cal E}_{\rm A}}{(1-e\cos{\cal E}_{\rm A})^3}.
\label{eq:tau_A}
\end{eqnarray}

Further, using expressions (\ref{eq:kepler-r})--(\ref{eq:E}) we determine the vector between the two spacecraft, ${\vec r}_{\rm AB}={\vec r}_{\rm B}-{\vec r}_{\rm A}$, as:
{}
\begin{eqnarray}
{\vec r}_{\rm AB}&=&
2a\sin\alpha_{\rm AB}\big(-\sin\beta_{\rm AB}, \sqrt{1-e^2}\cos\beta_{\rm AB}\big),
\label{eq:rAB}
\end{eqnarray}
where we introduced two angles, $\alpha_{\rm AB}$ and $\beta_{\rm AB}$, defined as
{}
\begin{eqnarray}
\alpha_{\rm AB}=\textstyle{\frac{1}{2}}({\cal E}_{\rm B}-{\cal E}_{\rm A}), ~~~\beta_{\rm AB}=\textstyle{\frac{1}{2}}({\cal E}_{\rm B}+{\cal E}_{\rm A}) ~~~~ {\rm or} ~~~~ {\cal E}_{\rm A}=\beta_{\rm AB}-\alpha_{\rm AB}, ~~~ {\cal E}_{\rm B}=\alpha_{\rm AB}+\beta_{\rm AB}.
\label{eq:r-albe}
\end{eqnarray}

Using Eqs.~(\ref{eq:rAB})--(\ref{eq:r-albe}), we develop the following expressions
{}
\begin{eqnarray}
r_{\rm AB}&=&2a\sin\alpha_{\rm AB}\,\sqrt{1-e^2\cos^2\beta_{\rm AB}}~~~{\rm and}~~~
{\vec n}_{\rm AB}=\frac{\vec{r}_{\rm AB}}{r_{\rm AB}}=
\frac{1}{\sqrt{1-e^2\cos^2\beta_{\rm AB}}}\Big(-\sin\beta_{\rm AB}, \sqrt{1-e^2}\cos\beta_{\rm AB}\Big).~~~
\label{eq:n-albe}
\end{eqnarray}
The equation for $r_{\rm AB}$ in Eq.~(\ref{eq:n-albe}) allows us to establish the exact expression for $\sin\alpha_{\rm AB}$:
{}
\begin{eqnarray}
\sin\alpha_{\rm AB} &=&\frac{r_{\rm AB}}{2a}\frac{1}{\sqrt{1-e^2\cos^2\beta_{\rm AB}}}.
\label{eq:al1}
\end{eqnarray}
Given the configuration of GRACE-FO, the angle $\alpha_{\rm AB}$ is small and is related to the orbital parameters of GRACE-FO as:
\begin{equation}
\alpha_{\rm AB}=\frac{d_{\rm AB}}{2a}+{\cal O}(d^3_{\rm AB},e^2)\approx 0.0198.
\label{eq:alpha}
\end{equation}
Such a small value would allow us to make simplifying approximations for various observable quantities on GRACE-FO. To estimate the numerical value for angle $\beta_{\rm AB}$, from Eq.~(\ref{eq:r-albe}) we will use the following expression:
{}
\begin{eqnarray}
\beta_{\rm AB}=\omega_{\rm G} t+{\cal O}(d_{\rm AB},e),
\label{eq:beta}
\end{eqnarray}
where $\omega_{\rm G}=\textstyle{\frac{1}{2}}(\omega_{\rm A}+\omega_{\rm B})+{\cal O}(e)\equiv (GM/a^3)^\frac{1}{2}+{\cal O}(e)$ being the mean orbital frequency of the GRACE-FO constellation.

Similarly to ${\vec r}_{\rm AB}$, we determine the velocity vector between the two spacecraft, ${\vec v}_{\rm AB}={\vec v}_{\rm B}-{\vec v}_{\rm A}$, as:
{}
\begin{eqnarray}
{\vec v}_{\rm AB}&=&
-\sqrt{\frac{GM}{a}}\frac{2\sin\alpha_{\rm AB}}{(1-e\cos {\cal E}_{\rm A})(1-e\cos {\cal E}_{\rm B})}\Big(\cos\beta_{\rm AB}-e\cos\alpha_{\rm AB},\, \sin\beta_{\rm AB}\Big)\approx 303~{\rm m/s}
\label{eq:vAB}
\end{eqnarray}

Expressions (\ref{eq:n-albe})--(\ref{eq:vAB}) allows us to compute the following dot product:
{}
\begin{eqnarray}
({\vec n}_{\rm AB}\cdot{\vec v}_{\rm AB})&=&
-\sqrt{\frac{GM}{a}}2\sin\alpha_{\rm AB}\Big(\frac{
e\cos\alpha_{\rm AB}\sin\beta_{\rm AB}+\big(\sqrt{1-e^2}-1\big)\sin\beta_{\rm AB}\cos\beta_{\rm AB}}{(1-e\cos {\cal E}_{\rm A})(1-e\cos {\cal E}_{\rm B})\sqrt{1-e^2\cos^2\beta_{\rm AB}}}\Big)\approx 30.3~{\rm cm/s}.
\label{eq:nAB.vAB}
\end{eqnarray}
and also
\begin{eqnarray}
({\vec r}_{\rm AB}\cdot{\vec v}_{\rm AB})&=&
-\sqrt{\frac{GM}{a}}4a\sin^2\alpha_{\rm AB}\Big(\frac{
e\cos\alpha_{\rm AB}\sin\beta_{\rm AB}+\big(\sqrt{1-e^2}-1\big)\sin\beta_{\rm AB}\cos\beta_{\rm AB}}{(1-e\cos {\cal E}_{\rm A})(1-e\cos {\cal E}_{\rm B})}\Big).
\label{eq:dAB.vAB}
\end{eqnarray}

Similarly, relying on the expressions derived above, we establish the following useful exact relations:
{}
\begin{eqnarray}
({\vec v}_{\rm A}\cdot{\vec v}_{\rm AB})&=&
-\frac{2GM}{a}\sin\alpha_{\rm AB}\Big(\frac{
\sin\alpha_{\rm AB}+e\cos\alpha_{\rm AB}\sin {\cal E}_{\rm A}+\big(\sqrt{1-e^2}-1\big)\sin\beta_{\rm AB}\cos {\cal E}_{\rm A}}{(1-e\cos {\cal E}_{\rm A})^2(1-e\cos {\cal E}_{\rm B})}\Big),
\label{eq:vA.vAB}\\
{}
({\vec a}_{\rm A}\cdot{\vec v}_{\rm AB})&=&
\frac{2GM}{a^2}\sqrt{\frac{GM}{a}}\frac{\sin\alpha_{\rm AB}}{(1-e\cos {\cal E}_{\rm A})^4(1-e\cos {\cal E}_{\rm B})}\Big(\cos\alpha_{\rm AB}(1+e^2)-\nonumber\\
&&\hskip 40pt -\,e\big(\cos\beta_{\rm AB}\sin {\cal E}_{\rm A}+\cos\alpha_{\rm AB}\cos {\cal E}_{\rm A}\big)+\big(\sqrt{1-e^2}-1\big)\sin\beta_{\rm AB}\sin {\cal E}_{\rm A}\Big),\\
\label{eq:aA.vAB}
{}
({\vec r}_{\rm A}\cdot{\vec n}_{\rm AB})&=&
-a
\frac{\sin\alpha_{\rm AB}(1-e^2\cos^2\beta_{\rm AB})-e\sin\beta_{\rm AB}(1-e\cos \alpha_{\rm AB}\cos\beta_{\rm AB})}{\sqrt{1-e^2\cos^2\beta_{\rm AB}}},
\label{eq:rA.nAB}\\
{}
({\vec v}_{\rm A}\cdot{\vec n}_{\rm AB})&=&
\sqrt{\frac{GM}{a}}
\frac{\cos\alpha_{\rm AB}-e^2\cos\beta_{\rm AB}\cos {\cal E}_{\rm A}}{(1-e\cos {\cal E}_{\rm A})\sqrt{1-e^2\cos^2\beta_{\rm AB}}},
\label{eq:nAB.vA}\\
{}
({\vec a}_{\rm A}\cdot{\vec n}_{\rm AB})&=&
\frac{GM}{a^2}\Big(\frac{
\sin\alpha_{\rm AB}\big(1-e^2\cos^2\beta_{\rm AB}\big)-e\sin\beta_{\rm AB}\big(1-e\cos\alpha_{\rm AB}\cos\beta_{\rm AB}\big)}{(1-e\cos {\cal E}_{\rm A})^3\sqrt{1-e^2\cos^2 \beta_{\rm AB}}}\Big),
\label{eq:aA.nAB}\\
{}
({\vec a}_{\rm A}\cdot{\vec r}_{\rm AB})&=&
\frac{2GM}{a}\sin\alpha_{\rm AB}\Big(\frac{
\sin\alpha_{\rm AB}\big(1-e^2\cos^2\beta_{\rm AB}\big)-e\sin\beta_{\rm AB}\big(1-e\cos\alpha_{\rm AB}\cos\beta_{\rm AB}\big)}{(1-e\cos {\cal E}_{\rm A})^3}\Big).
\label{eq:aA.rAB}
\end{eqnarray}

We also derive the expression for a relative acceleration between the two spacecraft ${\vec a}_{\rm AB}={\vec a}_{\rm B}-{\vec a}_{\rm A}$:
{}
\begin{eqnarray}
{\vec a}_{\rm AB}&=&-\frac{GM}{a^2}\Big(\frac{\cos {\cal E}_{\rm B}-e}{(1-e\cos {\cal E}_{\rm B})^3}-\frac{\cos {\cal E}_{\rm A}-e}{(1-e\cos {\cal E}_{\rm A})^3}, \frac{\sqrt{1-e^2}\sin {\cal E}_{\rm B}}{(1-e\cos {\cal E}_{\rm B})^3}-\frac{\sqrt{1-e^2}\sin {\cal E}_{\rm A}}{(1-e\cos {\cal E}_{\rm A})^3}\Big)\approx\nonumber\\
&\approx&
\frac{GM}{a^2}2\sin\alpha_{\rm AB}\big(\sin\beta_{\rm AB}, -\cos\beta_{\rm AB}\big)+
{\cal O}(e)\approx
\frac{GM}{a^2}\frac{d_{\rm AB}}{a}\big(\sin\omega_{\rm G}t, -\cos\omega_{\rm G}t\big)+
{\cal O}(e).
\label{eq:aAB}
\end{eqnarray}

Note that with the orbital eccentricity of $e=0.001$, both GRACE-FO spacecraft will have nearly circular orbits around the Earth. The form of the expressions above easily yields series of expansion with respect to $e$ (as we did in Eq.~(\ref{eq:aAB})), which, in addition to the smallness of $\alpha_{\rm AB}\approx0.0198$, as demonstrated by (\ref{eq:alpha}), will be useful to establish magnitudes of various effects contributing to the range and range rate observables of the GRACE-FO mission.

\end{document}